\documentclass{aa}
\usepackage{graphicx}
\usepackage{amsmath}
\usepackage{txfonts}
\usepackage{lipsum}
\usepackage{subcaption}         
\usepackage{lscape}             
\usepackage{placeins}          
\usepackage{color}

\usepackage{natbib}
\bibpunct{(}{)}{;}{a}{}{,} 

\newcommand{\be}{\begin{equation}}
\newcommand{\ee}{\end{equation}}

\def\M11{M_{11}}
\def\V100{V_{100}}
\def\R1{R_{Mpc}}
\def\T6{T_6}

\begin{document}

\title{Energy deposited by black holes in the hot X-ray gas of \\ elliptical galaxies, groups and clusters}

\author{A.~Cattaneo\inst{1} \and S.~Ettori\inst{2,3,1}}
\institute{Observatoire de Paris, LUX, PSL University, 61 avenue de l'Observatoire, 75014 Paris, France \and
INAF, Osservatorio di Astrofisica e Scienza dello Spazio, via Piero Gobetti 93/3, 40129 Bologna, Italy \and
INFN, Sezione di Bologna, viale Berti Pichat 6/2, 40127 Bologna, Italy}

\abstract
{Galaxy groups show entropy excesses $\Delta S$ with respect to theoretical expectations from models and simulations with purely gravitational heating.}
{We determine the heat $Q_{\rm heat}\simeq T\Delta S$ deposited by non-gravitational sources into the hot gas of elliptical galaxies, groups and clusters, and to compare it with the energy output of their central black holes (BHs).}
{We adopt a simple model for the hot gas with only one free parameter: the core entropy. We calibrate the core entropy on the observed relation between X-ray luminosity and halo mass, and we use it
to determine the entropy $S$ of the hot gas. We determine $\Delta S=S-S_0$ by taking the difference between $S$ and the entropy $S_0$ found in cosmological adiabatic simulations for haloes of the same mass. The temperature $T$ of the hot gas during the heating phase is determined from semi-analytic/semi-empirical modelling by assuming that heat absorption  traces the accretion histories of supermassive BHs.}
{Our findings suggest that  supermassive BHs thermalise 1 to $3\%$ of their energy output in the surrounding gas.} 
{Our results suggest two regimes:
(i) in elliptical galaxies and groups, active galactic nuclei (AGN) heat the gas much more rapidly than it can cool, generating the observed entropy excesses and quenching star formation when the accumulated heat is large enough to unbind the gas reservoir in the host halo; (ii) in galaxy clusters (at $M_{200}>10^{14}{\rm\,M}_\odot$), heating and cooling are in self-regulated equilibrium.}

\maketitle
\nolinenumbers

\section{Introduction}
\label{sect:introduction}

The hot gas in elliptical galaxies and in the intergalactic medium of galaxy groups and clusters radiates in X-rays with bolometric emissivity $n^2\Lambda$, where  $n$ is the particle number density. 
The cooling function $\Lambda$ depends on the temperature $T$ and the metallicity $Z$ of the hot gas, and it is the sum of two terms: a term from line cooling and a bremsstrahlung term proportional to $T^{1/2}$. 
The bolometric luminosity of the hot gas  within the characteristic  radius $r_{200}$ of the dark matter (DM) halo is:
\begin{equation}
L_{\rm bol}=4\pi\int_0^{r_{200}}n^2\Lambda(T,Z)r^2{\rm\,d}r
\label{Lbol}
\end{equation}
(see Appendix~\ref{sect:over} for definitions of all quantities with the subscript 200, 500, or vir, and for a glossary of all quantities in the article).

For purely gravitational heating,  
the properties of the hot gas on all mass scales, from elliptical galaxies to galaxy clusters, should obey a {\it self-similar} model, in which they are all simply rescaled versions of one another. In the self-similar model: i) the mass fraction $f_{\rm g}$ in hot gas is constant; ii) the dimensionless density profile $n(r)/n_{200}$ and the dimensionless temperature profile $T(r)/T_{200}$ are universal, in the sense that they are functions of $x=r/r_{200}$ only.
With these assumptions:
\begin{equation}
L_{\rm bol}=4\pi n_{200}^2r_{200}^3\int_0^1\left({n\over n_{200}}\right)^2\Lambda(T,Z)x^2{\rm\,d}x.
\label{Lbol2}
\end{equation}

At high $T$ (as in galaxy clusters), bremsstrahlung is the main cooling mechanism. For pure bremsstrahlung cooling, $\Lambda \propto T^{1/2}$. Hence:
\begin{equation}
L_{\rm bol}\propto n_{200}^2 r_{200}^3 T_{200}^{1\over 2}\propto f_{\rm g}^2T_{200}^2,
\label{Lbol3}
\end{equation}
since $n_{200}\propto f_{\rm g}$ and $r_{200}\propto T_{200}^{1/2}$ (Appendix~\ref{sect:over}). 

If we let $f_{\rm g}$ depend on $M_{\rm 200}\propto T_{200}^{3/2}$ through a power law $f_{\rm g}\propto M_{200}^\nu$, as in \citet{ettori15}, Eq.~(\ref{Lbol3}) becomes:
\begin{equation}
 L_{\rm bol}\propto T_{200}^{2+3\nu}\propto M_{200}^{{4\over 3}+2\nu}.
 \label{Lbol4}
 \end{equation}
The dependence of $f_{\rm g}$ on $M_{200}$ breaks the self-similarity of groups and clusters, but is not the only source of departure from self-similarity.
The dependence of $n/n_{200}$ on $x$ can vary, too.
We therefore take $\nu$ as an effective measure of the departure of groups and clusters from the self-similar relation, regardless of the precise way in which self-similarity is broken.

\begin{figure}
\begin{center}
\includegraphics[width=0.99\hsize]{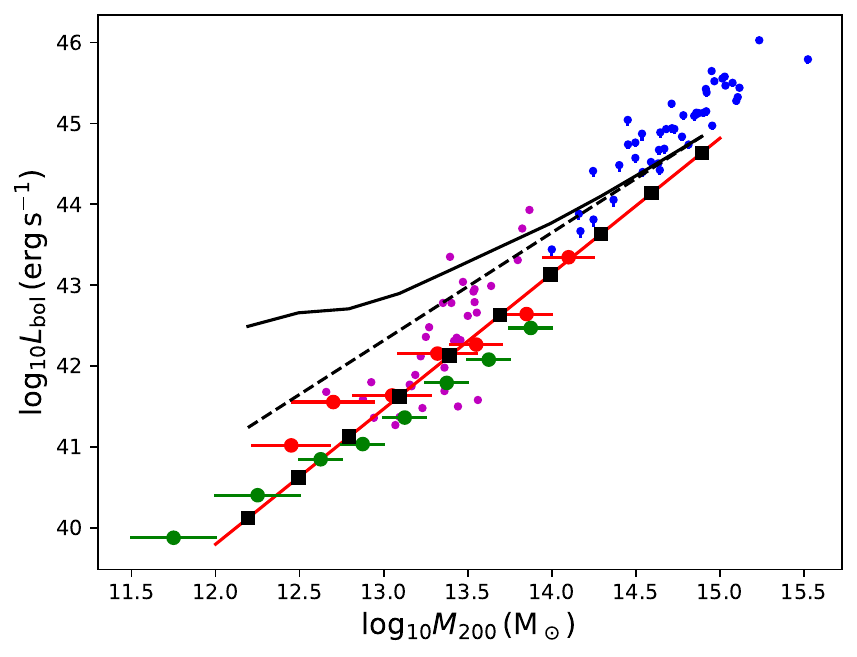} 
\end{center}
\caption{$L_{\rm bol}$--$M_{200}$ relation from stacked X-ray observations with eRosita. The red and green circles with error bars are the data points from \citet{popesso_etal24} and \citet{zhang_etal24}, respectively,
after applying the bolometric corrections $C_{0.5-2{\rm\,keV}}$  in Section~\ref{sect:LMR}.
The red line, referred throughout the article as the LMR, is the best fit to the red circles and corresponds to $\nu=0.17$.
The small blue circles show the clusters from \citet{zhu_etal21}, after applying the appropriate bolometric correction $C_{1-50{\rm\,keV}}$. 
The small magenta circles show the groups from \citet{heldson_ponman03}.
The solid black curve is the theoretical prediction of the self-similar model for the cooling function of \citet{sutherland_dopita93} with $Z=0.3{\rm\,Z}_\odot$, and $\eta=0.19$ at all masses.
The dashed black line $L_{\rm bol}\propto M_{200}^{4/3}$ shows the same model for cooling by bremsstrahlung alone.
The black squares show the $L_{\rm bol}$--$M_{200}$ relation predicted by our model after tuning $\eta$ to reproduce the mean LMR.
The $L_{200}$--$M_{200}$ relation for X-ray selected groups and clusters (magenta and blue circles, respectively)
has the same exponent on cluster scales (within the observational errors)
as the mean relation from stacked X-ray observations of optically selected systems
 but is steeper on group scales.
Already \citet{anderson_etal15} had remarked that there is no evidence for such steepening in optically selected samples.
}
\label{Fig1}
\end{figure}

Forty years of X-ray studies (\citealp{kaiser86}; \citealp{markevitch98}; \citealp{chen_etal07})
have consistently shown that that the $L_{\rm bol}$--$M_{200}$ relation for galaxy clusters is steeper 
than the self-similar scaling $L_{\rm bol}\propto M_{200}^{4/3}$ {(Fig.~\ref{Fig1}; also see \citealp{donahue_voit22} for a review)}.

The clusters in a recent study by Zhu et al. (2021; Fig.~\ref{Fig1}, {blue} circles) correspond to $\nu\simeq 0.16$.
In groups, \citet{heldson_ponman03} found an even steeper  relation with $\nu\simeq 0.34$ (Fig.~\ref{Fig1}, {magenta} points).
For elliptical galaxies, \citet{osullivan_etal03}, \citet{boroson_etal11} and \citet{goulding_etal16} found  similar exponents to \citet{heldson_ponman03}'s  but lower normalisations, 
which may derive from the difficulty of measuring X-ray fluxes from the outer regions of faint systems, where the surface brightness is lower than the X-ray background.

The discrepancy between the data and the predictions of the self-similar model is known as the group ``entropy problem'' \citep{ponman_etal99}.
To understand the origin of this name, consider that the entropy per particle of an ideal monoatomic gas is:
\begin{equation}
s={3\over 2}k{\rm\,ln}K={3\over 2}k{\rm\,ln}\left(Tn^{-{2\over 3}}\right).
\label{s} 
\end{equation}
Under the condition that $T\sim T_{200}$ (verified spectroscopically), lower 
number densities
correspond to higher specific entropies.
X-ray astronomers often call ``entropy'' the argument $K=Tn^{-2/3}$ of the logarithm in Eq.~(\ref{s}).
We shall always use inverted commas when we use ``entropy'' in this improper sense.
The total entropy $S$ is the integral of $s$ over all particles.

The importance of entropy (and the reason why this is known as the entropy rather than the density problem) is that the entropy of a system at thermodynamic equilibrium\footnote{The Joule expansion is a classic example of an irreversible thermodynamic transformation where the entropy increases but no heat is absorbed.
This is possible because there is a transient during which the gas is outside thermodynamic equilibrium and the equation of state for an ideal gas cannot be applied.}  
increases by ${\rm d}S$ if and only if the system
absorbs an amount of heat $\delta Q=T{\rm\,d}S$.
Entropy retains a record of the gas's thermal history.
If the entropy of the X-ray emitting gas is higher than expected from purely gravitational heating, there must be additional non-gravitational heating mechanisms (AGN, supernovae).
AGN and supernova feedback may cause a local, temporary increase in the gas temperature, but the gas will eventually relax to a temperature set by the gravitational potential well.
Ultimately, heating does not raise $T$. It causes the gas to expand, so that $n$ and $L_{\rm bol}$ decrease at nearly constant temperature (e.g. \citealp{mccarthy_etal10}).

Since \cite{voit_donahue05}, AGN are widely regarded as the most likely heat source.
To generate the same heat with supernovae, galaxy groups and clusters would have to form many more stars than observations allow,
whereas simulations with BH heating reproduce the entropies, stellar masses, and star formation rates of groups simultaneously \citep{mccarthy_etal10}.
One can thus use the observed entropy excesses to infer how much heat BHs have deposited into the hot gas.

\citet{zhu_etal21} studied the
``entropy'' profiles of 47 clusters.
They found that the heat injection required to increase the entropy from the predicted to the observed level is
 $Q_{\rm heat}\sim 0.02M_\bullet c^2$, where $M_\bullet$ is the mass of the central BH and $c$ is the speed of light. 
 Their study's
main limitations were: i) the {mass range}, limited to clusters, while the largest entropy excesses are on group and galactic scales (it is  hard to measure ``entropy'' profiles for low-mass systems, many of which are not even detected in X-rays), and
ii) the {X-ray selection}, which favours X-ray bright low-entropy cool-core clusters over X-ray dimmer high-entropy non-cool-core clusters 
(see \citealp{popesso07} and \citealp{andreon_etal24}
on how the selection affects the typical X-ray luminosity of the considered sample).

We overcome these limitations through the use of stacked eRosita X-ray observations
\citep{zhang_etal24,popesso_etal24}
 of optically selected systems from the GAMA galaxy group catalog \citep{robotham_etal11}.
 Using measurements derived from stacked X-ray observations, we can determine the $L_{\rm bol}$--$M_{200}$ relation over the entire mass range from elliptical galaxies to clusters ($10^{12}{\rm\,M}_\odot\lesssim M_{200}\lesssim10^{15}{\rm\,M}_\odot$).
Because it is derived from optically selected systems, this relation is free from X-ray selection biases and indeed has a lower normalization than the relations from X-ray studies (Fig.~\ref{Fig1}).
The exponents we derive from stacked X-ray observations of optically selected systems \citep{popesso_etal24} and X-ray selected samples \citep{zhu_etal21}
after the appropriate bolometric corrections (Section~\ref{sect:deltaS}) are, however, very similar.
They correspond to Eq.~(\ref{Lbol4}) with $\nu=0.17$ and $\nu=0.16$, respectively.

In this article, we use the $L_{\rm bol}$--$M_{200}$ relation that we derive from \citet{popesso_etal24}, hereafter simply referred to as the luminosity--mass relation (LMR):
\begin{enumerate}
    \item  To measure the entropy injection $\Delta S(M_{200})$ required for $L_{\rm bol}$ to descend from the self-similar relation onto the observed one (Section~\ref{sect:deltaS}).
    \item To infer the heat $Q_{\rm heat}$ injected by AGN into the hot gas by inverting the equation $\Delta S=\int{\delta Q/T}$, which provides the accumulated and residual entropy after accounting for all heat gains and losses (Section~\ref{sect:Qheat}).
\end{enumerate}
In Section~\ref{sect:results}, we present the results for
$Q_{\rm heat}$ obtained with the method detailed in Section~\ref{sect:Qheat}  and compare them
with the energy output of supermassive BHs
to determine the heating efficiency of AGN.
In Section~\ref{sect:disc}, we examine the uncertainties affecting our results and the broader implications of our findings for the role 
of BHs in the formation and evolution of galaxies. Finally, Section~\ref{sect:conc} summarises our conclusions.

\section{The calculation of $\Delta S$}
\label{sect:deltaS}

\subsection{Method}
\label{sect:method}

To compute $\Delta S=S-S_0$, we must determine the entropy $S$ associated to the systems in
\citet{popesso_etal24} and 
the baseline entropy $S_0$ for purely gravitational heating (both as functions of $M_{200}$).
In both cases, the entropy is calculated by integrating the entropy per particle over spherical shells:
\begin{equation}
S=4\pi\int snr^2{\rm\,d}r=6\pi k\int(\ln K) n r^2{\rm\,d}r.
\label{integrated_S}
\end{equation}
Equation~(\ref{integrated_S}) uses Eq.~(\ref{s}) to pass from $s$ to $K$.
Evaluating $S$ requires knowledge
of $K(r)$ and $n(r)$.

\citet{voit_etal05} looked at the radial ``entropy'' distribution  in 
cosmological adiabatic hydrodynamic simulations with the smoothed-particle-hydrodynamics (SPH) code {\sc GADGET} and the adaptive-mesh-refinement (AMR) code {\sc ENZO}. They could fit a baseline ``entropy'' profile $K_0(r)$ with the functional form:
\begin{equation}
K_0(r)=K_{200}\left[\eta_0+a\left({r\over r_{200}}\right)^b\right],
\label{K0}
\end{equation}
where $K_{200}=T_{200}n_{200}^{-2/3}$, $a=1.32$ and $b=1.1$ 
The ``central-entropy parameter''  $\eta_0=K_0(0)/K_{200}$ had a median value of $\eta_0\simeq 0.07$ in SPH simulations and $\eta_0\simeq 0.19$ in AMR simulations, with significant scatter in both cases.
{\citet{mitchell_etal09}}, \citet{valdarnini12}, \citet{power_etal14}, and \citet{biffi_valdarnini15} showed that the tension arose because SPH simulations without numerical dissipation struggle
to capture the development of Kelvin-Helmholtz instabilities.
Here, we assume a baseline value of $\eta_0=0.19$, consistent with AMR simulations. 

We further assume that the functional form in Eq.~(\ref{K0}), with the same $a$ and $b$ as in \citet{voit_etal05} but a different ``central-entropy parameter'' $\eta$, also applies 
to the real ``entropy'' profiles $K_\eta(r)$ of groups and clusters.
This assumption has both theoretical and observational justifications.
Theoretically, BH heating affects the core region but is not expected to modify the ``entropy'' distribution at large radii.
Observationally, the functional form in Eq.~(\ref{K0}) provides a reasonably good fit
to the ``entropy'' profiles of galaxy clusters, and the best-fitting $a$ and $b$ are fairly similar to those of Voit et al. \citep{zhu_etal21}. \citet{ghirardini_etal19} found
that the exponent of $K$ is consistent with $b=1.1$ at large radii.

With our assumption, all that AGN feedback does is to increase the ``central-entropy parameter'' $\eta$.
Hence, the calculation of $\Delta S$ boils down to determining how high 
$\eta$ is compared to the baseline value $\eta_0=0.19$.

The problem, brushed aside by previous studies, is that entropy variations affect
not only $K(r)$ but also $n(r)$. Previous studies (e.g. \citealp{zhu_etal21,eckert_etal25})
had usually assumed that the gas is heated isochorically, so that
the observed $n(r)$ also applies to the baseline configuration.
With this assumption, $\Delta S$ is readily found from Eq.~(\ref{integrated_S}).
All one has to do is to replace $\ln K$ with $\ln(K/K_0)$.
In reality, heating causes the gas to expand. Hence, the gas in the spherical shell between $r$ and $r+{\rm d}r$ is not the same before and after the heat injection.

To overcome this problem, we need a model for $n_\eta(r)$, i.e. a model for the density distribution of the hot gas that considers its dependence on $\eta$.
This model will also give us $T_\eta(r)=K_\eta(r)n_\eta^{2/3}(r)$ and allow us to compute $L_{\rm bol}$  
(Eq.~\ref{Lbol2}).

\subsection{The density and temperature distribution}

Our calculation of the gas density $n(r)$ is based on the assumption that the gas follows a polytropic equation of state, $T\propto n^{\gamma-1}$, where $\gamma$ is not the adiabatic index $5/3$ for an ideal monoatomic gas but a {\it phenomenological} polytropic index to be determined from observations.
A physical polytropic index indicates how the temperature of a gas element changes when that element is compressed. Instead, $\gamma$  tells us how $T$ increases when we move from lower $n$ elements at larger radii to denser elements at smaller radii.

Our approach goes back to \citet{komatsu_seljak01}, who solved the hydrostatic equilibrium equation for a polytropic gas in the gravitational potential of a DM halo described by the density distribution of \citet*[NFW]{navarro_etal97}. 
\citet{komatsu_seljak01} constrained $\gamma$ by requiring that the gas should follow the same profile as the DM at large radii and
found $\gamma\simeq 1.15+0.01(c_{\rm vir}-6.5)$ where $c_{\rm vir}$ is the virial concentration of the NFW model.

Direct measurements of $\gamma$ are now available from the pressure-density relation in the intracluster medium (ICM).
\citet{ghirardini_etal19} found that the ICM of the X-COP clusters follows a polytropic behaviour $p\propto n^\gamma$ that spans over three orders of magnitude in density.
They obtained the best joint overall fit to the density, pressure, temperature, and entropy profiles for  $\gamma = 1.19\pm 0.02$, although $\gamma$ appears to decrease in the central region, where measurements are more consistent with $\gamma\simeq 1.13$ (see the green data points with error bars on Fig.~4 of \citealp{ghirardini_etal19}). 
In this article, we assume $\gamma = 1.19\pm 0.02$ based on \citet{ghirardini_etal19}, but we  also consider an observationally disfavoured isothermal model with $\gamma=1$ as a way to test the sensitivity
of our results to our assumed polytropic model.

The polytropic assumption gives $K\propto n^{\gamma-5/3}$, $n\propto K^{3/(3\gamma-5)}$ and $T\propto K^{(3\gamma-3)/(3\gamma-5)}$.
By substituting Eq.~(\ref{K0}) into these scaling relations, we find:
\begin{equation}
n_\eta(r)=n_{200}(\eta+ax^b)^{3\over 3\gamma-5},
\label{n}
\end{equation}
\begin{equation}
T_\eta(r)=T_{200}(\eta+ax^b)^{3\gamma-3\over 3\gamma-5},
\label{T}
\end{equation}
\begin{equation}
K_\eta(r)=Tn^{-{2\over 3}}=K_{200}(\eta+ax^b),
\label{K}
\end{equation}
where $x=r/r_{200}$, and where we have made explicit the dependence of $n$, $T$ and
$K$ on $\eta$
 
\begin{table}
\begin{center}
\caption{Model predictions and observations at $r_{500}$ and $r_{200}$}
\setlength{\tabcolsep}{2pt}
\begin{tabular}{lcccc}
\hline
\hline 
Quantity  & Model & Obs. ($\epsilon$; $\sigma$) &   Model & Obs. ($\epsilon$; $\sigma$)  \\
        & \multicolumn{2} {c} {$r_{500}$} &
        \multicolumn{2} {c} {$r_{200}$} \\
  \hline
$K/K_{500}$ & $1.75$ & $1.23$ (2\%; 17\%) & $2.61$ & $1.75$ (3\%; 21\%)\\
$T/T_{500}$ &  $0.93$ & $0.64$ (2\%; 10\%) & $0.80$ & $0.53$ (4\%; 9\%) \\   
$n_e\,\,\,\,(10^{-5} {\rm cm}^{-3})$ & $15$ & $14.0$ (2\%; 23\%) & $6.5$ &  $5.0$ (4\%; 31\%) \\   
\hline
\hline
\label{model_parameters}
\end{tabular}
\end{center}
\tablefoot{
Theoretical predictions from Eqs.~(\ref{K}), (\ref{T}), (\ref{n}) for a baseline model with $\eta=0.19$ compared with observational data for the X-COP clusters at $r_{500}$ and $r_{200}$ \citep[see Table~3 of][]{ghirardini_etal19}; $\epsilon$ and $\sigma$ are the relative error and the internal scatter in the observational data, respectively.  
The electron densities were computed assuming a fully ionised plasma that is $75\%$ hydrogen and
$25\%$ helium in mass.}
\end{table}

In Eqs.~(\ref{n}) to (\ref{K}), the variable $\eta$
determines the core density, temperature, and ``entropy'',
while the constant $a$ and $b$ determine the behaviour
at large radii. 
As $a$ and $b$ were calibrated on simulations,
we run a coherence test to check if Eqs.~(\ref{n}) to (\ref{K}) are consistent
with the densities, temperatures, and entropies of the 
X-COP clusters at large radii (that is, $r_{200}$ and $r_{500}$).

\citet{ghirardini_etal19} fitted $K(r)$ with the functional form in Eq.~(\ref{K}).
The best overall fit was for $b=0.83$, but $b=1.1$ did indeed provide the best fit
at $r>0.6r_{500}$.
Table~\ref{model_parameters} shows that the electron densities from  Eq.~(\ref{n}) are in excellent agreement with the data of \citet{ghirardini_etal19}, especially when one considers how simple our model is and that all parameters except $\gamma$ were calibrated on adiabatic simulations and not real data.
This agreement is important because an accurate reproduction of the density is critical for reconstructing the luminosity.

Equations~(\ref{T}) and (\ref{K}) give higher temperatures and ``entropies'' than
those found in the X-COP clusters at large radii.
The discrepancy with \citet{ghirardini_etal19}'s best overall fit
is $\sim 40\%$ and $\sim 50\%$ at $r_{500}$ and $r_{200}$ , respectively
(Table~\ref{model_parameters}), but
the discrepancy with the actual data point is lower (see Fig.~3 of 
\citealp{ghirardini_etal19}).

We are therefore reassured that our simple model provides a reasonable description of the thermodynamic properties of the hot gas in the outer regions of galaxy clusters.
The properties in the core region depend on $\eta$,  which we calibrate on observations (Section~\ref{sect:calibration}), and therefore do not provide an independent test of the model.

\subsection{The luminosity--mass relation}
\label{sect:LMR}

In Section~\ref{sect:calibration}, we determine $\eta(M_{200})$ by requiring that our model reproduces the LMR $L_{\rm bol}(M_{200})$ (the red line in Fig.~\ref{Fig1}). 
Here, we explain how this relation is derived.

\citet{popesso_etal24} measured the mean X-ray luminosity $L_{0.5-2{\rm\,keV}}$ between 0.5 and 2\,keV in bins of $M_{200}$ and performed a linear fit of $\log_{10}L_{0.5-2{\rm\,keV}}$ as a function of $\log_{10}M_{200}$.
To pass from $L_{0.5-2{\rm\,keV}}$ to $L_{\rm bol}$, we must apply a bolometric correction:
\begin{equation} 
L_{\rm bol}=C_{0.5-2{\rm\,keV}}(T,Z)L_{0.5-2{\rm\,keV}}.
\label{bol_corr1}
\end{equation}
We compute the bolometric correction $C_{0.5-2{\rm\,keV}}$ with  the thermal model {\tt apec} in XSPEC \citep{arnaud96}. $C_{0.5-2{\rm\,keV}}$ depends on the X-ray temperature $T$ and the metallicity $Z$ of the hot gas. We assume that $Z$ is equal to $30\%$ of the Solar value tabulated in \citet{anders_grevesse89}\footnote{Observationally, the metallicity of the ICM exhibits considerable scatter from one cluster to another \citep{mantz_etal17,ghizzardi_etal21}. The average metallicity decreases with increasing radius, from $Z=0.61{\rm\,Z}_\odot$ at $r<0.1r_{500}$ to $Z=0.24{\rm\,Z}_\odot$ at $r>0.5r_{500}$, and  shows little dependence on temperature.
In fact, $Z$ decreases with temperature in the central region ($r<0.1r_{500}$), but the trend (if any) is in the opposite direction at larger radii  ($r>0.1r_{500}$).}.

Because there can be no X-ray temperatures for objects that are not individually detected in X-rays, we assign them a mean temperature based on the empirical relation from \citet{babyk_mcnamara23}:
\begin{equation}
M_{200}=\left({T\over 1{\rm\,keV}}\right)^{1.61}{M}_{\rm 1\,keV}.
\label{babyk}
\end{equation}
 \citet{babyk_mcnamara23} found  $M_{\rm 1\,keV}=10^{13.4}{\rm\,M}_\odot$.
 Eq.~(\ref{babyk}) fits the clusters of \citet{zhu_etal21} for $M_{\rm 1\,keV}=10^{13.6}{\rm\,M}_\odot$
 Here, we use the intermediate normalisation $M_{\rm 1\,keV}=10^{13.5}{\rm\,M}_\odot$.
 An error of $0.1\,$dex on $M_{\rm 1\,keV}$ corresponds to a $2\%$ error on $C_{0.5-2{\rm\,keV}}$, which is completely irrelevant when all other uncertainties are considered.

\subsection{The calibration of the ``entropy'' floor}
\label{sect:calibration}

\begin{figure}
\begin{center}
\includegraphics[width=0.99\hsize]{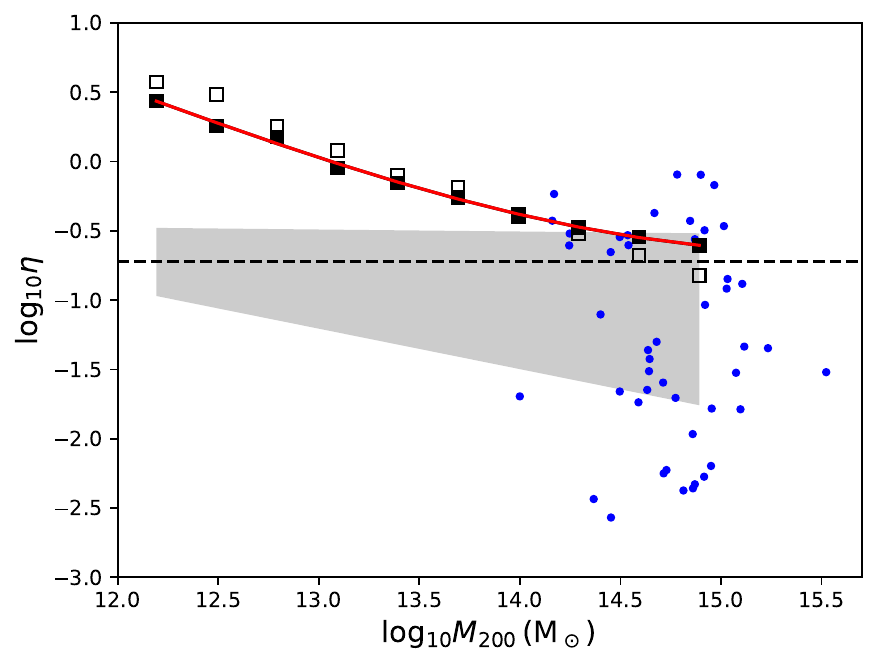} 
\end{center}
\caption{$\eta=K(0)/K_{200}$ vs. $M_{200}$ for $\gamma=1.19\pm 0.02$ (filled squares) and $\gamma=1$ (open squares).
The solid red curve shows a cubic fit to the black squares (Eq.~\ref{etafit}). The small blue circles show $K(0)/K_{200}$ vs. $M_{200}$ for the clusters of \citet{zhu_etal21}.
The horizontal dashed line shows the baseline value $\eta=0.19$.
The gray shaded area shows the equilibrium entropy according to the model of \citet[Section~\ref{sect:implications_maintenance}]{weinberger_pfrommer26} {adjusted for $\epsilon_{\rm accr}=0.1$, $\epsilon_{\rm heat}^{\rm maint}=1$ and our LMR}. Its upper and lower boundaries correspond to the BH masses of \citet{gaspari_etal19} and \citet{phipps_etal19}, respectively. The heating rate at $z=0$ is lower than the cooling rate where the red curve lies above the shaded area.  Systems in which this occurs have $Q_{\rm heat}\gg Q_{\rm cool}$ over the Hubble time (Fig.~\ref{QM}), however.
}
\label{Fig2}
\end{figure}

We compute $L_{\rm bol}(\eta)$ by substituting Eqs.~(\ref{n}) and (\ref{T}) into Eq.~(\ref{Lbol}),
and by using the cooling function of \citet{sutherland_dopita93} with an average metallicity of $0.3{\rm\,Z}_\odot$.
For ten values of $M_{200}$, we use a bisection algorithm to determine the $\eta$ for which $L_{\rm bol}(\eta)$ equals the $L_{\rm bol}$ from our LMR. The solid black squares in Fig.~\ref{Fig2} show our results for $\eta(M_{200})$.
Our calculations consider that $T$ depends on $r$, but we note that using the approximation $\Lambda(T)\sim\Lambda(T_{200})$ would not significantly alter the 
dependence of $L_{\rm bol}$ on $M_{200}$.

The black squares are well fitted by:
\begin{equation} 
\eta=10^{ 0.544-0.557\log_{10}M_{12}+0.029(\log_{10}M_{12})^2+0.009(\log_{10}M_{12})^3}
\label{etafit}
\end{equation}
(Fig.~\ref{Fig2} , red curve), where $M_{12}\equiv M_{200}/(10^{12}{\rm\,M}_\odot)$.
In contrast, assuming
$\eta=0.19$ yields bolometric luminosities that exceed the eRosita measurements (Fig.~\ref{Fig1}, solid black curve). At $M_{200}\le 10^{12.5}{\rm\,M}_\odot$, the self-similar model predicts luminosities that exceed the observations by a factor $>100$.

The error bars on $\eta$ due to the uncertainty on 
$\gamma=1.19\pm 0.02$ are smaller than
the symbols used to show the $\eta$--$M_{200}$ relation
in Fig.~\ref{Fig2}.
To test the sensitivity of $\eta$ to larger variations of $\gamma$, we repeated the same analysis for $\gamma=1$, even though we do not consider that value observationally plausible.  
The results for $\gamma=1$  (Fig.~\ref{Fig2}, open squares) are similar to those for $\gamma=1.19$, except at the highest masses, where the difference is not only quantitative, but also qualitative. For $\gamma=1.19$, $\eta>\eta_0$ at all masses.
For $\gamma=1$, $\eta$ drops below $\eta_0$
at $M_{200}\sim 10^{15}{\rm\,M}_\odot$:
the most massive clusters have entropy deficits rather than excesses.

The dependence of $\eta$ on $M_{200}$ in 
Eq.~(\ref{etafit}) is a prediction of our model, which we can test in two ways: 
via a direct comparison with observational measurements of  $\eta=K(0)/K_{200}$ and
through its implications for the gas content of DM haloes.

Direct measurements of $K(r)$ are possible only in massive systems, i.e. galaxy clusters, and show a huge scatter even in a narrow interval of $M_{200}$ \citep{cavagnolo_etal09}.
The clusters of \citet{zhu_etal21} span the whole interval
$10^{-2.5}\le\eta<1$ (Fig.~\ref{Fig2}).
Our prediction for the corresponding mass range is 
$\eta\simeq 10^{-0.6}$ (Eq.~\ref{etafit}),
which is about the upper quartile of the $\eta$ distribution for the clusters of \citet{zhu_etal21}.
Our interpretation of this finding is that our
$\eta$ reflects the average $K(0)/K_{200}$ of the whole cluster population, while the distribution of \citet{zhu_etal21} is for an X-ray selected sample that preferentially selects cool-core clusters with lower than average $K(0)$.

\begin{figure}
\begin{center}
\includegraphics[width=0.98\hsize]{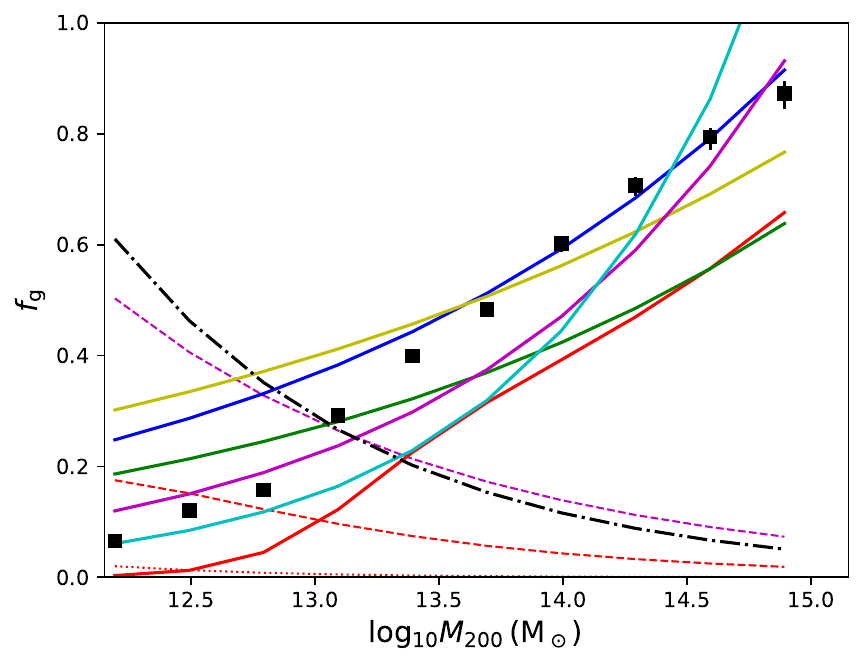} 
\end{center}
\caption{The hot gas fraction $f_{\rm g}=M_\eta(r_{200})/(f_{\rm b}M_{200})$ as a function of $M_{200}$.
The black squares show the predictions of our model for the $\eta$--$M_{200}$ relation in Fig.~\ref{Fig2}.
The solid curves show $f_{\rm g}$ according to the observational estimates by \citet[green]{pratt_etal09}, \citet[yellow]{andreon10}, \citet[blue]{ettori15}, \citet[magenta]{chiu_etal18}, \citet[cyan]{akino_etal22}
and \citet[red]{dev_etal24}.
The dashed magenta and red curves show $f_\star=M_\star/(f_{\rm b}M_{200})$ according to  \citet{chiu_etal16} and \citet{dev_etal24}, respectively.
The dotted red curve $f_{\rm HI}=M_{\rm HI}/(f_{\rm b}M_{200})$ is from \citet{dev_etal24}. It shows that neutral gas makes a negligible contribution to the baryonic mass of haloes above $10^{12}{\rm\,M}_\odot$.
The dotted-dashed black curve shows 
$f_{\rm cooled}=M_{\rm cooled}/(f_{\rm b}M_{200})$ in the GalICS semi-analytic model \citep{cattaneo_etal25}.
}
\label{Fig3}
\end{figure}

The mass of hot gas within radius $r$ is:
\begin{equation}
M_\eta(r)=4\pi {f_{\rm b}\rho_{200}}r_{200}^3\int_0^{r\over {r_{200}}}(\eta+ax^b)^{3\over 3\gamma-5}x^2{\rm\,d}x.
\label{M}
\end{equation}
(Eqs.~\ref{n} and \ref{n200}).
$M_\eta(r_{200})$ decreases with $\eta$
since $3/(3\gamma-5)<0$.
From Eq.~(\ref{M}), 
the hot gas fraction within $r_{200}$ relative to the universal baryon  fraction is:
\begin{equation}
f_{\rm g}={M_\eta(r_{200})\over f_{\rm b}M_{200}}=3\int_0^1(\eta+ax^b)^{3\over 3\gamma-5}x^2{\rm\,d}x.
\label{fhot}
\end{equation} 
Eq.~(\ref{fhot}) gives $f_{\rm g}= 1$ for $\eta=0.19$: in the baseline model, which contains no cooling, no star formation and no feedback,
the gas fraction within $r_{200}$ equals the universal baryon fraction. Increasing $\eta$ above the baseline value causes the gas to expand beyond $r_{200}$ and $f_{\rm g}$ to drop below unity. By inserting  Eq.~(\ref{etafit}) into Eq.~(\ref{fhot}), 
we obtain a prediction for $f_{\rm g}$ that
lies within the range of the observed distributions
(Fig.~\ref{Fig3}, black symbols with error bars).

There are two reasons why $M_\eta(r_{200})<f_{\rm b}M_{200}$. The first is that some hot gas has expanded at $r>r_{200}$. The total mass of the hot gas is thus $M_{\rm hot}>M_\eta(r_{200})$. The second is that some baryons are not in the hot gas because they have cooled or because they have 
been accreted through cold flows (e.g. \citealp{dekel_etal09}). Let $M_{\rm cooled}$ be their mass. 

These baryons may have met one of four possible fates:
i) they are still present in the halo as cold atomic or molecular gas,
ii) they have formed stars,
iii) they have been blown out of the halo by galactic winds, or 
iv) they have been heated and (re)incorporated into the hot gas.
Observations show that cold gas accounts for a small fraction of the baryons in haloes with $M_{200}>10^{12}{\rm M}_\odot$ \citep{dev_etal24}.
Hence, the contribution of the first path to $M_{\rm cooled}$ is negligible.
\citet[NIHAO simulations]{tollet_etal22}
found that the fourth path accounts for the lowest-entropy gas in the hot circumgalactic medium.
However, baryons that followed the fourth path do not contribute to $M_{\rm cooled}$ because they are in hot gas at $z=0$.
$M_{\rm cooled}$ is thus effectively the sum of two terms: the stellar mass $M_\star$ and the mass $M_{\rm wind}$ of the baryons that cooled but are now gone with the wind. 

$M_\star(M_{200})$ can be constrained observationally, but published determinations   \citep{andreon10,chiu_etal16,kravtsov_etal17,lin_etal17,akino_etal22,dev_etal24} vary by up to a factor of three, partly owing to differences in the adopted stellar-mass definitions and halo-mass calibrations.
Semi-empirical \citep{papastergis_etal12,rodriguez_etal17,moster_etal18,behroozi_etal19,fu_etal25}  and semi-analytic (e.g. GalICS, \citealp{cattaneo_etal25} models
favour stellar mass estimates in the lower range (e.g. \citealp{dev_etal24}).
High $M_\star/M_{200}$ ratios are in tension with the stellar mass function of galaxies.

$M_{\rm wind}$ is not an observable quantity\footnote{One can measure $f_{\rm b}M_{200}-M(r_{200})$, where $M(r)$ if the mass of hot gas within $r$, but 
$f_{\rm b}M_{200}-M(r_{200})>M_{\rm wind}$ due to the hot gas that spreads outside $r_{200}$.
One should not confuse this gas with the cooled baryons blown out by galactic winds. Both thermal expansion and winds reduce the baryonic mass within $r_{200}$, but their implications for the entropy of the gas are quite different.
Galactic winds remove the lowest-entropy baryons from the centres of haloes. Thermal expansion causes the highest-entropy gas to spill out of $r_{200}$.}.
However, both our semi-analytic model and hydrodynamic simulations (NIHAO, \citealp{tollet_etal19}; EAGLE, \citealp{mitchell_schaye22})
predict it to be the dominant contribution to $M_{\rm cooled}$. 

Here, we follow two complementary approaches. We use GalICS to compute $M_{\rm cooled}^{\rm GalICS}=M_\star+M_{\rm cold}+M_{\rm wind}$,
where $M_{\rm cold}$ is the total mass of cold gas within the halo,
 and the observations from \citet[Fig.~\ref{Fig3}]{dev_etal24} to compute the 
lower limit $M_{\rm cooled}^{\rm Dev}=M_\star+1.3M_{\rm HI}$ 
where the factor $1.3$ accounts for the helium mass.

\subsection{The entropy excesses}
\label{sect:entr_exc}

We have determined $K_\eta(r)$ and $K_0(r)$. To compute $\Delta S=S-S_0$, we must evaluate the integral in Eq.~(\ref{integrated_S}) 
for both.

When the gas is heated, its volume increases, but its mass remains constant.  
Therefore, it makes sense to invert the relation $M(r)$ from Eq.~(\ref{M})
and to use the gaseous mass $M$ enclosed by a spherical shell,
which does not vary when the shell expands, as the integration variable.
Substituting  $4\pi n r^2{\rm\,d}r={\rm d}M/(\mu m_{\rm p})$ into
Eq.~(\ref{integrated_S}) casts the integral  into a much simpler form
because the integrand depends only on $K$
and no longer on $n$.
We then find:
\begin{equation}
\Delta S=S-S_0=
\label{DS}
\end{equation}
$$=\frac{3k}{2\mu m_{\rm p}}\left[\int_0^{f_{\rm b}M_{200}-M_{\rm cooled}} \ln K_\eta(M){\rm\, d}M-\int_{M_{\rm cooled}}^{f_{\rm b}M_{200}} \ln K_0(M){\rm\, d}M\right],$$
where $K_\eta(M)$ and $K_0(M)$ are computed numerically from $K_\eta(r)$ (Eq.~\ref{K}) 
and $M_\eta(r)$ (Eq.~\ref{n}).
The integration intervals used to compute $S$ and $S_0$ have the same length $M_{\rm hot}=f_{\rm b}M_{200}-M_{\rm cooled}$ because $S$ and $S_0$ refer to the same gas.
The integral for $S$ starts from $M=0$ because $K(M)$ refers to the gas observed in X-rays; $K(0)$ is the real central ``entropy''.
$K_0(M)$ is a theoretical distribution in the absence of heating and cooling. Its lowest-entropy baryons have cooled and formed stars
\citep{tollet_etal22}. We must exclude them from the calculation of $S_0$  to determine the baseline entropy of the baryons in the form of hot gas today.

\begin{figure}
\begin{center}
\includegraphics[width=1.00\hsize]{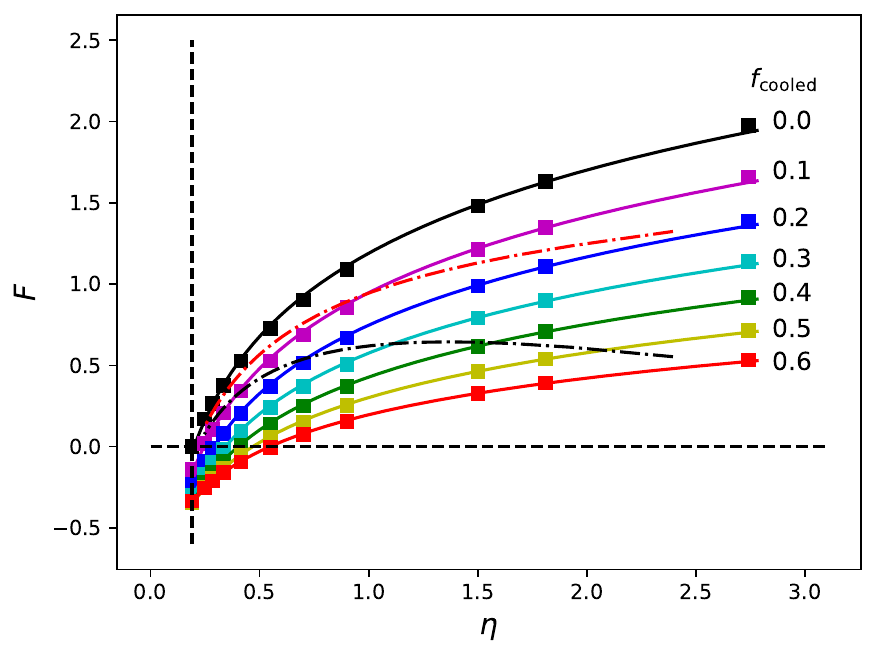} 
\end{center}
\caption{$F$ gives $\Delta S/(f_{\rm b}M_{200})$ in units of $3k/(2\mu m_{\rm p})$.
Symbols of different colours show $F(\eta)$ for different values of $f_{\rm cooled}$. From top to bottom: black, magenta, blue, green, yellow and red squares correspond to
$f_{\rm cooled}=0.1,\,0.2,\,0.3,\,0.4,\,0.5$, respectively.
The symbols are from Eq.~(\ref{Feta}). 
The solid curves are computed with the fitting formulae in Appendix~\ref{sect:Specificentropy};
{ $\eta_{\rm crit}(f_{\rm cooled})$ is the abscissa at which they intersect the horizontal dashed line $\Delta S=0$.
The vertical dashed line corresponds to $\eta=\eta_0$.
The dotted-dashed curves show $F(\eta)$ when $f_{\rm cooled}$ is not held fixed but constrained imposing  $f_{\rm cooled}=M_{\rm cooled}^{\rm GalICS}/(f_{\rm b}M_{200})$ (black curve) 
or $f_{\rm cooled}=M_{\rm cooled}^{\rm Dev}/(f_{\rm b}M_{200})$ (magenta curve).}
This figure is for $\gamma=1.19$. A different $\gamma$ would give different curves.
}
\label{Specificentropy}
\end{figure}

If we change the integration variable from $M$ to $m=M/(f_{\rm b}M_{200})$, we can rewrite
Eq.~(\ref{DS}) in the alternative form:
\begin{equation}
\frac{\Delta S}{f_{\rm b} M_{200}} =
\frac{3k}{2\mu m_{\rm p}} F(\eta,f_{\rm cooled}),
\label{DS2}
\end{equation}
where:
\begin{equation}
F(\eta,f_{\rm cooled})=\int_0^{1-f_{\rm cooled}} \ln K_\eta(m){\rm\, d}m-\int_{f_{\rm cooled}}^1 \ln K_0(m){\rm\, d}m
\label{Feta}
\end{equation}
and $f_{\rm cooled}=M_{\rm cooled}/(f_{\rm b}M_{200})$.
$F$ grows with $\eta$ at constant $f_{\rm cooled}$ and decreases with $f_{\rm cooled}$ at constant $\eta$
(Fig.~\ref{Specificentropy}).
$F=0$ for $\eta=\eta_{\rm crit}(f_{\rm cooled})\ge\eta_0$
and $\eta_{\rm crit}\rightarrow\eta_0$ for $f_{\rm cooled}\rightarrow 0$.
Hence, the condition for the presence of an entropy excess
is $\eta>\eta_{\rm crit}$ and not $\eta>\eta_0$.
Appendix~\ref{sect:Specificentropy} elaborates on how 
Eq.~(\ref{Feta}) produces this behaviour 
and gives analytic approximations for $F(\eta,f_{\rm cooled})$.

The fact that both $\eta$ and $f_{\rm cooled}$ decrease with $M_{200}$ establishes a relation between them.
The  dotted-dashed curves in Fig.~\ref{Specificentropy} show $F(\eta)=F(\eta,f_{\rm cooled}(\eta))$ for $M_{\rm cooled}=M_{\rm cooled}^{\rm GalICS}$ and $M_{\rm cooled}=M_{\rm cooled}^{\rm Dev}$.
The curve for $M_{\rm cooled}=M_{\rm cooled}^{\rm GalICS}$ has a maximum corresponding to $M_{200}\simeq 6\times 10^{12}{\rm\,M}_\odot$.
$F$ decreases at lower masses because low-mass haloes contain less hot gas even in proportion to their mass,
and at higher masses because $\eta$ is lower.

\begin{figure}
\begin{center}
\includegraphics[width=1.00\hsize]{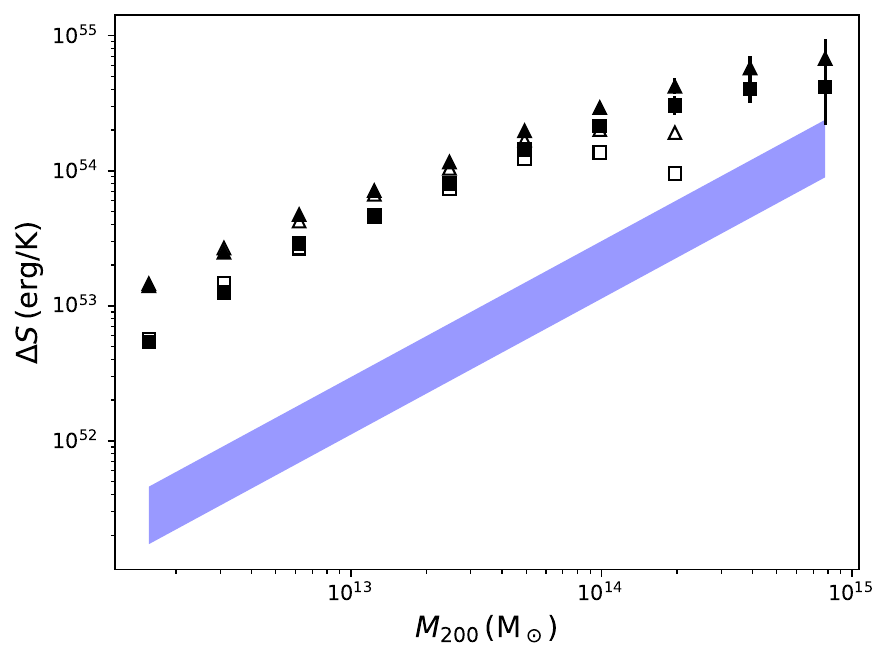} 
\end{center}
\caption{$\Delta S$ 
for $\gamma=1.19\pm 0.02$ (filled symbols with error bars) and  $\gamma=1$ (empty symbols). 
{Squares and triangles are for $M_{\rm cooled}=M_{\rm cooled}^{\rm GalICS}$ and $M_{\rm cooled}=M_{\rm cooled}^{\rm Dev}$, respectively. The open symbols at $M_{200}=4\times 10^{14}{\rm\,M}_\odot$ and $M_{200}=8\times 10^{14}{\rm\,M}_\odot$ do not appear on a logarithmic diagram because they correspond to $\Delta S<0$.}
The blue shaded area shows the allowable range for $|Q_{\rm cool}|/T_{200}=\zeta L_{\rm bol}t_{\rm H}/T_{200}$. Its lower and upper boundaries correspond
to $\zeta=0.3$ and $\zeta=0.8$, respectively.}
\label{T200DS}
\end{figure}

We saw that the uncertainty on
$\gamma=1.19\pm 0.02$ has little impact on $\eta$ (Fig.~\ref{Fig2}). The effect on $\Delta S$ is more obvious (Fig.~\ref{T200DS})
but still limited to cluster scales, where $\eta$ is only marginally larger than $\eta_{\rm crit}$. 
The uncertainty in $M_{\rm cooled}$ is much more significant. Even large variations of $\gamma$ have very little impact on group scales. 
Only in clusters (for $M_{200}>10^{14}{\rm\,M}_\odot$) does the difference between $\gamma=1.19$ and $\gamma=1$ become significant.
For $\eta<\eta_{\rm crit}$, which occurs only in clusters and only for $\gamma=1$, the model predicts entropy deficits ($\Delta S<0$) rather than excesses.

\section{The calculation of $Q_{\rm heat}$ and $Q_{\rm cool}$}
\label{sect:Qheat}

Let a system exchange an amount of heat $Q_i$ at temperature $T_i$ with $N$ sources. $\Delta S_i=Q_i/T_i$ is the entropy variation at each exchange.
$\Delta S=\sum_{i=1}^N \Delta S_i$ is the total entropy variation. $Q=\sum_{i=1}^N Q_i$ is the net heat absorption.
$\overline{T}\equiv Q/\Delta S$ is, by definition, the  characteristic temperature at which $Q$ is absorbed.

In our case, the system (the hot gas) absorbs heat from BHs and loses heat by radiating.
When we group all the heat gains under $Q_{\rm heat}>0$
and all the heat losses under $Q_{\rm cool}<0$, the accumulated and residual entropy of the hot gas at $z=0$ is:
\begin{equation}
\Delta S={Q_{\rm heat}\over\overline{T}_{\rm heat}}+{Q_{\rm cool}\over\overline{T}_{\rm cool}},
\label{DeltaS_thermo}
\end{equation}
where $\overline{T}_{\rm heat}$ and $\overline{T}_{\rm cool}$ are the characteristic temperatures associated with the heat gains and the heat losses, respectively (defined as above). $Q_{\rm heat}$ is found by solving
 Eq.~(\ref{DeltaS_thermo}):
\begin{equation}
Q_{\rm heat}=\overline{T}_{\rm heat}\left(\Delta S-{Q_{\rm cool}\over\overline{T}_{\rm cool}}\right).
\label{Qheat}
\end{equation}

\begin{figure}
\begin{center}
\includegraphics[width=0.99\hsize]{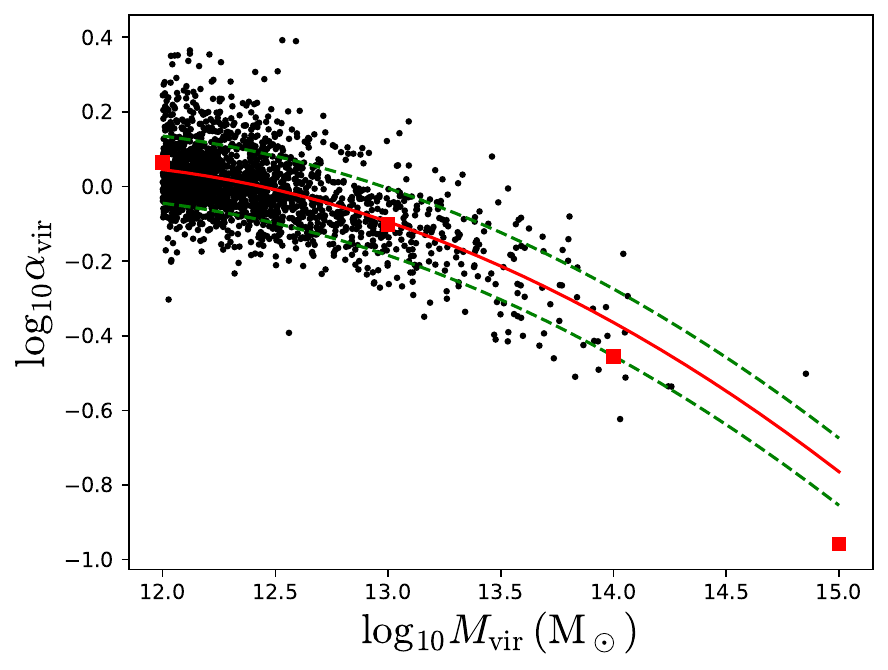} 
\end{center}
\caption{The coefficient
$\alpha_{\rm vir}$
in the GalICS semi-analytic model (black points), 
and in the TRINITY semi-empirical model (red squares).
The red curve corresponds to the fit in
Eq.~(\ref{alphavir_GalICS}).
The dashed green curves show the root-mean-square deviation of the logarithms of the black points from the fit.
In this figure, we show virial (instead of 200) quantities (see Appendix~\ref{sect:over})
because these are the quantities used by GalICS and TRINITY.
In the main body of the article and Table~2, we have converted $\alpha_{\rm vir}$ and
$M_{\rm vir}$ to values for a density contrast $\Delta=200$ by using average concentrations computed with
the formulae in \citet{dutton_maccio14}.
The conversions are $M_{200}\simeq 0.85M_{\rm vir}$ and $\alpha_{200}\simeq 0.9\alpha_{\rm vir}$.}
\label{alpha_fig}
\end{figure}

Previous studies (e.g. \citealp{zhu_etal21,eckert_etal25}) neglected the cooling term and assumed
that $\overline{T}_{\rm heat}$ is the current temperature
of the hot gas. This assumption gives $Q_{\rm heat}=\xi T_{200}\Delta S$, where $\xi\sim 1$ is a factor slightly larger than unity, which depends on the observed temperature profile of the gas.

A more accurate estimate requires determining at what epochs AGN generated the observed entropy excesses
and what was the temperature of the hot gas back then.
This calculation entails three aspects:
the redshift dependence of $\dot{Q}_{\rm heat}$ and $\dot{Q}_{\rm cool}$;
the redshift dependence of $T_{200}$; 
the relation
between $T_{200}(z)$, ${T}_{\rm heat}(z)$ and ${T}_{\rm cool}(z)$.
We start from the last point because it is the simplest, although, ultimately, the least significant.

Observationally, $T_{\rm cool}$ is the emission-weighted temperature.
$T_{\rm cool}>T_{200}$ because most of the X-ray emission comes from gas at $r\ll r_{200}$.
\citet{mazzotta_etal04} found that $T_{\rm cool}$ is also higher than the spectroscopic temperature
$T_{\rm spec}>T_{200}$.
For the clusters of \citet{vikhlinin_etal06}, $T_{\rm spec}/T_{200}\simeq 1.1$
and $T_{\rm peak}/T_{200}\simeq 1.2$, where $T_{\rm peak}$ is the maximum of the radial temperature distribution.  Hence
$1.1<T_{\rm cool}/T_{200}\lesssim 1.2$. This interval is for galaxy clusters at $z=0$ but is indicative of the plausible range for $\xi=T_{\rm cool}/T_{200}$, knowing that $\xi$ must be within the range $1<\xi\le 1.9$ to be consistent with our model (Eq.~\ref{T} with $\eta\ge 0.19$).

We expect that most of the heat absorption also occurs in the central region and thus that
$T_{\rm heat}>T_{200}$.
$T_{\rm heat}$ may differ from $T_{\rm cool}$ if heating and cooling  have different radial distributions. In practice, however, the radial distribution of the heat absorption is so uncertain that it makes sense to assume $T_{\rm heat}(z)=T_{\rm cool}(z)=\xi T_{200}(z)$. That does not imply $\overline{T}_{\rm heat}=\overline{T}_{\rm cool}$ because heating and cooling may have different redshift distributions even when the radial distribution is the same.

We parametrised $\overline{T}_{\rm heat}$ and $\overline{T}_{\rm cool}$ by defining:
\begin{equation}
\alpha_{200}\equiv \frac{\overline{T}_{\rm heat}}{\xi T_{200}} =
\frac{\int_\infty^0 T_{200}(z){\rm\,d}S_{\rm heat}(z)}{T_{200}\Delta S_{\rm heat}}
= \frac{Q_{\rm heat}}{T_{200}\int_\infty^0{\frac{\delta Q_{\rm heat}(z)}{T_{200}(z)}}}
\label{alpha1}
\end{equation}
and
\begin{equation}
\beta_{200}\equiv \frac{\overline{T}_{\rm cool}}{\xi T_{200}} = \frac{\int_\infty^0 T_{200}(z){\rm\,d}S_{\rm cool}(z)}{T_{200}\Delta S_{\rm cool}} = 
\frac{Q_{\rm cool}}{T_{200}\int_\infty^0{ \frac{\delta Q_{\rm cool}(z)}{T_{200}(z)}}}.
\label{beta1}
\end{equation}
Here
${\rm d}S_{\rm heat}(z)$ and $\delta Q_{\rm heat}(z)$ are the
entropy and heat gained through AGN feedback between $z+{\rm d}z$ and $z$, whereas
${\rm d}S_{\rm cool}(z)$ and $\delta Q_{\rm cool}(z)$ are 
the entropy and heat lost through radiative cooling between $z+{\rm d}z$ and $z$. With the parametrisation in Eqs.~(\ref{alpha1}) and (\ref{beta1}), Eq.~(\ref{Qheat}) becomes:
\begin{equation}
Q_{\rm heat}=\alpha_{200}\left(\xi{T}_{200}\Delta S-{Q_{\rm cool}\over\beta_{200}}\right).
\label{Qheat2}
\end{equation}

We compute $\alpha_{200}$ by assuming that $\dot{Q}_{\rm heat}(z)\propto\dot{M}_\bullet(z)$, and $\beta_{200}$ from
$\dot{Q}_{\rm cool}(z)=-L_{\rm bol}(z)$ by assuming the luminosity evolution:
\begin{equation}
L_{\rm bol}(z)=L_{\rm bol}(M_{200})\left[\frac{M_{200}(z)}{M_{200}}\right]^{4/3+2\nu}E^\lambda(z).
\label{Lbolz_short}
\end{equation}
$L_{\rm bol}(M_{200})$ comes from the LMR,
$\nu=0.17$ as per Section~\ref{sect:introduction}, and
$E(z)$ is defined in Appendix~\ref{sect:over}.

{Observationally, the exponent $\lambda$ is considerably uncertain ($0.6\lesssim\lambda\lesssim 1.9$; \citealp{vikhlinin_etal09,reichert_etal11,sereno_etal15,chiu_etal22}).
From a theoretical perspective, our semi-analytic model favours $\lambda\sim 2$ (Appendix~\ref{sect:calculation_lambda}).
To be on the safe side, we consider the extreme values $\lambda=0$ and $\lambda=2$, which give a lower and an upper limit for $L_{\rm bol}(z)$, respectively.
$T_{200}(z)$ follows from $M_{200}(z)$.}

\begin{table*}
\begin{center}
\caption{Predictions based on TRINITY halo-growth histories at different halo masses}
\begin{tabular}{cccccccccccc}
\hline
\hline 
$M_{200}(z=0)$ & $z_{\rm q}$ & $M_{200}(z_{\rm q})$&$\alpha_{200}$&$\beta_{200}$&$\zeta$&$\beta_{200}$&$\zeta$
&$Q_{\rm heat}$&$U$&$E_{\rm bind}$&$E_{\rm bind}(z_{\rm q})$\\
$({\rm M}_\odot)$ &  & $({\rm M}_\odot)$&&$(\lambda=0)$&$(\lambda=0)$&$(\lambda=2)$&$(\lambda=2)$
&$({\rm M}_\odot c^2)$&$({\rm M}_\odot c^2)$&$({\rm M}_\odot c^2)$&$({\rm M}_\odot c^2)$\\
  \hline
$8.5\cdot10^{11}$& $0.7$&$6\times 10^{11}$&$1.03$&$0.98$&$0.44$&$0.94$&$0.81$&2--$7\cdot 10^4$&0.5--$1.7\cdot 10^4$&$1.4\cdot 10^4$&$1.0\cdot 10^4$\\
$8.5\cdot 10^{12}$& $1.8$&$2\times 10^{12}$&$0.74$&$0.94$&$0.39$&$0.88$&$0.65$&6--$9\cdot 10^5$&6.8--$8.7\cdot 10^{5}$&$6.5\cdot 10^5$&$1.1\cdot 10^5$\\
$8.5\cdot 10^{13}$& $2.7$&$4\times 10^{12}$&$0.33$&$0.90$&$0.33$&$0.84$&$0.49$&7--$9\cdot 10^6$&4.0--$4.4\cdot 10^7$&$3.0\cdot 10^7$&$4.7\cdot 10^4$ \\
$8.5\cdot 10^{14}$& $3.6$&$5\times 10^{12}$&$0.10$&$0.92$&$0.28$&$0.90$&$0.35$&2--$4\cdot 10^7$&2.0--$2.1\cdot 10^9$&$1.4\cdot 10^9$&$8.4\cdot 10^5$\\
\hline
\hline
\label{tab:qheat}
\end{tabular}
\end{center}
\tablefoot{The redshift $z_{\rm q}$ at which $\dot{M}_\bullet$ has a maximum;
the halo mass $M_{200}(z_{\rm q})$ at $z_{\rm q}$; 
$\alpha_{200}$ (Eq.~\ref{alphavir_TRINITY});
$\beta_{200}$ and $\zeta$ for $\lambda=0$ and $\lambda=2$  (Appendix~D); 
$Q_{\rm heat}$ (Eq.~\ref{Qheat4}); 
the internal energy $U$ of the hot gas at $z=0$ (Eq.~\ref{int_en});  the binding energy $E_{\rm bind}$ at $z=0$ and $z=z_{\rm q}$ (Eq.~\ref{quenching_criterion}) normalised for $\Psi=1$.
All quantities in this Table are shown for four values of $M_{200}$ at $z=0$ and are computed based on the TRINITY semi-empirical model  \citep{zhang_etal23}.
When we give an interval, the lower bound is for $M_{\rm cooled}=M_{\rm cooled}^{\rm GalICS}$ and the upper bound is for $M_{\rm cooled}=M_{\rm cooled}^{\rm Dev}$.}
\end{table*}

Equations~(\ref{alpha1}), (\ref{beta1}) and (\ref{Lbolz_short}) show that computing $\alpha_{200}$ and $\beta_{200}$  boils down
to determining the evolution  of  $\dot{M}_\bullet$ and $M_{200}$ with $z$.
We have explored two models for $\dot{M}_{\bullet}(z)/M_\bullet$ and $M_{200}(z)$: the GalICS semi-analytic model \citep{cattaneo_etal25} and the TRINITY semi-empirical model \citep{zhang_etal23}. GalICS assumes that 
$\dot{M}_\bullet$ is proportional to the star formation rate in merger-driven starbursts. TRINITY constrains the evolution of galaxies, bulges and BHs with observations.

GalICS models galaxy formation inside DM merger trees from a 
cosmological N-body simulation. We have used GalICS to
evaluate Eq.~(\ref{alpha1}) halo by halo. The results are well-fitted by:
\begin{equation}
\alpha_{200}= 10^{-8.358 +1.476{\rm log}_{10}(M_{200}/{\rm M}_\odot) -0.065[{\rm log}_{10}(M_{200}/{\rm M}_\odot)]^2}.
\label{alphavir_GalICS}
\end{equation}

TRINITY is calibrated to quasar statistics and should, by construction, reproduce the average growth history of BHs better.
Assuming that BHs accrete their mass instantaneously at the peak accretion redshift $z_{\rm q}$ (the ``quasar epoch'') gives:
\begin{equation}
\alpha_{200}\sim{T_{200}(z_{\rm q})\over T_{200}}=[E(z_{\rm  q})]^{2\over 3}
\left[{M_{200}(z_{\rm q})\over M_{200}}\right]^{2\over 3}.
\label{alphavir_TRINITY}
\end{equation}
With $z_{\rm q}$ and $M_{200}(z_{\rm q})$
from TRINITY (Table~\ref{tab:qheat}), the results of
Eq.~(\ref{alphavir_TRINITY}) are in good agreement with those of GalICS,
 especially at low masses, where the semi-analytic model contains many haloes (Fig.~\ref{alpha_fig}).

The only significant discrepancy occurs on cluster scales. At $M_{200}\sim 10^{15} {\rm M}_\odot$, GalICS gives $\alpha_{200}\simeq0.15$, TRINITY $\alpha_{200}\simeq0.10$. This may reflect: (i) poor statistics in GalICS, which contains only one massive cluster; (ii) extended accretion tails at low redshift in the most massive systems \citep{zhang_etal23}, implying a mean accretion redshift lower than $z_{\rm q}$ and thus larger $\alpha_{200}$ than those inferred from
Eq.~(\ref{alphavir_TRINITY}); and iii) TRINITY's halo mass growth histories, which give lower values of 
$M_{200}(z)/M_{200}$ at high $z$ compared to other studies
(e.g. \citealp{mcbride_etal09,vandenbosch_etal14,correa_etal15}).

We  use Eq.~(\ref{Lbolz_short}) with  mass growth histories from TRINITY to compute
\begin{equation}
    Q_{\rm cool}=\int_\infty^0\delta Q_{\rm cool}(z)=\int_0^\infty L_{\rm bol}(z){{\rm d}t\over{\rm d}z}{\rm\,d}z,
\label{Qcool_int}
\end{equation}
and $\beta_{200}$.
The details of these calculations are in Appendix~\ref{sect:calculation_details}.
Here we note that the 
luminosity evolution in Eq.~(\ref{Lbolz_short}) is 
assumed to hold at $z<z_{\rm q}$.
At $z>z_{\rm q}$, we assume $L_{\rm bol}(z)=0$.
The justification for this assumption is that
$M_{200}(z_{\rm q})$ (Table~\ref{tab:qheat})
falls in the same mass range as the critical halo mass
above which haloes begin to fill with hot gas
\citep{keres_etal05,dekel_birnboim06,ocvirk_etal08,cattaneo_etal20}.
We follow this approach because we consider it more physical but note that applying Eq.~(\ref{Lbolz_short}) over the entire redshift range from zero to infinity makes little difference because
$M_{200}$ is a strongly decreasing function of $z$.

We find it convenient to express $Q_{\rm cool}$ via the dimensionless
parameter
$\zeta=-Q_{\rm cool}/(L_{\rm bol}t_{\rm H})$, where $L_{\rm bol}$
is the bolometric luminosity at $z=0$ and $t_{\rm H}=H_0^{-1}$ is
the Hubble time.
Table~\ref{tab:qheat} shows our findings for $\beta_{200}$ and 
$\zeta$ when $\lambda=0$ and $\lambda=2$. Despite the huge uncertainty on $\lambda$ (the exponent of $E$ in Eq.~\ref{Lbolz_short}),  $\beta_{200}$ is constrained reasonably well.

As a sanity check, we repeated the same calculation  with the halo-growth histories from \citet{mcbride_etal09}
instead of those from TRINITY. We still found values in the range $0.84\le\beta_{200}\le 1.0$. The parameter
 $\zeta$ is more sensitive to the assumed mass-growth histories.
Those from TRINITY give $0.28<\zeta<0.35$ for $M_{200}=8.5\times 10^{14}{\rm\,M}_\odot$.
Those from \citet{mcbride_etal09} give $0.40<\zeta<0.70$ for the same halo mass.
The heat radiated  across the Hubble time  is higher if protoclusters  were already  massive at high redshift.
Despite these uncertainties,
we always found values in the range $0.3\lesssim\zeta\lesssim 0.8$.

\section{Results}
\label{sect:results}

\subsection{Absorbed heat}
\label{subsect:abs_heat}

Let us rewrite Eq.~(\ref{Qheat2}) by factorizing the coefficient $\xi$:
\begin{equation}
    Q_{\rm heat}=\xi\alpha_{200}\left(T_{200}\Delta S -\frac{\xi^{-1}}{\beta_{200}} Q_{\rm cool}\right),
    \label{Qheat3}
\end{equation}
where $0.83\lesssim \xi^{-1}<0.91$ (Section~\ref{sect:Qheat}).
At high masses, $0.84\lesssim\beta_{200}\lesssim 0.92$ (Table~\ref{tab:qheat}). Hence:
\begin{equation}
    Q_{\rm heat}    \simeq \xi\alpha_{200}({T}_{200}\Delta S-Q_{\rm cool})=\xi\alpha_{200}({T}_{200}\Delta S+\zeta L_{\rm bol}t_{\rm H}).
    \label{Qheat4}
\end{equation}
We compute $Q_{\rm heat}$ by evaluating Eq.~(\ref{Qheat}) with $\xi=1.2$ and $\zeta=0.55\pm 0.25$ but note that
$Q_{\rm heat}\simeq\xi\alpha_{200}T_{200}\Delta S$ provides a reasonable approximation at all masses.

Figure~\ref{QM} compares three energies: 
the heat $Q_{\rm heat}$ deposited by BHs into the hot gas; 
the heat $|Q_{\rm cool}|$ radiated by the hot gas across the Hubble time; 
the mass energy $M_\bullet c^2$ of supermassive BHs.
This comparison prompts two considerations: 
\begin{enumerate}
\item $M_\bullet c^2\gg Q_{\rm heat}$ on all scales. Even though BH masses are considerably uncertain,
$Q_{\rm heat}/(M_\bullet c^2)$ is clearly lower than the canonical $10\%$ energetic efficiency of BH accretion. There is no doubt that BHs are energetically capable of generating the
observed entropy excesses.
\item Comparing $Q_{\rm heat}$ and $|Q_{\rm cool}|$ shows two regimes. At $M_{200}\lesssim 10^{14}{\rm\,M}_\odot$, $Q_{\rm heat}\gg |Q_{\rm cool}|$: in elliptical galaxies and groups, 
the gas stores the energy accumulated through AGN
feedback and remains in a high-entropy state
long after the quasar epoch. At $M_{200}\gtrsim 3\times 10^{14}{\rm\,M}_\odot$, $Q_{\rm heat}\sim |Q_{\rm cool}|$: in clusters,  most of the energy that was injected into the gas has been radiated.
\end{enumerate}

This finding does not contradict our previous statement that $Q_{\rm cool}$
has no significant effect on the value of $Q_{\rm heat}$.
When cooling is taken into account, the value of $Q_{\rm heat}$ increases by $+\xi\alpha_{200}|Q_{\rm cool}|$
(Eq.~\ref{Qheat4})
with $\xi\alpha_{200}\simeq 0.18$ at $M_{200}\sim  10^{15}{\rm\,M}_\odot$.
Therefore, the impact of $Q_{\rm cool}$ on $Q_{\rm heat}$ is small
even when $|Q_{\rm cool}|\sim Q_{\rm heat}$.

\subsection{Efficiency of BH heating}
\label{sect:effBH}

BHs thermalise a fraction:
\begin{equation}
    \epsilon_{\rm heat}\equiv{Q_{\rm heat}\over \epsilon_{\rm accr}M_\bullet c^2}=\xi\alpha_{200}{T_{200}\Delta S+\zeta L_{\rm bol}t_{\rm H}\over \epsilon_{\rm accr}M_\bullet c^2},
    \label{epsilonn_bullet}
\end{equation}
of their energy output in the surrounding gas,
where $\epsilon_{\rm accr}\sim 0.1$ is the energetic efficiency of BH accretion.

The uncertainty on the average BH mass $M_\bullet$
at a given $M_{200}$ affects $\epsilon_{\rm heat}$
as much as the uncertainty on $Q_{\rm heat}$
(compare 
\citealp{gaspari_etal19}, \citealp{phipps_etal19},
\citealp{marasco_etal21}, and \citealp{zhang_etal23}).
Several aspects contribute to the difficulty of establishing
the relation between $M_\bullet$ and $M_{200}$:
\begin{itemize}
    \item The relation contains significant scatter.
    \item Very few clusters with $M_{200}\gg 10^{14}{\rm\,M}_\odot$
    have dynamically confirmed central BH masses.
    \item There is evidence that the relation steepens at low masses \citep{dutton_etal10}.  Fitting a power law to datasets that contain many objects at low masses, where the relation is steeper,
leads to fits that overestimate $M_\bullet$ at $M_{200}\sim 10^{15}{\rm\,M}_\odot$.
\end{itemize}

\begin{figure}
\begin{center}
\includegraphics[width=0.99\hsize]{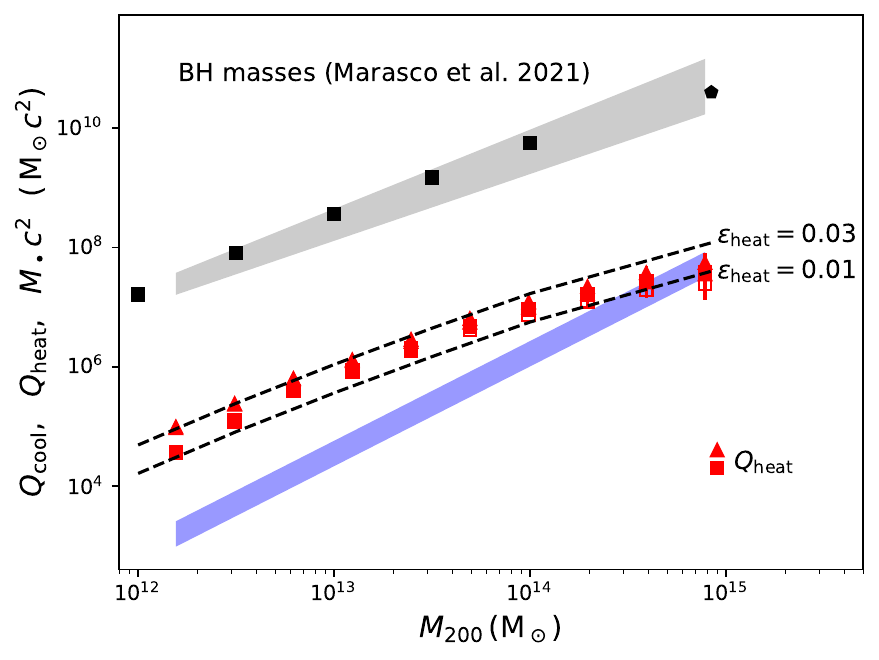} 
\end{center}
\caption{The red symbols with error bars
show our findings for $Q_{\rm heat}$.
{Red squares and triangles are for $M_{\rm cooled}=M_{\rm cooled}^{\rm GalICS}$ and $M_{\rm cooled}=M_{\rm cooled}^{\rm Dev}$, respectively}. 
Filled and open red symbols correspond to values of 
$\alpha_{200}$ from GalICS and TRINITY, respectively.
The blue shaded area shows $Q_{\rm cool}$.
Its lower and upper boundaries are for $\zeta=0.3$ and $\zeta=0.8$, respectively.
The black symbols and the gray-shaded area show different BH mass estimates.
The upper envelope of the gray shaded area
corresponds to the BH masses from
\citet{gaspari_etal19}.
The lower envelope is based on a combined fit of the data of \citet{phipps_etal19} at high masses and \citet{bogdan_etal18} at low masses.
The black squares show the $M_\bullet c^2$--$M_{200}$ relation from \citet{marasco_etal21}.
The black pentagon corresponds to Abell~85 \citep{mehrgan_etal19}. 
Its error bar  is so small that it is barely visible.
The dashed black curves correspond to the black symbols times $0.001$ and $0.003$.}
\label{QM}
\end{figure}

We estimate $\epsilon_{\rm heat}$ using the data of \citet[Fig.~\ref{QM}, black squares]{marasco_etal21}  because they are based on dynamically estimated BH and halo masses, and because \citet{marasco_etal21} fitted the data at $10^{12}{\rm\,M}_\odot\le M_{200}\le 10^{14}{\rm\,M}_\odot$ with a more complicated model
that avoids the pitfalls of power-law fits.
We complete the data by \citet{marasco_etal21} with
an additional point at 
$M_{200}=8.4\times 10^{14}{\rm\,M}_\odot$ (Fig.~\ref{QM}, black pentagon). 
This data point is for a single object, Abell~85, the only massive cluster we know of with a dynamically confirmed 
mass estimate for the central BH: $M_\bullet=(4.0\pm 0.8) \times 10^{10}{\rm\,M}_\odot$ \citep{mehrgan_etal19}.
For these BH masses and $\epsilon_{\rm accr}=0.1$, our findings are consistent with an AGN heating efficiency of $0.01\le\epsilon_{\rm heat}\le 0.03$ at all masses
(Fig.~\ref{QM}). 

\section{Discussion}
\label{sect:disc}

In this section, we discuss the uncertainties that affect our estimates and explore the implications of our results. The comparison with previous studies is discussed in Appendix~\ref{sect:comparison}.

\subsection{Uncertainties}

We adopted an effective polytropic index $\gamma = 1.19$, inferred from X-ray and Sunyaev--Zeldovich observations of galaxy clusters \citep{ghirardini_etal19}, where the ``central-entropy parameter'' $\eta$ is close to its baseline value $\eta_0$. If $\gamma$ depended on $\eta$, groups and, even more, individual galaxies could have effective polytropic indices significantly different from $\gamma = 1.19$.

A posteriori, even relatively large variations in $\gamma$ have little effect on $\Delta S$ in elliptical galaxies and groups (Fig.~\ref{T200DS}). Only clusters show sensitivity, but cluster observations strongly constrain departures from $\gamma=1.19$. Hence, the error on $Q_{\rm heat}$  from the uncertainty on $\gamma$ is moderate.

A greater uncertainty is the mass of the cooled baryons. The theoretically motivated assumption $M_{\rm cooled}=M_{\rm cooled}^{\rm GalICS}$ and the observational lower limit $M_{\rm cooled}=M_{\rm cooled}^{\rm Dev}$
give values of $Q_{\rm heat}$ that differ by up to a factor of three.

Physically, the red triangles in Fig.~\ref{QM} correspond to assuming that the X-ray luminosities of low-mass systems ($M_{200}\sim 10^{12}{\rm\,M}_\odot$) are entirely due to AGN heating, while the red squares correspond to a picture in which supernovae eject
a significant fraction of the baryonic mass before the quasar phase. As expected, AGN heating must be more efficient if it is required to do all the work.

Other uncertainties concern how heat is distributed over the gas and the redshift of absorption. These enter our estimates for $Q_{\rm heat}$ through the values of $\xi$ and $\alpha_{200}$ (Eq.~\ref{Qheat3}). The value of $\xi$ is uncertain at the $\sim 10\%$ level; $\alpha_{200}$ is well constrained at low masses, whereas in clusters the uncertainties in $\alpha_{200}$ can be as large as $50\%$ (Fig.~\ref{alpha_fig}). The parameter $\zeta$, which determines the value of $Q_{\rm cool}$ ($Q_{\rm cool}=-\zeta L_{\rm bol}t_{\rm H}$), has very little effect on $Q_{\rm heat}$ because $|Q_{\rm cool}|\ll T_{200}\Delta S$ (Fig.~\ref{T200DS}).
When all uncertainties are combined, the lower and upper estimates for $Q_{\rm heat}$ differ by a factor of three, dominated by the uncertainty in $M_{\rm cooled}/M_{200}$.

Our estimate $0.01\le\epsilon_{\rm heat}=Q_{\rm heat}/(0.1 M_\bullet c^2)\le 0.03$ is based on the $M_\bullet$--$M_{200}$ relation from \citet{marasco_etal21} and considers only the error on $Q_{\rm heat}$. In fact, the uncertainty on $M_\bullet$ is as large as that on $Q_{\rm heat}$. {Combining the $Q_{\rm heat}$ for $M_{\rm cooled}=M_{\rm cooled}^{\rm Dev}$ with the $M_\bullet$ from \citet{phipps_etal19} gives an upper limit $\epsilon_{\rm heat}\sim 0.06$. 

Figure~\ref{QM} suggests that $\epsilon_{\rm heat}$ decreases by a factor of two  from $M_{200}=10^{13}{\rm\,M}_\odot$ to $M_{200}=10^{15}{\rm\,M}_\odot$.
This may be a genuine astrophysical result or a sign that we overestimate the redshift at which the ICM was heated. Perhaps a non-negligible fraction of the heat absorption
 occurred in the accretion tails after the quasar epoch.

The halo masses at $z_{\rm q}$ in Table~\ref{tab:qheat} show that the mass range considered in this article (dictated by the stacked eRosita data of \citealp{popesso_etal24}) corresponds to systems 
that have been in the hot mode\footnote{The slow-cooling regime (in the language of \citealp{blumenthal_etal84}) or hot mode (in that of \citealp{dekel_birnboim06}) is a regime in which the radiative cooling time is longer than the dynamical/gravitational compression time.} 
at least since $z_{\rm q}$ \citep{blumenthal_etal84,dekel_birnboim06,cattaneo_etal20}.
Volume-filling hot atmospheres may have coexisted with cold gas at high redshift \citep{dekel_birnboim06}.
Some of the feedback energy may have been used to blow the cold gas out through the winds.
The red squares in Fig.~\ref{QM} do not account for this part of the feedback energy
if the winds have not mixed with the hot gas (e.g. because they have escaped).

\subsection{Implications for the physics of AGN feedback}
\label{sect:implications_feedback}

\citet{cattaneo_best09} estimated that AGN release $2\%$ of their power mechanically  {through jets}, although they were unable to rule out values as high as $10\%$. This estimate included not only the jet power from low-accretion-rate, radiatively inefficient ``radio-mode'' AGN, which release most of their power mechanically, but also the jet power from luminous ``quasar-mode'' AGN, which release only a small fraction of their power mechanically but have much higher total power.\footnote{The jet power fraction in quasars ranges from being negligible in most radio-quiet objects (e.g. \citealp{inoue_etal17}) to being comparable to the luminous power \citep{lopez_perucho12} or even higher \citep{punsly_etal20} in the most extreme radio-loud objects.}
{\citet{heckman_best23} reassessed the observational evidence and concluded that BHs release $3.4\%$ of their accretion power mechanically, mainly through jets ($\sim 3\%$) but also through winds ($\sim 0.5\%$).

Despite the apparently similar result, our work differs from these previous studies in its goal and not only in its methodology. \citet{cattaneo_best09} and \citet{heckman_best23} focused on AGN and how their power is released.
We focus on the hot gas and how much heat it has absorbed. The two approaches give complementary results.}

Photons carry most of the energy output from AGN but, unlike jets, are not thermalised efficiently (see \citealp{cattaneo_etal09}). It is therefore unclear whether AGN feedback is predominantly radiative or mechanical.
Our estimate for the heating efficiency of AGN, $0.01\le\epsilon_{\rm heat}\le 0.03$, is close to the mechanical power fractions found by \citet{cattaneo_best09} and \citet{heckman_best23}. This does not prove that AGN
heat the gas mechanically, but shows that mechanical feedback alone is energetically sufficient to explain the observed entropy excesses.

\subsection{Implications for galaxy formation}

Theorists of galaxy formation have invoked AGN feedback to solve two problems:
\begin{itemize}
    \item {\it The quenching problem.} The distribution in colour and specific star-formation rate \citep{schawinski_etal14,lian_etal16,darvish_etal18,bravo_etal23}, and the $[\alpha/{\rm Fe}]$ ratios of giant ellipticals \citep{thomas_etal05,yan_etal19} imply that massive E/S0 galaxies formed their stars on short timescales. Massive E/S0 galaxies contain old stellar populations \citep{sandage86} and were already in place at $z>2$ (e.g. \citealp{ilbert_etal13,huertas_etal15}). What physical mechanism abruptly terminated star formation in these galaxies at high redshift?
    \item {\it The maintenance problem.} Massive E/S0 galaxies display extended haloes of hot X-ray emitting gas. What prevents the hot gas in these systems from cooling and reactivating star formation? In galaxy clusters, this is also known as the cooling-flow problem \citep{peterson_fabian06,mcnamara_nulsen07}.
\end{itemize}
{We discuss these two aspects separately.}

\subsubsection{Implications for quenching}
\label{sect:implications_quenching}

\citet{chen_etal20} remarked that star-forming and passive galaxies occupy different loci on a stellar mass--surface density diagram. They argued that this observational finding 
is consistent with a scenario in which BHs grow and deposit energy in the surrounding gas until the accumulated energy is higher than a multiple of the binding energy $E_{\rm bind}$ of the baryons within the halo\footnote{For a gas in hydrostatic equilibrium, the binding energy is the sum of the internal energy and the gravitational energy. The binding energy is a negative quantity. $E_{\rm bind}$ is  its absolute value.}.

\citet{koutsouridou_cattaneo22} and \citet{cattaneo_etal25} tested this picture by implementing the quenching condition:
\begin{equation}
Q_{\rm heat}=\epsilon_{\rm heat}\epsilon_{\rm accr} M_\bullet c^2>E_{\rm bind}= {\Psi\over 2} f_{\rm b}M_{\rm vir}v_{\rm vir}^2
\label{quenching_criterion}
\end{equation} 
in the GalICS semi-analytic model. where $\Psi$ is an uncertainty factors that
depends on the DM's gravitational potential and the gas density distribution.
The uncertainty on $E_{\rm bind}$ is degenerate with the uncertainty on the critical $Q_{\rm heat}/E_{\rm bind}$ ratio for quenching.
The results of GalICS 
are sensitive to the product $\epsilon_{\rm eff}\equiv\epsilon_{\rm heat}\epsilon_{\rm accr}/\Psi$ but not to the individual parameters that compose it.

The results of GalICS are also sensitive to the BH masses used to calibrate the semi-analytic model.
\citet{koutsouridou_cattaneo22} and \citet{cattaneo_etal25}  obtained the best fit to the mass functions of star-forming and passive galaxies separately  for $\epsilon_{\rm eff}\sim 0.0012$ for the BH masses from  \citet{davis_etal18} and \citet{sahu_etal19}.
The BH mass estimates from  \citet{marasco_etal21}, which are larger by a factor of two, yield:
 \begin{equation}
 \epsilon_{\rm heat}\sim 6\times 10^{-4}{\Psi\over\epsilon_{\rm accr}}.
 \label{epsilon_heat_galics}
 \end{equation}

Further progress requires assumptions on $\Psi$. Let us start by assuming that the critical $Q_{\rm heat}/E_{\rm bind}$ ratio for quenching is unity, so that $\Psi$ purely parametrises the uncertainty on the binding energy.

Assuming that all the baryons are in hot gas and neglecting cooling, \citet{ostriker_etal05} found $2.5\lesssim\Psi\lesssim 3.4$. The lower and upper bounds correspond to halo concentrations of
$c=4$  and $c=10$, respectively.
Cooling affects $\Psi$ in two ways. In the beginning, the gas loses energy, sinks to lower radii and becomes more gravitationally bound, so $\Psi$ increases. Then star formation removes the most gravitationally bound gas with the effect that the gravitational binding energy of the remaining reservoir decreases, so $\Psi$ becomes lower. The second effect alone, which is likely to dominate, gives $\Psi=0.76$ for low-mass systems with $f_{\rm cooled}=0.5$ and $c=10$, and
$\Psi=2.0$ for cluster haloes with $f_{\rm cooled}=0.1$ and $c=4$ \citep{ostriker_etal05}.
As quenching occurs for $M_{200}\sim 10^{12}{\rm\,M}_\odot$ (e.g. \citealp{kauffmann_etal03,dekel_birnboim06}),
it is not unreasonable to assume $\Psi\sim 1$ on physical grounds.

An alternative, more phenomenological approach is to determine the $\Psi$ for which $Q_{\rm heat}=E_{\rm bind}$ at $z=z_{\rm q}$.
This second approach gives an effective $\Psi$ that depends on the critical $Q_{\rm heat}/E_{\rm bind}$ ratio for quenching.

$E_{\rm bind}(z_{\rm q})$ can be estimated semi-empirically.
The problem is  $Q_{\rm heat}(z_{\rm q})$.
In this study, $Q_{\rm heat}$ quantifies the total heat absorbed from all progenitors, not just the heat deposited in the main progenitor at $z_{\rm q}$. 
In central cluster galaxies, which have experienced substantial merger growth since $z_{\rm q}$,
$Q_{\rm heat}\gg Q_{\rm heat}(z_{\rm q})$ and thus $Q_{\rm heat}\gg E_{\rm bind}(z_{\rm q})$ (Table~\ref{tab:qheat}).
Measuring $Q_{\rm heat}(z_{\rm q})$ would require X-ray observations at $z_{\rm q}$, which are not possible. 
We can, however, take $Q_{\rm heat}$ as a proxy for $Q_{\rm heat}(z_{\rm q})$ in low-mass systems that have not grown much since $z_{\rm q}$.

At $M_{200}=8.5\times 10^{11}{\rm\,M}_\odot$,  $Q_{\rm heat}\sim E_{\rm bind}(z_{\rm q})$ for
$2\lesssim\Psi\lesssim 7$ (the values of $E_{\rm bind}$ in Table~\ref{tab:qheat} are normalised for $\Psi=1$). 
For $2\lesssim\Psi\lesssim 7$ and $\epsilon_{\rm accr}=0.1$, GalICS predicts $0.01\lesssim\epsilon_{\rm heat}\lesssim 0.04$ (Eq.~\ref{epsilon_heat_galics}), in excellent agreement with our findings.}

\subsubsection{Implications for maintenance}
\label{sect:implications_maintenance}

We start by examining the maintenance accretion rate $\dot{M}_\bullet^{\rm maint}$ required to keep the gas in thermal equilibrium.
Then we discuss whether the central BHs of elliptical galaxies, groups and clusters accrete at rates comparable, larger or smaller than $\dot{M}_\bullet^{\rm maint}$.

Let $\epsilon_{\rm heat}^{\rm maint}$ be the heating efficiency of AGN in the maintenance mode, which may be different from the cosmic average $\epsilon_{\rm heat}$. The condition:
\begin{equation}
\epsilon_{\rm heat}^{\rm maint}\epsilon_{\rm accr}\dot{M}_\bullet^{\rm maint} c^2=L_{\rm bol}
\label{maintenance1}
\end{equation}
determines the equilibrium accretion rate required to offset the radiative losses.
Evaluating $L_{\rm bol}$ with our LMR and $M_\bullet$ with the $M_\bullet\propto M_{200}^{1.33}$ relation from \citet{gaspari_etal19} gives:
\begin{equation}
\dot{m}_{\rm maint}\equiv{\dot{M}_\bullet^{\rm maint}\over\dot{M}_\bullet^{\rm Edd}}
\sim {2\times 10^{-6}\over\epsilon_{\rm heat}^{\rm maint}}M_{12}^{0.34},
\label{mEdd}
\end{equation}
{where $\dot{M}_\bullet^{\rm Edd}$ is the BH accretion rate required to sustain the Eddington luminosity (assuming $\epsilon_{\rm accr}=0.1$).}
Accretion rates $\dot{m}\lesssim 0.01$ correspond to a radiatively inefficient regime in which most of the power is released mechanically \citep{yuan_narayan14} and $\epsilon_{\rm heat}^{\rm maint}\sim 1$. Hence, the accretion rates required for thermal balance are very low, especially in low-mass haloes.

In order to investigate to what extent $\dot{M}_\bullet\sim\dot{M}_\bullet^{\rm maint}$, we follow \citet{weinberger_pfrommer26},
who evaluated $\dot{M}_\bullet$ based on  \citet{bondi52}'s spherical accretion model, which gives:
\begin{equation}
\dot{M}_\bullet\propto [K_\eta(0)]^{-{3\over 2}}M_\bullet^2\propto\eta^{-{3\over 2}}K_{200}^{-{3\over 2}}M_\bullet^2.
\label{bondi}
\end{equation} 
Substituting Eq.~(\ref{bondi}) into Eq.~(\ref{maintenance1}) and solving for $\eta$ gives the equilibrium ``central-entropy parameter'' $\eta_{\rm eq}$ for which heating balances cooling:
\begin{equation}
\eta_{\rm eq}\propto K_{200}^{-1}M_\bullet^{4\over 3} L_{\rm bol}^{-{2\over 3}}.
\label{maintenance2}
\end{equation}
Eq.~(\ref{bondi}) implies $\dot{M}_\bullet/\dot{M}_\bullet^{\rm maint}=(\eta/\eta_{\rm eq})^{-3/2}$. The BH accretion rate is lower than the maintenance rate if $\eta$ is higher than the equilibrium value.

Figure~\ref{Fig2} compares our findings for $\eta(M_{200})$ (red curve) and $\eta_{\rm eq}(M_{200})$ (gray shaded area; the uncertainty comes from $M_\bullet$).
This comparison shows two regimes: a {\it post-quenching} regime with $\eta\gg \eta_{\rm eq}$ for $M_{200}<10^{14}{\rm\,M}_\odot$ and a {\it self-regulated} regime with $\eta\sim \eta_{\rm eq}$ for $M_{200}>10^{14}{\rm\,M}_\odot$.
In the {post-quenching} regime,
central entropies are high and cooling times are long (e.g., \citealp{donahue_voit22}), but the cooling rate exceeds the heating rate.
The post-quenching regime lasts until $M_{200}\sim 10^{14}{\rm\,M}_\odot$, when haloes enter the self-regulated regime.

We can estimate the BH accretion rate $\dot{m}=(\eta/\eta_{\rm eq})^{-3/2}\dot{m}_{\rm maint}$ 
from Eqs.~(\ref{etafit}) and (\ref{mEdd}). We find:
\begin{equation}
\dot{m}=10^{-7.6+1.2\log_{10}M_{12}-0.04(\log_{10}M_{12})^2}.
\label{BHAR}
\end{equation}
On galactic scales, Eq.~(\ref{BHAR}) predicts accretion rates a hundred times lower than the maintenance values from Eq.~(\ref{maintenance1}).

There are two caveats. First, the Bondi model may not describe BH accretion correctly \citep{prasad_etal17}. \citet{voit_etal15} argued that the precipitation of cold clouds formed through thermal instability provides a more realistic picture of gas supply to supermassive BHs. Second, even if the Bondi model were correct, the density at the Bondi radius is likely higher than the core density of the hot gas \citep{allen_etal06,russell_etal15,bambic_etal23}. This would raise $\eta_{\rm eq}$ and lower the halo mass at which accretion becomes self-regulated.

\subsubsection{Ratio of absorbed heat to internal energy}

Our discussion of quenching in Section~\ref{sect:implications_quenching} was framed in terms of binding energy: quenching occurs when the heat accumulated in the gas exceeds its binding energy.
For a thermodynamic analysis and to compare our findings with those of \citet{donahue_voit22}, it is more convenient to reason in terms of the internal energy:
\begin{equation}
U={3\over 2}{M_{\rm hot}\over \mu m_{\rm p}}kT.
\label{int_en} \end{equation}
For a homogeneous gas, Eqs.~(\ref{s}) and (\ref{int_en}) give $Q_{\rm heat}=T\Delta S= U\ln(K/K_0)$.
$M_{\rm cooled}$ affects both $Q_{\rm heat}$ and $U$, but has little effect on their ratio (Table~\ref{tab:qheat}),  which has a direct link to entropy.   $K\gg K_0$ to the extent that $Q_{\rm heat}/U$ is significant.

Low-mass systems with $\eta\gg\eta_0$ and  high-mass systems with $\eta\sim\eta_0$ correspond to haloes with  $Q> U$ and $Q\ll U$, respectively. 
This difference corresponds to the two regimes discussed in Section~\ref{sect:implications_maintenance}.

The post-quenching regime is, by necessity, a state with $Q_{\rm heat}> U$, since the internal energy is always comparable to the binding energy.
However, $U$ continues to grow through the hierarchical growth of the DM halo after theheat injection ceases.
When $Q_{\rm heat}/U\ll 1$, the heat injection is so diluted that the entropy excesses are no longer measurable, $\eta$ converges to $\eta_{\rm eq}$, and haloes enter the self-regulated regime.

\citet{donahue_voit22}  argued that $Q_{\rm heat}/U$ must be of order unity: $Q_{\rm heat}\gg U$ would completely unbind the gas; $Q_{\rm heat}\ll U$ would not be able to lift it and produce any measurable effects.
They then used the condition $Q_{\rm heat}=\epsilon_{\rm heat}\epsilon_{\rm accr}M_\bullet c^2=U$ to infer $\epsilon_{\rm heat}\sim 0.05$.

Behind the similarity of our results  lies a difference: 
\begin{equation}
 {U\over M_\bullet c^2}={U\over Q_{\rm heat}}\cdot{Q_{\rm heat}\over M_\bullet c^2}.
 \end{equation}
In \citet{donahue_voit22}, $\epsilon_{\rm heat}\propto U/M_\bullet$. Here, $\epsilon_{\rm heat}\propto Q_{\rm heat}/M_\bullet$.
The two approaches give similar efficiencies at $M_{200}\sim 10^{13}{\rm\,M}_\odot$,  where $U/Q_{\rm heat}\sim 1$ (Table~\ref{tab:qheat}),
but we  find that $U/Q_{\rm heat}$ increases by two orders of magnitude from $M_{200}=10^{12}{\rm\,M}_\odot$ to $M_{200}=10^{15}{\rm\,M}_\odot$.
When applied to this entire mass range, the method of  \citet{donahue_voit22}  gives $0.01\lesssim\epsilon_{\rm heat}\lesssim 0.1$.

Assuming $Q_{\rm heat}\sim U$ provides the correct order of magnitude for the heat deposited into the gas on all mass scales, but there are significant deviations from  
this characteristic value at low masses (where $Q_{\rm heat}\gg U$) and at high masses (where $Q_{\rm heat}\ll U$).
Our approach has allowed us to measure $Q_{\rm heat}$ and explore how $Q_{\rm heat}/M_\bullet$ varies across the entire mass range from individual elliptical galaxies to galaxy clusters.

\section{Conclusions}
\label{sect:conc}

We used the entropy excesses observed in elliptical galaxies, groups, and clusters to measure the heat $Q_{\rm heat}$ deposited by AGN into the hot gas, and we have compared $Q_{\rm heat}$ with the energy output from supermassive BHs (Fig.~\ref{QM}). For a canonical energetic efficiency of BH accretion of $10\%$, our findings suggest that 1 to $3\%$ of the energy output of supermassive BHs is thermalised into the surrounding gas.
As BHs release more than $\sim3\%$ of their energy output in the form of jets \citep{heckman_best23}, mechanical feedback alone could explain the observed excess entropy.

\citet{koutsouridou_cattaneo22} and \citet{cattaneo_etal25}
presented a  semi-analytic model in 
which BHs grow until the feedback energy that has accumulated in
the surrounding gas is high enough to unbind it, 
with the effect of quenching
star formation 
and the growth of the BH themselves.
Our findings are consistent
with  the predictions of the semi-analytic model ($0.01\lesssim\epsilon_{\rm heat}\lesssim 0.04$).
A semi-empirical analysis that compares $Q_{\rm heat}$ with the binding energy $E_{\rm bind}(z_{\rm q})$ at the quenching epoch
further supports this scenario.

The comparison of $Q_{\rm heat}$  with the internal energy $U$ of the hot gas at $z=0$ (Table~\ref{tab:qheat}) and the heat $Q_{\rm cool}$ radiated by the hot gas throughout the cosmic lifetime (Fig.~\ref{QM})
shows two regimes:
\begin{enumerate}
\item Elliptical galaxies and groups ($10^{12}{\rm\,M}_\odot\lesssim M_{200}<10^{14}{\rm\,M}_\odot$) are in a {\it post-quenching} regime with
$Q_{\rm heat}\sim U\gg Q_{\rm cool}$. The thermal energy accumulated at the quasar epoch
is still present in the hot gas today and maintains a strong imprint on its thermodynamic state, reflected in large entropy excesses (Fig.~\ref{Fig2}), low hot-gas fractions  (Fig.~\ref{Fig3}), and low
maintenance requirements (Eq.~\ref{mEdd}).
Our work suggests even lower BH accretion rates than the already low maintenance rates (Eq.~\ref{BHAR}).
This prediction can be tested observationally from the demography of AGN.

\item In clusters ($M_{200}> 10^{14}{\rm\,M}_\odot$), $Q_{\rm heat}\sim Q_{\rm cool}\ll U$: the thermal energy deposited by BHs into the hot gas at the quasar epoch has been radiated. 
Even if it had not, it would be low compared to the huge internal energy of the ICM.
Clusters retain no memory of the original energy injection.
They have transitioned to a {\it self-regulated} regime with $\dot{Q}_{\rm heat}\sim\dot{Q}_{\rm cool}$,
as shown by X-ray studies \citep{birzan_etal04,mcnamara_etal06,rafferty_etal06,fornasiero_etal25} and hydrodynamic simulations of the interaction of AGN with the ICM  \citep{cattaneo_teyssier07,li_bryan14,yang_reynolds16}.
Star formation might even be reactivated if AGN did not fully compensate for the energy lost via radiation \citep{bildfell_etal08}.

\end{enumerate}

Our findings have implications for the BH heating efficiencies that  semi-analytic models and
cosmological hydrodynamic simulations can reasonably assume.
The accretion histories of supermassive BHs determine the evolution of
$\Delta S=\epsilon_{\rm heat}\epsilon_{\rm accr}\int_0^{t(z)}T^{-1}\dot{M}_\bullet c^2{\rm\,d}t$.
Our model provides a way to pass from $\Delta S$ 
to $\eta$ and predict X-ray luminosities not only at $z=0$ but also across the Hubble time
(Appendix~\ref{sect:Specificentropy}).
Spatially and spectroscopically resolved observations of haloes at different masses and redshifts with upcoming X-ray instruments (from HUBS\footnote{\url{https://hubs-mission.cn/en/}} to AXIS\footnote{\url{https://axis.umd.edu/}} and NewAthena\footnote{\url{https://www.the-athena-x-ray-observatory.eu/en}})  will probe the implications of the  scenario outlined here in quantitative details.

\begin{acknowledgements}
We thank: the {\it Centre de Conf\'erences Jules Janssen} for supporting Stefano Ettori’s visit to the Observatoire de Paris; 
Vittorio Ghirardini for providing the values quoted in Table~\ref{model_parameters}; Mark Voit and Megan Donahue for comments on the manuscript.
S.E. acknowledges the financial contribution from the {\it Bando INAF per la Ricerca Fondamentale 2024} with a {\it Theory Grant} on
``Constraining the non-thermal pressure in galaxy clusters with high-resolution X-ray spectroscopy'' (1.05.24.05.10).
\end{acknowledgements}

\bibliographystyle{aa} 

\bibliography{ref_av}

@ARTICLE{vikhlinin_etal09,
       author = {{Vikhlinin}, A. and {Burenin}, R.~A. and {Ebeling}, H. and {Forman}, W.~R. and {Hornstrup}, A. and {Jones}, C. and {Kravtsov}, A.~V. and {Murray}, S.~S. and {Nagai}, D. and {Quintana}, H. and {Voevodkin}, A.},
        title = "{Chandra Cluster Cosmology Project. II. Samples and X-Ray Data Reduction}",
      journal = {\apj},
         year = 2009,
        month = feb,
       volume = {692},
       number = {2},
        pages = {1033-1059},
          doi = {10.1088/0004-637X/692/2/1033},
archivePrefix = {arXiv},
       eprint = {0805.2207},
 primaryClass = {astro-ph},
       adsurl = {https://ui.adsabs.harvard.edu/abs/2009ApJ...692.1033V}
}

@INPROCEEDINGS{arnaud96,
       author = {{Arnaud}, K.~A.},
        title = "{XSPEC: The First Ten Years}",
    booktitle = {Astronomical Data Analysis Software and Systems V},
         year = 1996,
       editor = {{Jacoby}, George H. and {Barnes}, Jeannette},
       series = {Astronomical Society of the Pacific Conference Series},
       volume = {101},
        month = jan,
        pages = {17},
       adsurl = {https://ui.adsabs.harvard.edu/abs/1996ASPC..101...17A}
}

@ARTICLE{anders_grevesse89,
       author = {{Anders}, E. and {Grevesse}, N.},
        title = "{Abundances of the elements: Meteoritic and solar}",
      journal = {\gca},
         year = 1989,
        month = jan,
       volume = {53},
       number = {1},
        pages = {197-214},
          doi = {10.1016/0016-7037(89)90286-X},
       adsurl = {https://ui.adsabs.harvard.edu/abs/1989GeCoA..53..197A}
}

@ARTICLE{babyk_mcnamara23,
       author = {{Babyk}, Iurii V. and {McNamara}, Brian R.},
        title = "{The Halo Mass-Temperature Relation for Clusters, Groups, and Galaxies}",
      journal = {\apj},
         year = 2023,
        month = mar,
       volume = {946},
       number = {1},
          eid = {54},
        pages = {54},
          doi = {10.3847/1538-4357/acbf4b},
archivePrefix = {arXiv},
       eprint = {2302.11247},
 primaryClass = {astro-ph.GA},
       adsurl = {https://ui.adsabs.harvard.edu/abs/2023ApJ...946...54B}
}

@ARTICLE{mitchell_schaye22,
       author = {{Mitchell}, Peter D. and {Schaye}, Joop},
        title = "{Baryonic mass budgets for haloes in the EAGLE simulation, including ejected and prevented gas}",
      journal = {\mnras},
         year = 2022,
        month = apr,
       volume = {511},
       number = {2},
        pages = {2600-2609},
          doi = {10.1093/mnras/stab3686},
archivePrefix = {arXiv},
       eprint = {2112.08244},
 primaryClass = {astro-ph.GA},
       adsurl = {https://ui.adsabs.harvard.edu/abs/2022MNRAS.511.2600M}
}

@ARTICLE{ettori26,
       author = {{Ettori}, S.},
        title = "{Beyond Self-Similarity: Reconciling X-Ray Scaling Relations in Galaxy Clusters and Groups}",
      journal = {arXiv e-prints},
         year = 2026,
        month = jun,
          eid = {arXiv:2606.02719},
        pages = {arXiv:2606.02719},
          doi = {10.48550/arXiv.2606.02719},
archivePrefix = {arXiv},
       eprint = {2606.02719},
 primaryClass = {astro-ph.CO},
       adsurl = {https://ui.adsabs.harvard.edu/abs/2026arXiv260602719E}
}

@ARTICLE{donahue_voit22,
       author = {{Donahue}, Megan and {Voit}, G. Mark},
        title = "{Baryon cycles in the biggest galaxies}",
      journal = {\physrep},
         year = 2022,
        month = aug,
       volume = {973},
        pages = {1-109},
          doi = {10.1016/j.physrep.2022.04.005},
archivePrefix = {arXiv},
       eprint = {2204.08099},
 primaryClass = {astro-ph.GA},
       adsurl = {https://ui.adsabs.harvard.edu/abs/2022PhR...973....1D}
}

@ARTICLE{heckman_best23,
       author = {{Heckman}, Timothy M. and {Best}, Philip N.},
        title = "{A Global Inventory of Feedback}",
      journal = {Galaxies},
         year = 2023,
        month = jan,
       volume = {11},
       number = {1},
          eid = {21},
        pages = {21},
          doi = {10.3390/galaxies11010021},
archivePrefix = {arXiv},
       eprint = {2301.11960},
 primaryClass = {astro-ph.GA},
       adsurl = {https://ui.adsabs.harvard.edu/abs/2023Galax..11...21H}
}

@ARTICLE{mitchell_etal09,
       author = {{Mitchell}, N.~L. and {McCarthy}, I.~G. and {Bower}, R.~G. and {Theuns}, T. and {Crain}, R.~A.},
        title = "{On the origin of cores in simulated galaxy clusters}",
      journal = {\mnras},
         year = 2009,
        month = may,
       volume = {395},
       number = {1},
        pages = {180-196},
          doi = {10.1111/j.1365-2966.2009.14550.x},
archivePrefix = {arXiv},
       eprint = {0812.1750},
 primaryClass = {astro-ph},
       adsurl = {https://ui.adsabs.harvard.edu/abs/2009MNRAS.395..180M}
}

@ARTICLE{robotham_etal11,
       author = {{Robotham}, A.~S.~G. and {Norberg}, P. and {Driver}, S.~P. and {Baldry}, I.~K. and {Bamford}, S.~P. and {Hopkins}, A.~M. and {Liske}, J. and {Loveday}, J. and {Merson}, A. and {Peacock}, J.~A. and {Brough}, S.},
        title = "{Galaxy and Mass Assembly (GAMA): the GAMA galaxy group catalogue (G$^{3}$Cv1)}",
      journal = {\mnras},
         year = 2011,
        month = oct,
       volume = {416},
       number = {4},
        pages = {2640-2668},
          doi = {10.1111/j.1365-2966.2011.19217.x},
archivePrefix = {arXiv},
       eprint = {1106.1994},
 primaryClass = {astro-ph.CO},
       adsurl = {https://ui.adsabs.harvard.edu/abs/2011MNRAS.416.2640R}
}

@ARTICLE{bambic_etal23,
       author = {{Bambic}, C.~J. and {Russell}, H.~R. and {Reynolds}, C.~S. and {Fabian}, A.~C. and {McNamara}, B.~R. and {Nulsen}, P.~E.~J.},
        title = "{AGN feeding and feedback in M84: from kiloparsec scales to the Bondi radius}",
      journal = {\mnras},
         year = 2023,
        month = jul,
       volume = {522},
       number = {3},
        pages = {4374-4391},
          doi = {10.1093/mnras/stad824},
archivePrefix = {arXiv},
       eprint = {2301.11937},
 primaryClass = {astro-ph.HE},
       adsurl = {https://ui.adsabs.harvard.edu/abs/2023MNRAS.522.4374B}
}

@ARTICLE{russell_etal15,
       author = {{Russell}, H.~R. and {Fabian}, A.~C. and {McNamara}, B.~R. and {Broderick}, A.~E.},
        title = "{Inside the Bondi radius of M87}",
      journal = {\mnras},
         year = 2015,
        month = jul,
       volume = {451},
       number = {1},
        pages = {588-600},
          doi = {10.1093/mnras/stv954},
archivePrefix = {arXiv},
       eprint = {1504.07633},
 primaryClass = {astro-ph.GA},
       adsurl = {https://ui.adsabs.harvard.edu/abs/2015MNRAS.451..588R}
}

@ARTICLE{prasad_etal17,
       author = {{Prasad}, Deovrat and {Sharma}, Prateek and {Babul}, Arif},
        title = "{AGN jet-driven stochastic cold accretion in cluster cores}",
      journal = {\mnras},
         year = 2017,
        month = oct,
       volume = {471},
       number = {2},
        pages = {1531-1542},
          doi = {10.1093/mnras/stx1698},
archivePrefix = {arXiv},
       eprint = {1611.02710},
 primaryClass = {astro-ph.GA},
       adsurl = {https://ui.adsabs.harvard.edu/abs/2017MNRAS.471.1531P}
}

@ARTICLE{weinberger_pfrommer26,
       author = {{Weinberger}, R. and {Pfrommer}, C.}, 
        title = "{How supermassive black holes shape central entropies in galaxy clusters}",
      journal = {\aap},
         year = 2026,
        month = feb,
       volume = {706},
          eid = {L13},
        pages = {L13},
          doi = {10.1051/0004-6361/202558710},
archivePrefix = {arXiv},
       eprint = {2601.20932},
 primaryClass = {astro-ph.GA},
       adsurl = {https://ui.adsabs.harvard.edu/abs/2026A&A...706L..13W}
}

@ARTICLE{yuan_narayan14,
       author = {{Yuan}, Feng and {Narayan}, Ramesh},
        title = "{Hot Accretion Flows Around Black Holes}",
      journal = {\araa},
         year = 2014,
        month = aug,
       volume = {52},
        pages = {529-588},
          doi = {10.1146/annurev-astro-082812-141003},
archivePrefix = {arXiv},
       eprint = {1401.0586},
 primaryClass = {astro-ph.HE},
       adsurl = {https://ui.adsabs.harvard.edu/abs/2014ARA&A..52..529Y}
}

@ARTICLE{zhang_erratum,
       author = {{Zhang}, Haowen and {Behroozi}, Peter and {Volonteri}, Marta and {Silk}, Joseph and {Fan}, Xiaohui and {Hopkins}, Philip F. and {Yang}, Jinyi and {Aird}, James},
        title = "{TRINITY I: self-consistently modelling the dark matter halo-galaxy-supermassive black hole connection from z = 0-10}",
      journal = {\mnras},
         year = 2023,
        month = jan,
       volume = {518},
       number = {2},
        pages = {2123-2163},
          doi = {10.1093/mnras/stac2633},
archivePrefix = {arXiv},
       eprint = {2105.10474},
 primaryClass = {astro-ph.GA},
       adsurl = {https://ui.adsabs.harvard.edu/abs/2023MNRAS.518.2123Z}
}

@ARTICLE{vandenbosch_etal14,
       author = {{van den Bosch}, Frank C. and {Jiang}, Fangzhou and {Hearin}, Andrew and {Campbell}, Duncan and {Watson}, Douglas and {Padmanabhan}, Nikhil},
        title = "{Coming of age in the dark sector: how dark matter haloes grow their gravitational potential wells}",
      journal = {\mnras},
         year = 2014,
        month = dec,
       volume = {445},
       number = {2},
        pages = {1713-1730},
          doi = {10.1093/mnras/stu1872},
archivePrefix = {arXiv},
       eprint = {1409.2750},
 primaryClass = {astro-ph.GA},
       adsurl = {https://ui.adsabs.harvard.edu/abs/2014MNRAS.445.1713V}
}

@ARTICLE{correa_etal15,
       author = {{Correa}, Camila A. and {Wyithe}, J. Stuart B. and {Schaye}, Joop and {Duffy}, Alan R.},
        title = "{The accretion history of dark matter haloes - I. The physical origin of the universal function}",
      journal = {\mnras},
         year = 2015,
        month = jun,
       volume = {450},
       number = {2},
        pages = {1514-1520},
          doi = {10.1093/mnras/stv689},
archivePrefix = {arXiv},
       eprint = {1409.5228},
 primaryClass = {astro-ph.GA},
       adsurl = {https://ui.adsabs.harvard.edu/abs/2015MNRAS.450.1514C}
}

@ARTICLE{popesso07,
       author = {{Popesso}, P. and {Biviano}, A. and {B{\"o}hringer}, H. and {Romaniello}, M.},
        title = "{RASS-SDSS galaxy cluster survey. V. The X-ray-underluminous Abell clusters}",
      journal = {\aap},
         year = 2007,
        month = jan,
       volume = {461},
       number = {2},
        pages = {397-410},
          doi = {10.1051/0004-6361:20054493},
archivePrefix = {arXiv},
       eprint = {astro-ph/0606191},
 primaryClass = {astro-ph},
       adsurl = {https://ui.adsabs.harvard.edu/abs/2007A&A...461..397P}
}

@ARTICLE{yan_etal19,
       author = {{Yan}, Zhiqiang and {Jerabkova}, Tereza and {Kroupa}, Pavel},
        title = "{The star formation timescale of elliptical galaxies. Fitting [Mg/Fe] and total metallicity simultaneously}",
      journal = {\aap},
         year = 2019,
        month = dec,
       volume = {632},
          eid = {A110},
        pages = {A110},
          doi = {10.1051/0004-6361/201936636},
archivePrefix = {arXiv},
       eprint = {1911.02568},
 primaryClass = {astro-ph.GA},
       adsurl = {https://ui.adsabs.harvard.edu/abs/2019A&A...632A.110Y}
}

@ARTICLE{bravo_etal23,
       author = {{Bravo}, Mat{\'\i}as and {Robotham}, Aaron S.~G. and {Lagos}, Claudia del P. and {Davies}, Luke J.~M. and {Bellstedt}, Sabine and {Thorne}, Jessica E.},
        title = "{Galaxy quenching time-scales from a forensic reconstruction of their colour evolution}",
      journal = {\mnras},
         year = 2023,
        month = jul,
       volume = {522},
       number = {3},
        pages = {4481-4498},
          doi = {10.1093/mnras/stad1234},
archivePrefix = {arXiv},
       eprint = {2301.03702},
 primaryClass = {astro-ph.GA},
       adsurl = {https://ui.adsabs.harvard.edu/abs/2023MNRAS.522.4481B}
}

@ARTICLE{darvish_etal18,
       author = {{Darvish}, Behnam and {Martin}, Christopher and {Gon{\c{c}}alves}, Thiago S. and {Mobasher}, Bahram and {Scoville}, Nick Z. and {Sobral}, David},
        title = "{Quenching or Bursting: The Role of Stellar Mass, Environment, and Specific Star Formation Rate to z\textbackslashsim 1}",
      journal = {\apj},
         year = 2018,
        month = feb,
       volume = {853},
       number = {2},
          eid = {155},
        pages = {155},
          doi = {10.3847/1538-4357/aaa5a4},
archivePrefix = {arXiv},
       eprint = {1801.02618},
 primaryClass = {astro-ph.GA},
       adsurl = {https://ui.adsabs.harvard.edu/abs/2018ApJ...853..155D}
}

@ARTICLE{lian_etal16,
       author = {{Lian}, Jianhui and {Yan}, Renbin and {Zhang}, Kai and {Kong}, Xu},
        title = "{The Quenching Timescale and Quenching Rate of Galaxies}",
      journal = {\apj},
         year = 2016,
        month = nov,
       volume = {832},
       number = {1},
          eid = {29},
        pages = {29},
          doi = {10.3847/0004-637X/832/1/29},
archivePrefix = {arXiv},
       eprint = {1609.04805},
 primaryClass = {astro-ph.GA},
       adsurl = {https://ui.adsabs.harvard.edu/abs/2016ApJ...832...29L}
}

@ARTICLE{bogdan_etal18,
       author = {{Bogd{\'a}n}, {\'A}kos and {Lovisari}, Lorenzo and {Volonteri}, Marta and {Dubois}, Yohan},
        title = "{Correlation between the Total Gravitating Mass of Groups and Clusters and the Supermassive Black Hole Mass of Brightest Galaxies}",
      journal = {\apj},
         year = 2018,
        month = jan,
       volume = {852},
       number = {2},
          eid = {131},
        pages = {131},
          doi = {10.3847/1538-4357/aa9ab5},
archivePrefix = {arXiv},
       eprint = {1711.09900},
 primaryClass = {astro-ph.GA},
       adsurl = {https://ui.adsabs.harvard.edu/abs/2018ApJ...852..131B}
}

@ARTICLE{punsly_etal20,
       author = {{Punsly}, Brian and {Hill}, Gary J. and {Marziani}, Paola and {Kharb}, Preeti and {Berton}, Marco and {Crepaldi}, Luca and {Indahl}, Briana L. and {Zeimann}, Greg},
        title = "{The Energetics of Launching the Most Powerful Jets in Quasars: A Study of 3C 82}",
      journal = {\apj},
         year = 2020,
        month = aug,
       volume = {898},
       number = {2},
          eid = {169},
        pages = {169},
          doi = {10.3847/1538-4357/aba1e8},
archivePrefix = {arXiv},
       eprint = {2007.00170},
 primaryClass = {astro-ph.GA},
       adsurl = {https://ui.adsabs.harvard.edu/abs/2020ApJ...898..169P}
}

@ARTICLE{lopez_perucho12,
       author = {{L{\'o}pez-Corredoira}, M. and {Perucho}, M.},
        title = "{Kinetic power of quasars and statistical excess of MOJAVE superluminal motions}",
      journal = {\aap},
         year = 2012,
        month = aug,
       volume = {544},
          eid = {A56},
        pages = {A56},
          doi = {10.1051/0004-6361/201219410},
archivePrefix = {arXiv},
       eprint = {1206.6282},
 primaryClass = {astro-ph.CO},
       adsurl = {https://ui.adsabs.harvard.edu/abs/2012A&A...544A..56L}
}

@ARTICLE{inoue_etal17,
       author = {{Inoue}, Yoshiyuki and {Doi}, Akihiro and {Tanaka}, Yasuyuki T. and {Sikora}, Marek and {Madejski}, Grzegorz M.},
        title = "{Disk-Jet Connection in Active Supermassive Black Holes in the Standard Accretion Disk Regime}",
      journal = {\apj},
         year = 2017,
        month = may,
       volume = {840},
       number = {1},
          eid = {46},
        pages = {46},
          doi = {10.3847/1538-4357/aa6b57},
archivePrefix = {arXiv},
       eprint = {1704.00123},
 primaryClass = {astro-ph.HE},
       adsurl = {https://ui.adsabs.harvard.edu/abs/2017ApJ...840...46I}
}

@ARTICLE{mehrgan_etal19,
       author = {{Mehrgan}, Kianusch and {Thomas}, Jens and {Saglia}, Roberto and {Mazzalay}, Ximena and {Erwin}, Peter and {Bender}, Ralf and {Kluge}, Matthias and {Fabricius}, Maximilian},
        title = "{A 40 Billion Solar-mass Black Hole in the Extreme Core of Holm 15A, the Central Galaxy of Abell 85}",
      journal = {\apj},
         year = 2019,
        month = dec,
       volume = {887},
       number = {2},
          eid = {195},
        pages = {195},
          doi = {10.3847/1538-4357/ab5856},
archivePrefix = {arXiv},
       eprint = {1907.10608},
 primaryClass = {astro-ph.GA},
       adsurl = {https://ui.adsabs.harvard.edu/abs/2019ApJ...887..195M}
}

@ARTICLE{marasco_etal21,
       author = {{Marasco}, A. and {Cresci}, G. and {Posti}, L. and {Fraternali}, F. and {Mannucci}, F. and {Marconi}, A. and {Belfiore}, F. and {Fall}, S.~M.},
        title = "{A universal relation between the properties of supermassive black holes, galaxies, and dark matter haloes}",
      journal = {\mnras},
         year = 2021,
        month = nov,
       volume = {507},
       number = {3},
        pages = {4274-4293},
          doi = {10.1093/mnras/stab2317},
archivePrefix = {arXiv},
       eprint = {2105.10508},
 primaryClass = {astro-ph.GA},
       adsurl = {https://ui.adsabs.harvard.edu/abs/2021MNRAS.507.4274M}
}

@article{chiu_etal22,
   title={The eROSITA Final Equatorial-Depth Survey (eFEDS): X-ray observable-to-mass-and-redshift relations of galaxy clusters and groups with weak-lensing mass calibration from the Hyper Suprime-Cam Subaru Strategic Program survey},
   volume={661},
   ISSN={1432-0746},
   url={http://dx.doi.org/10.1051/0004-6361/202141755},
   DOI={10.1051/0004-6361/202141755},
   journal={\aap},
   publisher={EDP Sciences},
   author={Chiu, I-Non and Ghirardini, Vittorio and Liu, Ang and Grandis, Sebastian and Bulbul, Esra and Emre Bahar, Y. and Comparat, Johan and Bocquet, Sebastian and Clerc, Nicolas and Klein, Matthias and Liu, Teng and Li, Xiangchong and Miyatake, Hironao and Mohr, Joseph},
   year={2022},
   month=may, pages={A11} }

@ARTICLE{mcbride_etal09,
       author = {{McBride}, James and {Fakhouri}, Onsi and {Ma}, Chung-Pei},
        title = "{Mass accretion rates and histories of dark matter haloes}",
      journal = {\mnras},
         year = 2009,
        month = oct,
       volume = {398},
       number = {4},
        pages = {1858-1868},
          doi = {10.1111/j.1365-2966.2009.15329.x},
archivePrefix = {arXiv},
       eprint = {0902.3659},
 primaryClass = {astro-ph.CO},
       adsurl = {https://ui.adsabs.harvard.edu/abs/2009MNRAS.398.1858M}
}

@ARTICLE{mazzotta_etal04,
       author = {{Mazzotta}, P. and {Rasia}, E. and {Moscardini}, L. and {Tormen}, G.},
        title = "{Comparing the temperatures of galaxy clusters from hydrodynamical N-body simulations to Chandra and XMM-Newton observations}",
      journal = {\mnras},
         year = 2004,
        month = oct,
       volume = {354},
       number = {1},
        pages = {10-24},
          doi = {10.1111/j.1365-2966.2004.08167.x},
archivePrefix = {arXiv},
       eprint = {astro-ph/0404425},
 primaryClass = {astro-ph},
       adsurl = {https://ui.adsabs.harvard.edu/abs/2004MNRAS.354...10M}
}

@ARTICLE{reichert_etal11,
       author = {{Reichert}, A. and {B{\"o}hringer}, H. and {Fassbender}, R. and {M{\"u}hlegger}, M.},
        title = "{Observational constraints on the redshift evolution of X-ray scaling relations of galaxy clusters out to z \raisebox{-0.5ex}\textasciitilde 1.5}",
      journal = {\aap},
         year = 2011,
        month = nov,
       volume = {535},
          eid = {A4},
        pages = {A4},
          doi = {10.1051/0004-6361/201116861},
archivePrefix = {arXiv},
       eprint = {1109.3708},
 primaryClass = {astro-ph.CO},
       adsurl = {https://ui.adsabs.harvard.edu/abs/2011A&A...535A...4R}
}

@ARTICLE{sereno_etal15,
       author = {{Sereno}, Mauro and {Ettori}, Stefano},
        title = "{CoMaLit - IV. Evolution and self-similarity of scaling relations with the galaxy cluster mass}",
      journal = {\mnras},
         year = 2015,
        month = jul,
       volume = {450},
       number = {4},
        pages = {3675-3695},
          doi = {10.1093/mnras/stv814},
       adsurl = {https://ui.adsabs.harvard.edu/abs/2015MNRAS.450.3675S}
}

@ARTICLE{schawinski_etal14,
       author = {{Schawinski}, Kevin and {Urry}, C. Megan and {Simmons}, Brooke D. and {Fortson}, Lucy and {Kaviraj}, Sugata and {Keel}, William C. and {Lintott}, Chris J. and {Masters}, Karen L. and {Nichol}, Robert C. and {Sarzi}, Marc and {Skibba}, Ramin and {Treister}, Ezequiel and {Willett}, Kyle W. and {Wong}, O. Ivy and {Yi}, Sukyoung K.},
        title = "{The green valley is a red herring: Galaxy Zoo reveals two evolutionary pathways towards quenching of star formation in early- and late-type galaxies}",
      journal = {\mnras},
         year = 2014,
        month = may,
       volume = {440},
       number = {1},
        pages = {889-907},
          doi = {10.1093/mnras/stu327},
archivePrefix = {arXiv},
       eprint = {1402.4814},
 primaryClass = {astro-ph.GA},
       adsurl = {https://ui.adsabs.harvard.edu/abs/2014MNRAS.440..889S}
}

@ARTICLE{li_bryan14,
       author = {{Li}, Yuan and {Bryan}, Greg L.},
        title = "{Modeling Active Galactic Nucleus Feedback in Cool-core Clusters: The Balance between Heating and Cooling}",
      journal = {\apj},
         year = 2014,
        month = jul,
       volume = {789},
       number = {1},
          eid = {54},
        pages = {54},
          doi = {10.1088/0004-637X/789/1/54},
archivePrefix = {arXiv},
       eprint = {1401.7695},
 primaryClass = {astro-ph.GA},
       adsurl = {https://ui.adsabs.harvard.edu/abs/2014ApJ...789...54L}
}

@ARTICLE{mcnamara_etal06,
       author = {{McNamara}, B.~R. and {Rafferty}, D.~A. and {B{\^\i}rzan}, L. and {Steiner}, J. and {Wise}, M.~W. and {Nulsen}, P.~E.~J. and {Carilli}, C.~L. and {Ryan}, R. and {Sharma}, M.},
        title = "{The Starburst in the Abell 1835 Cluster Central Galaxy: A Case Study of Galaxy Formation Regulated by an Outburst from a Supermassive Black Hole}",
      journal = {\apj},
         year = 2006,
        month = sep,
       volume = {648},
       number = {1},
        pages = {164-175},
          doi = {10.1086/505859},
archivePrefix = {arXiv},
       eprint = {astro-ph/0604044},
 primaryClass = {astro-ph},
       adsurl = {https://ui.adsabs.harvard.edu/abs/2006ApJ...648..164M}
}

@ARTICLE{fornasiero_etal25,
       author = {{Fornasiero}, I. and {Ubertosi}, F. and {Gitti}, M.},
        title = "{Investigating AGN feedback in H{\ensuremath{\alpha}}-luminous galaxy clusters: First Chandra X-ray analysis of Abell 2009}",
      journal = {\aap},
         year = 2025,
        month = mar,
       volume = {695},
          eid = {A265},
        pages = {A265},
          doi = {10.1051/0004-6361/202453315},
archivePrefix = {arXiv},
       eprint = {2503.07781},
 primaryClass = {astro-ph.GA},
       adsurl = {https://ui.adsabs.harvard.edu/abs/2025A&A...695A.265F}
}

@ARTICLE{fu_etal25,
       author = {{Fu}, Hao and {Boco}, Lumen and {Shankar}, Francesco and {Lapi}, Andrea and {Ayromlou}, Mohammadreza and {Roberts}, Daniel and {Peng}, Yingjie and {Rodr{\'\i}guez-Puebla}, Aldo and {Yuan}, Feng and {Cleland}, Cressida and {Mei}, Simona and {Menci}, Nicola},
        title = "{Shedding light on the star formation rate-halo accretion rate connection and halo quenching mechanism via DECODE, the Discrete statistical sEmi-empiriCal mODEl}",
      journal = {\aap},
         year = 2025,
        month = mar,
       volume = {695},
          eid = {A252},
        pages = {A252},
          doi = {10.1051/0004-6361/202453218},
archivePrefix = {arXiv},
       eprint = {2502.06942},
 primaryClass = {astro-ph.GA},
       adsurl = {https://ui.adsabs.harvard.edu/abs/2025A&A...695A.252F}
}

@ARTICLE{voit_etal15,
       author = {{Voit}, G.~M. and {Donahue}, M. and {Bryan}, G.~L. and {McDonald}, M.},
        title = "{Regulation of star formation in giant galaxies by precipitation, feedback and conduction}",
      journal = {\nat},
         year = 2015,
        month = mar,
       volume = {519},
       number = {7542},
        pages = {203-206},
          doi = {10.1038/nature14167},
archivePrefix = {arXiv},
       eprint = {1409.1598},
 primaryClass = {astro-ph.GA},
       adsurl = {https://ui.adsabs.harvard.edu/abs/2015Natur.519..203V}
}

@ARTICLE{rodriguez_etal17,
       author = {{Rodr{\'\i}guez-Puebla}, Aldo and {Primack}, Joel R. and {Avila-Reese}, Vladimir and {Faber}, S.~M.},
        title = "{Constraining the galaxy-halo connection over the last 13.3 Gyr: star formation histories, galaxy mergers and structural properties}",
      journal = {\mnras},
         year = 2017,
        month = sep,
       volume = {470},
       number = {1},
        pages = {651-687},
          doi = {10.1093/mnras/stx1172},
archivePrefix = {arXiv},
       eprint = {1703.04542},
 primaryClass = {astro-ph.GA},
       adsurl = {https://ui.adsabs.harvard.edu/abs/2017MNRAS.470..651R}
}

@ARTICLE{moster_etal18,
       author = {{Moster}, Benjamin P. and {Naab}, Thorsten and {White}, Simon D.~M.},
        title = "{EMERGE - an empirical model for the formation of galaxies since z {\ensuremath{\sim}} 10}",
      journal = {\mnras},
         year = 2018,
        month = jun,
       volume = {477},
       number = {2},
        pages = {1822-1852},
          doi = {10.1093/mnras/sty655},
archivePrefix = {arXiv},
       eprint = {1705.05373},
 primaryClass = {astro-ph.GA},
       adsurl = {https://ui.adsabs.harvard.edu/abs/2018MNRAS.477.1822M}
}

@ARTICLE{cattaneo_best09,
       author = {{Cattaneo}, A. and {Best}, P.~N.},
        title = "{On the jet contribution to the active galactic nuclei cosmic energy budget}",
      journal = {\mnras},
         year = 2009,
        month = may,
       volume = {395},
       number = {1},
        pages = {518-523},
          doi = {10.1111/j.1365-2966.2009.14557.x},
archivePrefix = {arXiv},
       eprint = {0812.1562},
 primaryClass = {astro-ph},
       adsurl = {https://ui.adsabs.harvard.edu/abs/2009MNRAS.395..518C}
}

@ARTICLE{ostriker_etal05,
       author = {{Ostriker}, Jeremiah P. and {Bode}, Paul and {Babul}, Arif},
        title = "{A Simple and Accurate Model for Intracluster Gas}",
      journal = {\apj},
         year = 2005,
        month = dec,
       volume = {634},
       number = {2},
        pages = {964-976},
          doi = {10.1086/497122},
archivePrefix = {arXiv},
       eprint = {astro-ph/0504334},
 primaryClass = {astro-ph},
       adsurl = {https://ui.adsabs.harvard.edu/abs/2005ApJ...634..964O}
}

@ARTICLE{behroozi_etal19,
       author = {{Behroozi}, Peter and {Wechsler}, Risa H. and {Hearin}, Andrew P. and {Conroy}, Charlie},
        title = "{UNIVERSEMACHINE: The correlation between galaxy growth and dark matter halo assembly from z = 0-10}",
      journal = {\mnras},
         year = 2019,
        month = sep,
       volume = {488},
       number = {3},
        pages = {3143-3194},
          doi = {10.1093/mnras/stz1182},
archivePrefix = {arXiv},
       eprint = {1806.07893},
 primaryClass = {astro-ph.GA},
       adsurl = {https://ui.adsabs.harvard.edu/abs/2019MNRAS.488.3143B}
}

@ARTICLE{bildfell_etal08,
       author = {{Bildfell}, Chris and {Hoekstra}, Henk and {Babul}, Arif and {Mahdavi}, Andisheh},
        title = "{Resurrecting the red from the dead: optical properties of BCGs in X-ray luminous clusters}",
      journal = {\mnras},
         year = 2008,
        month = oct,
       volume = {389},
       number = {4},
        pages = {1637-1654},
          doi = {10.1111/j.1365-2966.2008.13699.x},
archivePrefix = {arXiv},
       eprint = {0802.2712},
 primaryClass = {astro-ph},
       adsurl = {https://ui.adsabs.harvard.edu/abs/2008MNRAS.389.1637B}
}

@ARTICLE{eckert_etal25,
       author = {{Eckert}, D. and {Gastaldello}, F. and {Lovisari}, L. and {McGee}, S. and {Pasini}, T. and {Brienza}, M. and {Kolokythas}, K. and {O'Sullivan}, E. and {Simionescu}, A. and {Sun}, M. and {Ayromlou}, M. and {Bourne}, M.~A. and {Chen}, Y. and {Cui}, W. and {Ettori}, S. and {Finoguenov}, A.},
        title = "{Extreme AGN feedback in the fossil galaxy group SDSSTG 4436}",
      journal = {\aap},
         year = 2025,
        month = sep,
       volume = {701},
          eid = {A127},
        pages = {A127},
          doi = {10.1051/0004-6361/202555212},
archivePrefix = {arXiv},
       eprint = {2506.13907},
 primaryClass = {astro-ph.GA},
       adsurl = {https://ui.adsabs.harvard.edu/abs/2025A&A...701A.127E}
}

@ARTICLE{zhang_etal23,
       author = {{Zhang}, Haowen and {Behroozi}, Peter and {Volonteri}, Marta and {Silk}, Joseph and {Fan}, Xiaohui and {Hopkins}, Philip F. and {Yang}, Jinyi and {Aird}, James},
        title = "{TRINITY I: self-consistently modelling the dark matter halo-galaxy-supermassive black hole connection from z = 0-10}",
      journal = {\mnras},
         year = 2023,
        month = jan,
       volume = {518},
       number = {2},
        pages = {2123-2163},
          doi = {10.1093/mnras/stac2633},
archivePrefix = {arXiv},
       eprint = {2105.10474},
 primaryClass = {astro-ph.GA},
       adsurl = {https://ui.adsabs.harvard.edu/abs/2023MNRAS.518.2123Z}
}

@ARTICLE{biffi_valdarnini15,
       author = {{Biffi}, V. and {Valdarnini}, R.},
        title = "{The role of the artificial conductivity in SPH simulations of galaxy clusters: effects on the ICM properties}",
      journal = {\mnras},
         year = 2015,
        month = jan,
       volume = {446},
       number = {3},
        pages = {2802-2822},
          doi = {10.1093/mnras/stu2278},
archivePrefix = {arXiv},
       eprint = {1410.8529},
 primaryClass = {astro-ph.CO},
       adsurl = {https://ui.adsabs.harvard.edu/abs/2015MNRAS.446.2802B}
}

@ARTICLE{power_etal14,
       author = {{Power}, C. and {Read}, J.~I. and {Hobbs}, A.},
        title = "{The formation of entropy cores in non-radiative galaxy cluster simulations: smoothed particle hydrodynamics versus adaptive mesh refinement}",
      journal = {\mnras},
         year = 2014,
        month = jun,
       volume = {440},
       number = {4},
        pages = {3243-3256},
          doi = {10.1093/mnras/stu418},
archivePrefix = {arXiv},
       eprint = {1307.0668},
 primaryClass = {astro-ph.CO},
       adsurl = {https://ui.adsabs.harvard.edu/abs/2014MNRAS.440.3243P}
}

@ARTICLE{valdarnini12,
       author = {{Valdarnini}, R.},
        title = "{Hydrodynamic capabilities of an SPH code incorporating an artificial conductivity term with a gravity-based signal velocity}",
      journal = {\aap},
         year = 2012,
        month = oct,
       volume = {546},
          eid = {A45},
        pages = {A45},
          doi = {10.1051/0004-6361/201219715},
archivePrefix = {arXiv},
       eprint = {1207.6980},
 primaryClass = {astro-ph.IM},
       adsurl = {https://ui.adsabs.harvard.edu/abs/2012A&A...546A..45V}
}

@ARTICLE{anderson_etal15,
       author = {{Anderson}, Michael E. and {Gaspari}, Massimo and {White}, Simon D.~M. and {Wang}, Wenting and {Dai}, Xinyu},
        title = "{Unifying X-ray scaling relations from galaxies to clusters}",
      journal = {\mnras},
         year = 2015,
        month = jun,
       volume = {449},
       number = {4},
        pages = {3806-3826},
          doi = {10.1093/mnras/stv437},
archivePrefix = {arXiv},
       eprint = {1409.6965},
 primaryClass = {astro-ph.CO},
       adsurl = {https://ui.adsabs.harvard.edu/abs/2015MNRAS.449.3806A}
}

@ARTICLE{andreon_etal24,
       author = {{Andreon}, S. and {Trinchieri}, G. and {Moretti}, A.},
        title = "{Observed abundance of X-ray low surface brightness clusters in optical, X-ray, and SZ selected samples}",
      journal = {\aap},
         year = 2024,
        month = jun,
       volume = {686},
          eid = {A284},
        pages = {A284},
          doi = {10.1051/0004-6361/202345900},
archivePrefix = {arXiv},
       eprint = {2404.12435},
 primaryClass = {astro-ph.CO},
       adsurl = {https://ui.adsabs.harvard.edu/abs/2024A&A...686A.284A}
}

@ARTICLE{markevitch98,
       author = {{Markevitch}, Maxim},
        title = "{The L$_{X}$-T Relation and Temperature Function for Nearby Clusters Revisited}",
      journal = {\apj},
         year = 1998,
        month = sep,
       volume = {504},
       number = {1},
        pages = {27-34},
          doi = {10.1086/306080},
archivePrefix = {arXiv},
       eprint = {astro-ph/9802059},
 primaryClass = {astro-ph},
       adsurl = {https://ui.adsabs.harvard.edu/abs/1998ApJ...504...27M}
}

@ARTICLE{mccarthy_etal10,
       author = {{McCarthy}, I.~G. and {Schaye}, J. and {Ponman}, T.~J. and {Bower}, R.~G. and {Booth}, C.~M. and {Dalla Vecchia}, C. and {Crain}, R.~A. and {Springel}, V. and {Theuns}, T. and {Wiersma}, R.~P.~C.},
        title = "{The case for AGN feedback in galaxy groups}",
      journal = {\mnras},
         year = 2010,
        month = aug,
       volume = {406},
       number = {2},
        pages = {822-839},
          doi = {10.1111/j.1365-2966.2010.16750.x},
archivePrefix = {arXiv},
       eprint = {0911.2641},
 primaryClass = {astro-ph.CO},
       adsurl = {https://ui.adsabs.harvard.edu/abs/2010MNRAS.406..822M}
}

@ARTICLE{birzan_etal04,
       author = {{B{\^\i}rzan}, L. and {Rafferty}, D.~A. and {McNamara}, B.~R. and {Wise}, M.~W. and {Nulsen}, P.~E.~J.},
        title = "{A Systematic Study of Radio-induced X-Ray Cavities in Clusters, Groups, and Galaxies}",
      journal = {\apj},
         year = 2004,
        month = jun,
       volume = {607},
       number = {2},
        pages = {800-809},
          doi = {10.1086/383519},
archivePrefix = {arXiv},
       eprint = {astro-ph/0402348},
 primaryClass = {astro-ph},
       adsurl = {https://ui.adsabs.harvard.edu/abs/2004ApJ...607..800B}
}

@ARTICLE{tollet_etal22,
       author = {{Tollet}, E. and {Cattaneo}, A. and {Macci{\`o}}, A.~V. and {Kang}, X.},
        title = "{Cold-mode and hot-mode accretion in galaxy formation: an entropy approach}",
      journal = {\mnras},
         year = 2022,
        month = sep,
       volume = {515},
       number = {3},
        pages = {3453-3471},
          doi = {10.1093/mnras/stac1867},
archivePrefix = {arXiv},
       eprint = {2206.12637},
 primaryClass = {astro-ph.GA},
       adsurl = {https://ui.adsabs.harvard.edu/abs/2022MNRAS.515.3453T}
}

@ARTICLE{zhu_etal21,
       author = {{Zhu}, Zhenghao and {Xu}, Haiguang and {Hu}, Dan and {Shan}, Chenxi and {Zhu}, Yongkai and {Fan}, Shida and {Zhao}, Yuanyuan and {Gu}, Liyi and {Wu}, Xiang-Ping},
        title = "{A Study of Gas Entropy Profiles of 47 Galaxy Clusters and Groups out to the Virial Radius}",
      journal = {\apj},
         year = 2021,
        month = feb,
       volume = {908},
       number = {1},
          eid = {17},
        pages = {17},
          doi = {10.3847/1538-4357/abd327},
archivePrefix = {arXiv},
       eprint = {2101.05947},
 primaryClass = {astro-ph.CO},
       adsurl = {https://ui.adsabs.harvard.edu/abs/2021ApJ...908...17Z}
}

@ARTICLE{zhang_etal24,
       author = {{Zhang}, Yi and {Comparat}, Johan and {Ponti}, Gabriele and {Merloni}, Andrea and {Nandra}, Kirpal and {Haberl}, Frank and {Truong}, Nhut and {Pillepich}, Annalisa and {Locatelli}, Nicola and {Zhang}, Xiaoyuan and {Sanders}, Jeremy and {Zheng}, Xueying and {Liu}, Ang and {Popesso}, Paola and {Liu}, Teng and {Predehl}, Peter and {Salvato}, Mara and {Shreeram}, Soumya and {Yeung}, Michael C.~H.},
        title = "{The hot circumgalactic medium in the eROSITA All-Sky Survey: II. Scaling relations between X-ray luminosity and galaxies' mass}",
      journal = {\aap},
         year = 2024,
        month = oct,
       volume = {690},
          eid = {A268},
        pages = {A268},
          doi = {10.1051/0004-6361/202449413},
archivePrefix = {arXiv},
       eprint = {2401.17309},
 primaryClass = {astro-ph.GA},
       adsurl = {https://ui.adsabs.harvard.edu/abs/2024A&A...690A.268Z}
}

@ARTICLE{lin_etal17,
       author = {{Lin}, Yen-Ting and {Hsieh}, Bau-Ching and {Lin}, Sheng-Chieh and {Oguri}, Masamune and {Chen}, Kai-Feng and {Tanaka}, Masayuki and {Chiu}, I. -Non and {Huang}, Song and {Kodama}, Tadayuki and {Leauthaud}, Alexie and {More}, Surhud and {Nishizawa}, Atsushi J. and {Bundy}, Kevin and {Lin}, Lihwai and {Miyazaki}, Satoshi},
        title = "{First Results on the Cluster Galaxy Population from the Subaru Hyper Suprime-Cam Survey. III. Brightest Cluster Galaxies, Stellar Mass Distribution, and Active Galaxies}",
      journal = {\apj},
         year = 2017,
        month = dec,
       volume = {851},
       number = {2},
          eid = {139},
        pages = {139},
          doi = {10.3847/1538-4357/aa9bf5},
archivePrefix = {arXiv},
       eprint = {1709.04484},
 primaryClass = {astro-ph.GA},
       adsurl = {https://ui.adsabs.harvard.edu/abs/2017ApJ...851..139L}
}

@ARTICLE{kravtsov_etal17,
       author = {{Kravtsov}, A.~V. and {Vikhlinin}, A.~A. and {Meshcheryakov}, A.~V.},
        title = "{Stellar Mass{\textemdash}Halo Mass Relation and Star Formation Efficiency in High-Mass Halos}",
      journal = {Astronomy Letters},
         year = 2018,
        month = jan,
       volume = {44},
       number = {1},
        pages = {8-34},
          doi = {10.1134/S1063773717120015},
archivePrefix = {arXiv},
       eprint = {1401.7329},
 primaryClass = {astro-ph.CO},
       adsurl = {https://ui.adsabs.harvard.edu/abs/2018AstL...44....8K}
}

@ARTICLE{chiu_etal16,
       author = {{Chiu}, I. and {Saro}, A. and {Mohr}, J. and {Desai}, S. and {Bocquet}, S. and {Capasso}, R. and {Gangkofner}, C. and {Gupta}, N. and {Liu}, J.},
        title = "{Stellar mass to halo mass scaling relation for X-ray-selected low-mass galaxy clusters and groups out to redshift z {\ensuremath{\approx}} 1}",
      journal = {\mnras},
         year = 2016,
        month = may,
       volume = {458},
       number = {1},
        pages = {379-393},
          doi = {10.1093/mnras/stw292},
archivePrefix = {arXiv},
       eprint = {1512.01244},
 primaryClass = {astro-ph.GA},
       adsurl = {https://ui.adsabs.harvard.edu/abs/2016MNRAS.458..379C}
}

@ARTICLE{andreon10,
       author = {{Andreon}, S.},
        title = "{The stellar mass fraction and baryon content of galaxy clusters and groups}",
      journal = {\mnras},
         year = 2010,
        month = sep,
       volume = {407},
       number = {1},
        pages = {263-276},
          doi = {10.1111/j.1365-2966.2010.16856.x},
archivePrefix = {arXiv},
       eprint = {1004.2785},
 primaryClass = {astro-ph.CO},
       adsurl = {https://ui.adsabs.harvard.edu/abs/2010MNRAS.407..263A}
}

@ARTICLE{akino_etal22,
       author = {{Akino}, Daichi and {Eckert}, Dominique and {Okabe}, Nobuhiro and {Sereno}, Mauro and {Umetsu}, Keiichi and {Oguri}, Masamune and {Gastaldello}, Fabio and {Chiu}, I. -Non and {Ettori}, Stefano and {Evrard}, August E.},
        title = "{HSC-XXL: Baryon budget of the 136 XXL groups and clusters}",
      journal = {Publications of the Astronomical Society of Japan},
         year = 2022,
        month = feb,
       volume = {74},
       number = {1},
        pages = {175-208},
          doi = {10.1093/pasj/psab115},
archivePrefix = {arXiv},
       eprint = {2111.10080},
 primaryClass = {astro-ph.CO},
       adsurl = {https://ui.adsabs.harvard.edu/abs/2022PASJ...74..175A}
}

@ARTICLE{pratt_etal09,
       author = {{Pratt}, G.~W. and {Croston}, J.~H. and {Arnaud}, M. and {B{\"o}hringer}, H.},
        title = "{Galaxy cluster X-ray luminosity scaling relations from a representative local sample (REXCESS)}",
      journal = {\aap},
         year = 2009,
        month = may,
       volume = {498},
       number = {2},
        pages = {361-378},
          doi = {10.1051/0004-6361/200810994},
archivePrefix = {arXiv},
       eprint = {0809.3784},
 primaryClass = {astro-ph},
       adsurl = {https://ui.adsabs.harvard.edu/abs/2009A&A...498..361P}
}

@ARTICLE{chiu_etal18,
       author = {{Chiu}, I. and {Mohr}, J.~J. and {McDonald}, M. and {Bocquet}, S. and {Desai}, S. and {Klein}, M. and {Israel}, H. and {Ashby}, M.~L.~N. and {Stanford}, A. and {Benson}, B.~A. and {Brodwin}, M. and {Abbott}, T.~M.~C.},
        title = "{Baryon content in a sample of 91 galaxy clusters selected by the South Pole Telescope at 0.2 <z < 1.25}",
      journal = {\mnras},
         year = 2018,
        month = aug,
       volume = {478},
       number = {3},
        pages = {3072-3099},
          doi = {10.1093/mnras/sty1284},
archivePrefix = {arXiv},
       eprint = {1711.00917},
 primaryClass = {astro-ph.CO},
       adsurl = {https://ui.adsabs.harvard.edu/abs/2018MNRAS.478.3072C}
}

@ARTICLE{ghizzardi_etal21,
       author = {{Ghizzardi}, Simona and {Molendi}, Silvano and {van der Burg}, Remco and {De Grandi}, Sabrina and {Bartalucci}, Iacopo and {Gastaldello}, Fabio and {Rossetti}, Mariachiara and {Biffi}, Veronica and {Borgani}, Stefano and {Eckert}, Dominique and {Ettori}, Stefano and {Gaspari}, Massimo and {Ghirardini}, Vittorio and {Rasia}, Elena},
        title = "{Iron in X-COP: Tracing enrichment in cluster outskirts with high accuracy abundance profiles}",
      journal = {\aap},
         year = 2021,
        month = feb,
       volume = {646},
          eid = {A92},
        pages = {A92},
          doi = {10.1051/0004-6361/202038501},
archivePrefix = {arXiv},
       eprint = {2007.01084},
 primaryClass = {astro-ph.CO},
       adsurl = {https://ui.adsabs.harvard.edu/abs/2021A&A...646A..92G}
}

@ARTICLE{ettori15,
       author = {{Ettori}, S.},
        title = "{The physics inside the scaling relations for X-ray galaxy clusters: gas clumpiness, gas mass fraction and slope of the pressure profile}",
      journal = {\mnras},
         year = 2015,
        month = jan,
       volume = {446},
       number = {3},
        pages = {2629-2639},
          doi = {10.1093/mnras/stu2292},
archivePrefix = {arXiv},
       eprint = {1410.8522},
 primaryClass = {astro-ph.CO},
       adsurl = {https://ui.adsabs.harvard.edu/abs/2015MNRAS.446.2629E}
}

@ARTICLE{dev_etal24,
       author = {{Dev}, Ajay and {Driver}, Simon P. and {Meyer}, Martin and {Robotham}, Aaron and {Obreschkow}, Danail and {Popesso}, Paola and {Comparat}, Johan},
        title = "{The baryon census and the mass-density of stars, neutral gas, and hot gas as a function of halo mass}",
      journal = {\mnras},
         year = 2024,
        month = dec,
       volume = {535},
       number = {3},
        pages = {2357-2374},
          doi = {10.1093/mnras/stae2485},
archivePrefix = {arXiv},
       eprint = {2411.00456},
 primaryClass = {astro-ph.CO},
       adsurl = {https://ui.adsabs.harvard.edu/abs/2024MNRAS.535.2357D}
}

@ARTICLE{popesso_etal24,
     author = {{Popesso}, P. and {Marini}, I. and {Dolag}, K. and {Lamer}, G. and {Csizi}, B. and {Biffi}, V. and {Robothan}, A. and {Bravo}, M. and {Biviano}, A. and {Vladutescu-Zopp}, S. and {Lovisari}, L. and {Ettori}, S. and {Angelinelli}, M. and {Driver}, S. and {Toptun}, V. },
        title = "{Average X-ray properties of galaxy groups: From Milky Way-like halos to massive clusters}",
      journal = {\aap},
         year = 2025,
        month = dec,
       volume = {704},
          eid = {A278},
        pages = {A278},
          doi = {10.1051/0004-6361/202453255},
archivePrefix = {arXiv},
       eprint = {2411.17120},
 primaryClass = {astro-ph.GA},
       adsurl = {https://ui.adsabs.harvard.edu/abs/2025A&A...704A.278P}
}

@ARTICLE{cattaneo_etal25,
       author = {{Cattaneo}, A. and {Dimauro}, P. and {Koutsouridou}, I.},
        title = "{On the origin of the {\ensuremath{\Sigma}}$_{1}$-M$_{{\ensuremath{\star}}}$ quenching boundary}",
      journal = {\mnras},
         year = 2025,
        month = mar,
       volume = {537},
       number = {4},
        pages = {3929-3942},
          doi = {10.1093/mnras/staf253},
archivePrefix = {arXiv},
       eprint = {2502.04319},
 primaryClass = {astro-ph.GA},
       adsurl = {https://ui.adsabs.harvard.edu/abs/2025MNRAS.537.3929C}
}

@ARTICLE{koutsouridou_cattaneo22,
       author = {{Koutsouridou}, I. and {Cattaneo}, A.},
        title = "{Probing the link between quenching and morphological evolution}",
      journal = {\mnras},
         year = 2022,
        month = nov,
       volume = {516},
       number = {3},
        pages = {4194-4211},
          doi = {10.1093/mnras/stac2240},
archivePrefix = {arXiv},
       eprint = {2209.12883},
 primaryClass = {astro-ph.GA},
       adsurl = {https://ui.adsabs.harvard.edu/abs/2022MNRAS.516.4194K}
}

@ARTICLE{chen_etal20,
       author = {{Chen}, Zhu and {Faber}, S.~M. and {Koo}, David C. and {Somerville}, Rachel S. and {Primack}, Joel R. and {Dekel}, Avishai and {Rodr{\'\i}guez-Puebla}, Aldo and {Guo}, Yicheng and {Barro}, Guillermo and {Kocevski}, Dale D. and {van der Wel}, A. and {Woo}, Joanna and {Bell}, Eric F.},
        title = "{Quenching as a Contest between Galaxy Halos and Their Central Black Holes}",
      journal = {\apj},
         year = 2020,
        month = jul,
       volume = {897},
       number = {1},
          eid = {102},
        pages = {102},
          doi = {10.3847/1538-4357/ab9633},
archivePrefix = {arXiv},
       eprint = {1909.10817},
 primaryClass = {astro-ph.GA},
       adsurl = {https://ui.adsabs.harvard.edu/abs/2020ApJ...897..102C}
}

@ARTICLE{phipps_etal19,
       author = {{Phipps}, Frederika and {Bogd{\'a}n}, {\'A}kos and {Lovisari}, Lorenzo and {Kov{\'a}cs}, Orsolya E. and {Volonteri}, Marta and {Dubois}, Yohan},
        title = "{Expanding the Sample: The Relationship between the Black Hole Mass of BCGs and the Total Mass of Galaxy Clusters}",
      journal = {\apj},
         year = 2019,
        month = apr,
       volume = {875},
       number = {2},
          eid = {141},
        pages = {141},
          doi = {10.3847/1538-4357/ab107c},
archivePrefix = {arXiv},
       eprint = {1903.09965},
 primaryClass = {astro-ph.GA},
       adsurl = {https://ui.adsabs.harvard.edu/abs/2019ApJ...875..141P}
}

@ARTICLE{gaspari_etal19,
       author = {{Gaspari}, M. and {Eckert}, D. and {Ettori}, S. and {Tozzi}, P. and {Bassini}, L. and {Rasia}, E. and {Brighenti}, F. and {Sun}, M. and {Borgani}, S. and {Johnson}, S.~D. and {Tremblay}, G.~R. and {Stone}, J.~M. and {Temi}, P. and {Yang}, H. -Y.~K. and {Tombesi}, F. and {Cappi}, M.},
        title = "{The X-Ray Halo Scaling Relations of Supermassive Black Holes}",
      journal = {\apj},
         year = 2019,
        month = oct,
       volume = {884},
       number = {2},
          eid = {169},
        pages = {169},
          doi = {10.3847/1538-4357/ab3c5d},
archivePrefix = {arXiv},
       eprint = {1904.10972},
 primaryClass = {astro-ph.GA},
       adsurl = {https://ui.adsabs.harvard.edu/abs/2019ApJ...884..169G}
}

@ARTICLE{cavagnolo_etal09,
       author = {{Cavagnolo}, Kenneth W. and {Donahue}, Megan and {Voit}, G. Mark and {Sun}, Ming},
        title = "{Intracluster Medium Entropy Profiles for a Chandra Archival Sample of Galaxy Clusters}",
      journal = {\apjs},
         year = 2009,
        month = may,
       volume = {182},
       number = {1},
        pages = {12-32},
          doi = {10.1088/0067-0049/182/1/12},
archivePrefix = {arXiv},
       eprint = {0902.1802},
 primaryClass = {astro-ph.CO},
       adsurl = {https://ui.adsabs.harvard.edu/abs/2009ApJS..182...12C}
}

@ARTICLE{mantz_etal17,
       author = {{Mantz}, Adam B. and {Allen}, Steven W. and {Morris}, R. Glenn and {Simionescu}, Aurora and {Urban}, Ondrej and {Werner}, Norbert and {Zhuravleva}, Irina},
        title = "{The metallicity of the intracluster medium over cosmic time: further evidence for early enrichment}",
      journal = {\mnras},
         year = 2017,
        month = dec,
       volume = {472},
       number = {3},
        pages = {2877-2888},
          doi = {10.1093/mnras/stx2200},
archivePrefix = {arXiv},
       eprint = {1706.01476},
 primaryClass = {astro-ph.CO},
       adsurl = {https://ui.adsabs.harvard.edu/abs/2017MNRAS.472.2877M}
}

@ARTICLE{chen_etal07,
       author = {{Chen}, Y. and {Reiprich}, T.~H. and {B{\"o}hringer}, H. and {Ikebe}, Y. and {Zhang}, Y. -Y.},
        title = "{Statistics of X-ray observables for the cooling-core and non-cooling core galaxy clusters}",
      journal = {\aap},
         year = 2007,
        month = may,
       volume = {466},
       number = {3},
        pages = {805-812},
          doi = {10.1051/0004-6361:20066471},
archivePrefix = {arXiv},
       eprint = {astro-ph/0702482},
 primaryClass = {astro-ph},
       adsurl = {https://ui.adsabs.harvard.edu/abs/2007A&A...466..805C}
}

@ARTICLE{heldson_ponman03,
       author = {{Helsdon}, Stephen F. and {Ponman}, Trevor J.},
        title = "{X-ray bright groups and their galaxies}",
      journal = {\mnras},
         year = 2003,
        month = apr,
       volume = {340},
       number = {2},
        pages = {485-498},
          doi = {10.1046/j.1365-8711.2003.06320.x},
archivePrefix = {arXiv},
       eprint = {astro-ph/0212046},
 primaryClass = {astro-ph},
       adsurl = {https://ui.adsabs.harvard.edu/abs/2003MNRAS.340..485H}
}

@ARTICLE{osullivan_etal03,
       author = {{O'Sullivan}, Ewan and {Ponman}, Trevor J. and {Collins}, Ross S.},
        title = "{X-ray scaling properties of early-type galaxies}",
      journal = {\mnras},
         year = 2003,
        month = apr,
       volume = {340},
       number = {4},
        pages = {1375-1399},
          doi = {10.1046/j.1365-8711.2003.06396.x},
archivePrefix = {arXiv},
       eprint = {astro-ph/0301153},
 primaryClass = {astro-ph},
       adsurl = {https://ui.adsabs.harvard.edu/abs/2003MNRAS.340.1375O}
}

@ARTICLE{boroson_etal11,
       author = {{Boroson}, Bram and {Kim}, Dong-Woo and {Fabbiano}, Giuseppina},
        title = "{Revisiting with Chandra the Scaling Relations of the X-ray Emission Components (Binaries, Nuclei, and Hot Gas) of Early-type Galaxies}",
      journal = {\apj},
         year = 2011,
        month = mar,
       volume = {729},
       number = {1},
          eid = {12},
        pages = {12},
          doi = {10.1088/0004-637X/729/1/12},
archivePrefix = {arXiv},
       eprint = {1011.2529},
 primaryClass = {astro-ph.HE},
       adsurl = {https://ui.adsabs.harvard.edu/abs/2011ApJ...729...12B}
}

@ARTICLE{goulding_etal16,
       author = {{Goulding}, Andy D. and {Greene}, Jenny E. and {Ma}, Chung-Pei and {Veale}, Melanie and {Bogdan}, Akos and {Nyland}, Kristina and {Blakeslee}, John P. and {McConnell}, Nicholas J. and {Thomas}, Jens},
        title = "{The MASSIVE Survey. IV. The X-ray Halos of the Most Massive Early-type Galaxies in the Nearby Universe}",
      journal = {\apj},
         year = 2016,
        month = aug,
       volume = {826},
       number = {2},
          eid = {167},
        pages = {167},
          doi = {10.3847/0004-637X/826/2/167},
archivePrefix = {arXiv},
       eprint = {1604.01764},
 primaryClass = {astro-ph.GA},
       adsurl = {https://ui.adsabs.harvard.edu/abs/2016ApJ...826..167G}
}

@ARTICLE{kaiser86,
       author = {{Kaiser}, N.},
        title = "{Evolution and clustering of rich clusters.}",
      journal = {\mnras},
         year = 1986,
        month = sep,
       volume = {222},
        pages = {323-345},
          doi = {10.1093/mnras/222.2.323},
       adsurl = {https://ui.adsabs.harvard.edu/abs/1986MNRAS.222..323K}
}

@ARTICLE{ponman_etal99,
       author = {{Ponman}, Trevor J. and {Cannon}, Damian B. and {Navarro}, Julio F.},
        title = "{The thermal imprint of galaxy formation on X-ray clusters}",
      journal = {\nat},
         year = 1999,
        month = jan,
       volume = {397},
       number = {6715},
        pages = {135-137},
          doi = {10.1038/16410},
archivePrefix = {arXiv},
       eprint = {astro-ph/9810359},
 primaryClass = {astro-ph},
       adsurl = {https://ui.adsabs.harvard.edu/abs/1999Natur.397..135P}
}

@ARTICLE{allen_etal06,
   author = {{Allen}, S.~W. and {Dunn}, R.~J.~H. and {Fabian}, A.~C. and 
	{Taylor}, G.~B. and {Reynolds}, C.~S.},
    title = "{The relation between accretion rate and jet power in X-ray luminous elliptical galaxies}",
  journal = {\mnras},
   eprint = {arXiv:astro-ph/0602549},
     year = 2006,
    month = oct,
   volume = 372,
    pages = {21-30},
      doi = {10.1111/j.1365-2966.2006.10778.x},
   adsurl = {http://adsabs.harvard.edu/abs/2006MNRAS.372...21A}
}

@ARTICLE{bryan_norman98,
   author = {{Bryan}, G.~L. and {Norman}, M.~L.},
    title = "{Statistical Properties of X-Ray Clusters: Analytic and Numerical Comparisons}",
  journal = {\apj},
   eprint = {arXiv:astro-ph/9710107},
     year = 1998,
    month = mar,
   volume = 495,
    pages = {80-99},
      doi = {10.1086/305262},
   adsurl = {http://adsabs.harvard.edu/abs/1998ApJ...495...80B}
}

@ARTICLE{blumenthal_etal84,
   author = {{Blumenthal}, G.~R. and {Faber}, S.~M. and {Primack}, J.~R. and 
	{Rees}, M.~J.},
    title = "{Formation of galaxies and large-scale structure with cold dark matter}",
  journal = {\nat},
     year = 1984,
    month = oct,
   volume = 311,
    pages = {517-525},
   adsurl = {http://adsabs.harvard.edu/cgi-bin/nph-bib_query?bibcode=1984Natur.311..517B&db_key=AST}
}

@ARTICLE{bondi52,
    author = {{Bondi}, H.},
    title = "{On spherically symmetrical accretion}",
    journal = {\mnras},
    year = 1952,
    volume = 112,
    pages = {195-+},
    adsurl = {http://cdsads.u-strasbg.fr/cgi-bin/nph-bib_query?bibcode=1952MNRAS.112..195B&amp;db_key=AST}
}

@ARTICLE{cattaneo_teyssier07,
   author = {{Cattaneo}, A. and {Teyssier}, R.},
    title = "{AGN self-regulation in cooling flow clusters}",
  journal = {\mnras},
     year = 2007,
    month = apr,
   volume = 376,
    pages = {1547-1556},
      doi = {10.1111/j.1365-2966.2007.11512.x},
   adsurl = {http://adsabs.harvard.edu/abs/2007MNRAS.376.1547C}
}

@ARTICLE{cattaneo_etal09,
   author = {{Cattaneo}, A. and {Faber}, S.~M. and {Binney}, J. and {Dekel}, A. and 
	{Kormendy}, J. and {Mushotzky}, R. and {Babul}, A. and {Best}, P.~N. and 
	{Br{\"u}ggen}, M. and {Fabian}, A.~C. and {Frenk}, C.~S. and 
	{Khalatyan}, A. and {Netzer}, H. and {Mahdavi}, A. and {Silk}, J. and 
	{Steinmetz}, M. and {Wisotzki}, L.},
    title = "{The role of black holes in galaxy formation and evolution}",
  journal = {\nat},
archivePrefix = "arXiv",
   eprint = {0907.1608},
 primaryClass = "astro-ph.CO",
     year = 2009,
    month = jul,
   volume = 460,
    pages = {213-219},
      doi = {10.1038/nature08135},
   adsurl = {http://adsabs.harvard.edu/abs/2009Natur.460..213C}
}

@ARTICLE{cattaneo_etal20,
       author = {{Cattaneo}, A. and {Koutsouridou}, I. and {Tollet}, E. and
         {Devriendt}, J. and {Dubois}, Y.},
        title = "{GalICS 2.1: a new semianalytic model for cold accretion, cooling, feedback and their roles in galaxy formation}",
      journal = {\mnras},
         year = 2020,
        month = jul,
          doi = {10.1093/mnras/staa1832},
archivePrefix = {arXiv},
       eprint = {2005.05958},
 primaryClass = {astro-ph.GA},
       adsurl = {https://ui.adsabs.harvard.edu/abs/2020MNRAS.tmp.2025C}
}

@ARTICLE{davis_etal18,
       author = {{Davis}, Benjamin L. and {Graham}, Alister W. and {Cameron}, Ewan},
        title = "{Black Hole Mass Scaling Relations for Spiral Galaxies. II. M $_{BH}$-M $_{*,tot}$ and M $_{BH}$-M $_{*,disk}$}",
      journal = {\apj},
         year = "2018",
        month = "Dec",
       volume = {869},
       number = {2},
          eid = {113},
        pages = {113},
          doi = {10.3847/1538-4357/aae820},
archivePrefix = {arXiv},
       eprint = {1810.04888},
 primaryClass = {astro-ph.GA},
       adsurl = {https://ui.adsabs.harvard.edu/abs/2018ApJ...869..113D}
}

@ARTICLE{sahu_etal19,
       author = {{Sahu}, Nandini and {Graham}, Alister W. and {Davis}, Benjamin L.},
        title = "{Black Hole Mass Scaling Relations for Early-type Galaxies. I. M $_{BH}$─M $_{*,}$ $_{sph}$ and M $_{BH}$─M $_{*,gal}$}",
      journal = {\apj},
         year = "2019",
        month = "May",
       volume = {876},
       number = {2},
          eid = {155},
        pages = {155},
          doi = {10.3847/1538-4357/ab0f32},
archivePrefix = {arXiv},
       eprint = {1903.04738},
 primaryClass = {astro-ph.GA},
       adsurl = {https://ui.adsabs.harvard.edu/abs/2019ApJ...876..155S}
}

@ARTICLE{dekel_etal09,
   author = {{Dekel}, A. and {Birnboim}, Y. and {Engel}, G. and {Freundlich}, J. and 
	{Goerdt}, T. and {Mumcuoglu}, M. and {Neistein}, E. and {Pichon}, C. and 
	{Teyssier}, R. and {Zinger}, E.},
    title = "{Cold streams in early massive hot haloes as the main mode of galaxy formation}",
  journal = {\nat},
archivePrefix = "arXiv",
   eprint = {0808.0553},
     year = 2009,
    month = jan,
   volume = 457,
    pages = {451-454},
      doi = {10.1038/nature07648},
   adsurl = {http://adsabs.harvard.edu/abs/2009Natur.457..451D}
}

@ARTICLE{dekel_birnboim06,
   author = {{Dekel}, A. and {Birnboim}, Y.},
    title = "{Galaxy bimodality due to cold flows and shock heating}",
  journal = {\mnras},
   eprint = {astro-ph/0412300},
     year = 2006,
    month = may,
   volume = 368,
    pages = {2-20},
      doi = {10.1111/j.1365-2966.2006.10145.x},
   adsurl = {http://adsabs.harvard.edu/abs/2006MNRAS.368....2D}
}

@ARTICLE{dutton_etal10,
   author = {{Dutton}, A.~A. and {Conroy}, C. and {van den Bosch}, F.~C. and 
	{Prada}, F. and {More}, S.},
    title = "{The kinematic connection between galaxies and dark matter haloes}",
  journal = {\mnras},
archivePrefix = "arXiv",
   eprint = {1004.4626},
 primaryClass = "astro-ph.CO",
     year = 2010,
    month = sep,
   volume = 407,
    pages = {2-16},
      doi = {10.1111/j.1365-2966.2010.16911.x},
   adsurl = {http://adsabs.harvard.edu/abs/2010MNRAS.407....2D}
}

@ARTICLE{dutton_maccio14,
   author = {{Dutton}, A.~A. and {Macci{\`o}}, A.~V.},
    title = "{Cold dark matter haloes in the Planck era: evolution of structural parameters for Einasto and NFW profiles}",
  journal = {\mnras},
archivePrefix = "arXiv",
   eprint = {1402.7073},
     year = 2014,
    month = jul,
   volume = 441,
    pages = {3359-3374},
      doi = {10.1093/mnras/stu742},
   adsurl = {http://adsabs.harvard.edu/abs/2014MNRAS.441.3359D}
}

@ARTICLE{ghirardini_etal19,
       author = {{Ghirardini}, V. and {Ettori}, S. and {Eckert}, D. and {Molendi}, S.},
        title = "{Polytropic state of the intracluster medium in the X-COP cluster sample}",
      journal = {\aap},
         year = 2019,
        month = jul,
       volume = {627},
          eid = {A19},
        pages = {A19},
          doi = {10.1051/0004-6361/201834875},
archivePrefix = {arXiv},
       eprint = {1906.00977},
 primaryClass = {astro-ph.CO},
       adsurl = {https://ui.adsabs.harvard.edu/abs/2019A&A...627A..19G}
}

@ARTICLE{huertas_etal15,
       author = {{Huertas-Company}, M. and {P{\'e}rez-Gonz{\'a}lez}, P.~G. and {Mei}, S. and
         {Shankar}, F. and {Bernardi}, M. and {Daddi}, E. and {Barro}, G. and
         {Cabrera-Vives}, G. and {Cattaneo}, A. and {Dimauro}, P. and
         {Gravet}, R.},
        title = "{The Morphologies of Massive Galaxies from z \raisebox{-0.5ex}\textasciitilde 3{\textemdash}Witnessing the Two Channels of Bulge Growth}",
      journal = {\apj},
         year = "2015",
        month = "Aug",
       volume = {809},
       number = {1},
          eid = {95},
        pages = {95},
          doi = {10.1088/0004-637X/809/1/95},
archivePrefix = {arXiv},
       eprint = {1506.03084},
 primaryClass = {astro-ph.GA},
       adsurl = {https://ui.adsabs.harvard.edu/abs/2015ApJ...809...95H}
}

@ARTICLE{ilbert_etal13,
   author = {{Ilbert}, O. and {McCracken}, H.~J. and {Le F{\`e}vre}, O. and 
	{Capak}, P. and {Dunlop}, J. and {Karim}, A. and {Renzini}, M.~A. and 
	{Caputi}, K. and {Boissier}, S. and {Arnouts}, S. and {Aussel}, H. and 
	{Comparat}, J. and {Guo}, Q. and {Hudelot}, P. and {Kartaltepe}, J.},
    title = "{Mass assembly in quiescent and star-forming galaxies since z {\sime} 4 from UltraVISTA}",
  journal = {\aap},
archivePrefix = "arXiv",
   eprint = {1301.3157},
 primaryClass = "astro-ph.CO",
     year = 2013,
    month = aug,
   volume = 556,
      eid = {A55},
    pages = {A55},
      doi = {10.1051/0004-6361/201321100},
   adsurl = {http://adsabs.harvard.edu/abs/2013A%26A...556A..55I}
}

@ARTICLE{kauffmann_etal03,
    author = {{Kauffmann}, G. and {Heckman}, T.~M. and {Tremonti}, C. and 
        {Brinchmann}, J. and {Charlot}, S. and {White}, S.~D.~M. and 
        {Ridgway}, S.~E. and {Brinkmann}, J. and {Fukugita}, M. and 
        {Hall}, P.~B. and {Ivezi{\' c}}, {\v Z}. and {Richards}, G.~T. and 
        {Schneider}, D.~P.},
    title = "{The host galaxies of active galactic nuclei}",
    journal = {\mnras},
    year = 2003,
    month = dec,
    volume = 346,
    pages = {1055-1077},
    adsurl = {http://cdsads.u-strasbg.fr/cgi-bin/nph-bib_query?bibcode=2003MNRAS.346.1055K&amp;db_key=AST}
}

@ARTICLE{keres_etal05,
   author = {{Kere{\v s}}, D. and {Katz}, N. and {Weinberg}, D.~H. and {Dav{\'e}}, R.
	},
    title = "{How do galaxies get their gas?}",
  journal = {\mnras},
     year = 2005,
    month = oct,
   volume = 363,
    pages = {2-28},
      doi = {10.1111/j.1365-2966.2005.09451.x},
   adsurl = {http://adsabs.harvard.edu/cgi-bin/nph-bib_query?bibcode=2005MNRAS.363....2K&db_key=AST}
}

@ARTICLE{komatsu_seljak01,
   author = {{Komatsu}, E. and {Seljak}, U.},
    title = "{Universal gas density and temperature profile}",
  journal = {\mnras},
   eprint = {astro-ph/0106151},
     year = 2001,
    month = nov,
   volume = 327,
    pages = {1353-1366},
      doi = {10.1046/j.1365-8711.2001.04838.x},
   adsurl = {http://adsabs.harvard.edu/abs/2001MNRAS.327.1353K}
}

@ARTICLE{mcnamara_nulsen07,
   author = {{McNamara}, B.~R. and {Nulsen}, P.~E.~J.},
    title = "{Heating Hot Atmospheres with Active Galactic Nuclei}",
  journal = {\araa},
archivePrefix = "arXiv",
   eprint = {0709.2152},
     year = 2007,
    month = sep,
   volume = 45,
    pages = {117-175},
      doi = {10.1146/annurev.astro.45.051806.110625},
   adsurl = {http://adsabs.harvard.edu/abs/2007ARA%26A..45..117M}
}

@ARTICLE{navarro_etal97,
   author = {{Navarro}, J.~F. and {Frenk}, C.~S. and {White}, S.~D.~M.},
    title = "{A Universal Density Profile from Hierarchical Clustering}",
  journal = {\apj},
     year = 1997,
    month = dec,  
   volume = 490,
    pages = {493-+},
      doi = {10.1086/304888},
   adsurl = {http://adsabs.harvard.edu/cgi-bin/nphbib_querybibcode=1997ApJ...490..493N&db_key=AST}
}

@ARTICLE{ocvirk_etal08,
   author = {{Ocvirk}, P. and {Pichon}, C. and {Teyssier}, R.},
    title = "{Bimodal gas accretion in the Horizon-MareNostrum galaxy formation simulation}",
  journal = {\mnras},
archivePrefix = "arXiv",
   eprint = {0803.4506},
     year = 2008,
    month = nov,
   volume = 390,
    pages = {1326-1338},
      doi = {10.1111/j.1365-2966.2008.13763.x},
   adsurl = {http://adsabs.harvard.edu/abs/2008MNRAS.390.1326O}
}

@ARTICLE{papastergis_etal12,
   author = {{Papastergis}, E. and {Cattaneo}, A. and {Huang}, S. and {Giovanelli}, R. and 
	{Haynes}, M.~P.},
    title = "{A Direct Measurement of the Baryonic Mass Function of Galaxies and Implications for the Galactic Baryon Fraction}",
  journal = {\apj},
archivePrefix = "arXiv",
   eprint = {1208.5229},
 primaryClass = "astro-ph.CO",
     year = 2012,
    month = nov,
   volume = 759,
      eid = {138},
    pages = {138},
      doi = {10.1088/0004-637X/759/2/138},
   adsurl = {http://adsabs.harvard.edu/abs/2012ApJ...759..138P}
}

@ARTICLE{peterson_fabian06,
   author = {{Peterson}, J.~R. and {Fabian}, A.~C.},
    title = "{X-ray spectroscopy of cooling clusters}",
  journal = {Phys. Rep.},
   eprint = {astro-ph/0512549},
     year = 2006,
    month = apr,
   volume = 427,
    pages = {1-39},
      doi = {10.1016/j.physrep.2005.12.007},
   adsurl = {http://adsabs.harvard.edu/abs/2006PhR...427....1P}
}

@ARTICLE{rafferty_etal06,
   author = {{Rafferty}, D.~A. and {McNamara}, B.~R. and {Nulsen}, P.~E.~J. and 
	{Wise}, M.~W.},
    title = "{The Feedback-regulated Growth of Black Holes and Bulges through Gas Accretion and Starbursts in Cluster Central Dominant Galaxies}",
  journal = {\apj},
   eprint = {arXiv:astro-ph/0605323},
     year = 2006,
    month = nov,
   volume = 652,
    pages = {216-231},
      doi = {10.1086/507672},
   adsurl = {http://adsabs.harvard.edu/abs/2006ApJ...652..216R}
}

@ARTICLE{sandage86,
   author = {{Sandage}, A.},
    title = "{Star formation rates, galaxy morphology, and the Hubble sequence}",
  journal = {\aap},
     year = 1986,
    month = jun,
   volume = 161,
    pages = {89-101},
   adsurl = {http://cdsads.u-strasbg.fr/cgi-bin/nph-bib_query?bibcode=1986A%26A...161...89S&db_key=AST}
}

@ARTICLE{sutherland_dopita93,
    author = {{Sutherland}, R.~S. and {Dopita}, M.~A.},
    title = "{Cooling functions for low-density astrophysical plasmas}",
    journal = {\apjs},
    year = 1993,
    month = sep,
    volume = 88,
    pages = {253-327},
    adsurl = {http://cdsads.u-strasbg.fr/cgi-bin/nph-bib_query?bibcode=1993ApJS...88..253S&amp;db_key=AST}
}

@ARTICLE{thomas_etal05,
   author = {{Thomas}, D. and {Maraston}, C. and {Bender}, R. and {Mendes de Oliveira}, C.
	},
    title = "{The Epochs of Early-Type Galaxy Formation as a Function of Environment}",
  journal = {\apj},
   eprint = {astro-ph/0410209},
     year = 2005,
    month = mar,
   volume = 621,
    pages = {673-694},
      doi = {10.1086/426932},
   adsurl = {http://cdsads.u-strasbg.fr/cgi-bin/nph-bib_query?bibcode=2005ApJ...621..673T&db_key=AST}
}

@ARTICLE{tollet_etal19,
   author = {{Tollet}, {\'E}. and {Cattaneo}, A. and {Macci{\`o}}, A.~V. and 
	{Dutton}, A.~A. and {Kang}, X.},
    title = "{NIHAO XIX: how supernova feedback shapes the galaxy baryon cycle}",
  journal = {\mnras},
archivePrefix = "arXiv",
   eprint = {1902.03888},
     year = 2019,
    month = may,
   volume = 485,
    pages = {2511-2531},
      doi = {10.1093/mnras/stz545},
   adsurl = {http://adsabs.harvard.edu/abs/2019MNRAS.485.2511T}
}

@ARTICLE{vikhlinin_etal06,
       author = {{Vikhlinin}, A. and {Kravtsov}, A. and {Forman}, W. and {Jones}, C. and
         {Markevitch}, M. and {Murray}, S.~S. and {Van Speybroeck}, L.},
        title = "{Chandra Sample of Nearby Relaxed Galaxy Clusters: Mass, Gas Fraction, and Mass-Temperature Relation}",
      journal = {\apj},
         year = 2006,
        month = apr,
       volume = {640},
       number = {2},
        pages = {691-709},
          doi = {10.1086/500288},
archivePrefix = {arXiv},
       eprint = {astro-ph/0507092},
 primaryClass = {astro-ph},
       adsurl = {https://ui.adsabs.harvard.edu/abs/2006ApJ...640..691V}
}

@ARTICLE{voit_donahue05,
   author = {{Voit}, G.~M. and {Donahue}, M.},
    title = "{An Observationally Motivated Framework for AGN Heating of Cluster Cores}",
  journal = {\apj},
   eprint = {astro-ph/0509176},
     year = 2005,
    month = dec,
   volume = 634,
    pages = {955-963},
      doi = {10.1086/497063},
   adsurl = {http://adsabs.harvard.edu/cgi-bin/nph-bib_query?bibcode=2005ApJ...634..955V&db_key=AST}
}

@ARTICLE{voit_etal05,
       author = {{Voit}, G. Mark and {Kay}, Scott T. and {Bryan}, Greg L.},
        title = "{The baseline intracluster entropy profile from gravitational structure formation}",
      journal = {\mnras},
         year = 2005,
        month = dec,
       volume = {364},
       number = {3},
        pages = {909-916},
          doi = {10.1111/j.1365-2966.2005.09621.x},
archivePrefix = {arXiv},
       eprint = {astro-ph/0511252},
 primaryClass = {astro-ph},
       adsurl = {https://ui.adsabs.harvard.edu/abs/2005MNRAS.364..909V}
}

@ARTICLE{yang_reynolds16,
   author = {{Yang}, H.-Y.~K. and {Reynolds}, C.~S.},
    title = "{Interplay Among Cooling, AGN Feedback, and Anisotropic Conduction in the Cool Cores of Galaxy Clusters}",
  journal = {\apj},
archivePrefix = "arXiv",
   eprint = {1512.05796},
     year = 2016,
    month = feb,
   volume = 818,
      eid = {181},
    pages = {181},
      doi = {10.3847/0004-637X/818/2/181},
   adsurl = {http://adsabs.harvard.edu/abs/2016ApJ...818..181Y}
}

\appendix

\section{Cosmological definitions}
\label{sect:over}

In this article, we assume a flat $\Lambda$ cold dark matter (LCDM) cosmology.
The assumed values of the cosmological parameters at $z=0$ are 
$\Omega_{\rm M}=0.3$ for the matter density parameter, 
$\Omega_\Lambda=0.7$ for the cosmological constant density parameter and 
$H_0=70{\rm\,km\,s^{-1}\,Mpc^{-1}}$ for the Hubble constant.
The Hubble constant at $z>0$ is given by $H=E(z)H_0$, where:
\begin{equation}
E(z)=[\Omega_{\rm M}(1+z)^3+\Omega_\Lambda]^{1\over 2}.
\label{Ez}
\end{equation}

Let $r_{\Delta}$ be the radius within which the mean density equals $\Delta$ times the critical density of the Universe
$\rho_{\rm c}=3H^2/(8\pi G)$, where $G$ is the gravitational constant. 
The mass within $r_{\Delta}$ is, by definition: 
\begin{equation}
M_{\Delta}={4\pi \over 3}(\Delta\cdot\rho_{\rm c})r_{\Delta}^3.
\label{M200}
\end{equation}

In the observers' community, it is common to define halo masses based on a fixed density contrast, e.g. $\Delta=200$ or $\Delta=500$.
Theorists are more likely to use virial masses defined with respect to a redshift-dependence density contrast $\Delta_{\rm vir}$, which is usually computed with the formulae of \citet{bryan_norman98}.
For our cosmology, $\Delta_{\rm vir}$ ranges from $\Delta_{\rm vir}=102$ at $z=0$ to $\Delta_{\rm vir}\rightarrow 18\pi^2=178$ for $z\rightarrow\infty$.
In this article, we adopt  $\Delta=200$ as our reference value. We occasionally consider a density contrast of $\Delta=500$ or $\Delta=\Delta_{\rm vir}$ when
that is appropriate for comparison with observations or theoretical models,
respectively.

The conversion between radii and masses defined for different $\Delta$ is not straightforward because it depends
on the assumed density distribution of DM haloes.
Even if one assumes the standard density profile of \citet*{navarro_etal97}, there is still a dependence on the DM concentration.
In this article, we have used the fitting formulae of
\citet{dutton_maccio14} to determine the mean
concentration of a halo of mass $M_\Delta$
at redshift $z$. 
We use this mean concentration to pass from 
$M_{\rm vir}$ to $M_{200}$ when necessary.
At $z=0$, this choice gives
$M_{200}\simeq 0.85M_{\rm vir}$
over the entire mass range
$10^{12}{\rm\,M}_\odot\lesssim M_{200}\lesssim10^{15}{\rm\,M}_\odot$.

The mean particle number density within $r_{200}$ is:
 \begin{equation}
 n_\Delta={ f_{\rm g}\Delta\cdot\rho_{\rm c}\over \mu {\rm m_p}},
 \label{n200}
 \end{equation}
where $f_{\rm g}$ is the mass fraction in hot gas within $r_{200}$ and $\mu m_{\rm p}$  is the mean particle mass in the hot gas ($m_{\rm p}$ is the mass of the proton).

We define the characteristic temperature of the hot gas as in \citet{voit_etal05}:
\begin{equation}
T_\Delta\equiv{\mu {\rm m_p}\over 2k}v_\Delta^2,
\label{T200}
\end{equation}
where $k$ is the Boltzmann constant and:
\begin{equation}
v_\Delta\equiv\sqrt{GM_\Delta\over r_\Delta}.
\label{v200}
\end{equation}
We note that \citet{komatsu_seljak01}
had used an alternative definition of $T_\Delta$ with $3$ instead of $2$ at the denominator.

Substituting Eq.~(\ref{M200}) into Eq.~(\ref{T200}) 
and then using Eq.~(\ref{T200}) again gives:
\begin{equation}
T_\Delta\propto (\Delta\cdot\rho_{\rm c})r_\Delta^2\propto \Delta\cdot E^2(z) r_\Delta^2\propto\Delta\cdot E^2(z)\left({M_\Delta\over T_\Delta}\right)^2.
\label{TDeltaz}
\end{equation}
Solving for $T_\Delta$ gives the temperature dependence on halo mass and redshift:
\begin{equation}
    T_\Delta \propto \Delta^{1\over 3}E^{2\over 3}M_\Delta^{2\over 3}.
    \label{TDelta}
\end{equation}

At a given redshift (at constant $E$ and $\Delta$),
Eqs.~(\ref{TDeltaz}) and (\ref{TDelta}) give 
$r_\Delta\propto T_\Delta^{1/2}$ and $M_\Delta \propto T_\Delta^{3/2}$, respectively.

Table~\ref{glossary1} to \ref{glossary3} provide a glossary of all constants, variables and parameters used in this article.

\begin{table}
\caption{Physical constants, cosmological parameters and halo definitions}
\footnotesize
\setlength{\tabcolsep}{3pt}
\renewcommand{\arraystretch}{0.95}
\begin{tabular}{ll}
\hline
\hline
$c$ & speed of light\\
$G$ &gravitational constant\\
$k$ &Boltzmann constant \\
$m_{\rm p}$ & Proton mass \\
$\mu m_{\rm p}$ & Mean particle mass ($\mu=0.59$) \\
$\Omega_{\rm M}$ & Matter density parameter \\
$\Omega_\Lambda$ & Cosmological constant density parameter \\
$f_{\rm b}$ & Cosmic baryon fraction \\
$H_0$ & Hubble constant at $z=0$ \\
$H(z)$ & Hubble constant at redshift $z$ \\
$E$ & $H(z)/H_0$ \\
$t_{\rm H}$ & Hubble time $H_0^{-1}$ \\
$\rho_{\rm c}$ & Critical density, $3H_0^2/(8\pi G)$ \\
$\Delta_{\rm vir}(z)$ & Virial overdensity contrast \\
$r_{200}$, $r_{500}$, $r_{\rm vir}$ & Radii enclosing $200\rho_{\rm c}$, $500\rho_{\rm c}$, and $\Delta_{\rm vir}\rho_{\rm c}$ \\
$M_{200}$, $M_{500}$, $M_{\rm vir}$ & Halo masses within $r_{200}$, $r_{500}$, $r_{\rm vir}$ \\
$M_{12}$ & $M_{200}/(10^{12}{\rm\,M}_\odot)$\\
$v_{200}$, $v_{500}$, $v_{\rm vir}$ & $(GM_\Delta/r_\Delta)^{1/2}$ \\
$T_{200}$, $T_{500}$, $T_{\rm vir}$ & $\mu m_{\rm p}v_\Delta^2/(2k)$ \\
$n_{200}$, $n_{500}$, $n_{\rm vir}$ & $f_{\rm b}\Delta\rho_{\rm c}/(\mu m_{\rm p})$ \\
$c_{200}$ & Halo concentration \\
\hline
\hline
\end{tabular}
\label{glossary1}
\end{table}

\begin{table}
\caption{Thermodynamic quantities and gas properties}
\footnotesize
\setlength{\tabcolsep}{3pt}
\renewcommand{\arraystretch}{0.95}
\begin{tabular}{ll}
\hline
\hline
$\gamma$ & Effective polytropic index \\
$Z$ & Gas metallicity \\
$\Lambda$ & Cooling function \\
$n_\eta(r)$ & Particle number density \\
$T_\eta(r)$ & Temperature profile \\
$K_\eta(r)$ & ``Entropy profile" $T_\eta n_\eta^{-2/3}$ \\
$n_0(r)$ & Baseline density profiles \\
$T_0$ & Baseline temperature profile\\
$K_0(r)$ & Baseline ``entropy'' profile\\
$n_{\rm e}$ & Electron number density \\
$\eta$ & Central entropy parameter $K(0)/K_{200}$ \\
$\eta_0$ & Baseline central entropy parameter \\
$L_{\rm bol}$ & Bolometric luminosity of the hot gas within $r_{200}$ \\
$L_{\rm 0.5-2\,keV}$ & X-ray luminosity radiated between 0.5 and 2\,keV \\
$C_{\rm 0.5-2\,keV}$ & Bolometric correction $L_{\rm bol}/L_{\rm 0.5-2\,keV}$\\
$Q_{\rm cool}$ & Radiative cooling power \\
$Q_{\rm heat}$ & Heating rate from feedback \\
$T_{\rm cool}$ & Emission-weighted temperature \\
$T_{\rm heat}$ & Temperature of the gas that absorbs the feedback energy\\
$T_{\rm spec}$ & Spectroscopic temperature \\
$M_{\rm 1\,keV}$ & Halo mass $M_{200}$ for which $T_{\rm spec}=1\,$keV\\
$M_{\rm hot}$ & Total hot gas mass \\
$M_\eta(r)$ & Hot gas mass within $r$ \\
$M_{\rm cooled}$ & Cooled baryonic mass \\
$M_\star$ & Stellar mass \\
$M_{\rm cold}$ & Total mass of cold neutral gas\\
$M_{\rm HI}$ & Atomic hydrogen mass \\
$M_{\rm wind}$ & Mass ejected by winds \\
$f_{\rm hot}$ & Hot gas fraction $M_{\rm hot}/(f_{\rm b}M_{200})$ \\
$f_{\rm g}$ & Hot gas fraction within $r_{200}$ $M_\eta(r_{200})/(f_{\rm b}M_{200})$ \\
$f_{\rm cooled}$ & Cooled baryon fraction $M_{\rm cooled}/(f_{\rm b}M_{200})$ \\
$f_\star$ & Stellar mass fraction  $M_\star/(f_{\rm b}M_{200})$ \\
$f_{\rm HI}$ & HI mass fraction  $M_{\rm HI}/(f_{\rm b}M_{200})$ \\
$s$ & Entropy per particle \\
$S$, $S_0$ & Observed and baseline entropy \\
$\Delta S$ & $S - S_0$ \\
$N$ & Total number of particles in the gas\\
$U$ & Internal energy of the gas $3NkT_{200}/2$\\
$E_{\rm bind}$ & Binding energy of halo gas $(\Psi/2)f_{\rm b}M_{200}v_{200}^2$\\
$\Psi$ & Model parameter that normalises $E_{\rm bind}$\\
$\nu$ & Slope parameter in $L_{\rm bol}\propto T_{200}^{2+3\nu}$ \\
$\lambda$ & Redshift scaling exponent $L_{\rm bol}\propto E^\lambda$ \\
$\zeta$ & $|Q_{\rm cool}|/(L_{\rm bol}t_{\rm H})$ \\
\hline
\hline
\end{tabular}
\label{glossary2}
\end{table}

\begin{table}
\caption{Black hole accretion and feedback parameters}
\footnotesize
\setlength{\tabcolsep}{3pt}
\renewcommand{\arraystretch}{0.95}
\begin{tabular}{ll}
\hline
\hline
$M_\bullet$ & Black hole mass \\
$\dot{M}_\bullet^{\rm Edd}$ & Eddington accretion rate \\
$\dot{M}_\bullet$, $\dot{m}$ & Accretion rate (physical)\\
$\dot{M}_\bullet^{\rm maint}$ & Accretion rate required to offset cooling \\
$\dot{m}$ & Accretion rate (relative to Eddington's)\\
$\dot{m}^{\rm maint}$ & Accretion rate required to offset cooling (relative to Eddington's)\\
$\epsilon_{\rm accr}$ & Accretion efficiency \\
$\epsilon_{\rm heat}$ & Heating efficiency (averaged over all accretion modes)\\
$\epsilon_{\rm heat}^{\rm maint}$ & Heating efficiency in the maintenance mode\\
$\epsilon_{\rm eff}$ & Effective heating efficiency \\
\hline
\hline
\end{tabular}
\label{glossary3}
\end{table}

\section{$\Delta S/M_{200}$: behaviour and analytic fits}
\label{sect:Specificentropy}

The function
\begin{equation}
F(\eta,f_{\rm cooled})=I_\eta-I_0
\label{Feta_App}
\end{equation}
with
\begin{equation}
I_\eta\equiv\int_0^{1-f_{\rm cooled}} \ln {K_\eta(m)\over K_{200}}{\rm\, d}m
\label{Feta_App1}
\end{equation}
and
\begin{equation}
I_0\equiv\int_{f_{\rm cooled}}^1 \ln {K_0(m)\over K_{200}}{\rm\, d}m
\label{Feta_App2}
\end{equation}
gives $\Delta S/(f_{\rm b}M_{200})$ in units of $3k/(2\mu m_{\rm p})$.
Expanding the logarithms in Eqs.~(\ref{Feta_App1}) and (\ref{Feta_App2}) shows that Eq.~(\ref{Feta_App}) is identical to Eq.~(\ref{Feta}).

Figure~\ref{Kex} shows $K_\eta(m)/K_{200}$ and $K_0(m)/K_{200}$  for a particular value of $\eta>\eta_0$ chosen by example. Our choice $\eta=0.415$ was such that $I_\eta=I_0$ for $f_{\rm cooled}=0.4$. For $f_{\rm cooled}=0.4$, $\eta<\eta_{\rm crit}=0.415$ gives $F<0$, even if $\eta>\eta_0=0.19$. 

Physically,
$\eta>\eta_0$ implies $K_\eta(0)>K_0(0)$:
the lowest-entropy baryons in the observed X-ray emitting gas have higher entropy than the lowest-entropy baryons in the baseline model.
However, those baryons are not in the X-ray emitting gas today.
The minimum ``entropy'' of the gas after the lowest-entropy baryons have cooled and formed stars is $K_0(f_{\rm cooled})$.
If $K_\eta(0)<K_0(f_{\rm cooled})$, then the gas with $K\sim K_\eta(0)$
has lost entropy rather than gaining it.

We fitted $F(\eta,f_{\rm cooled})$ with the functional form:
\begin{equation}
F=A+B\ln\eta,
\label{F_fit}
\end{equation}
where the fitting parameters $A$ and $B$ depend on $f_{\rm cooled}$. 
Figure~\ref{A_and_B_fits} shows the best fit $A$ and $B$ for eleven values of $f_{\rm cooled}$
(triangles and squares, respectively).

\begin{figure}
\begin{center}
\includegraphics[width=0.99\hsize]{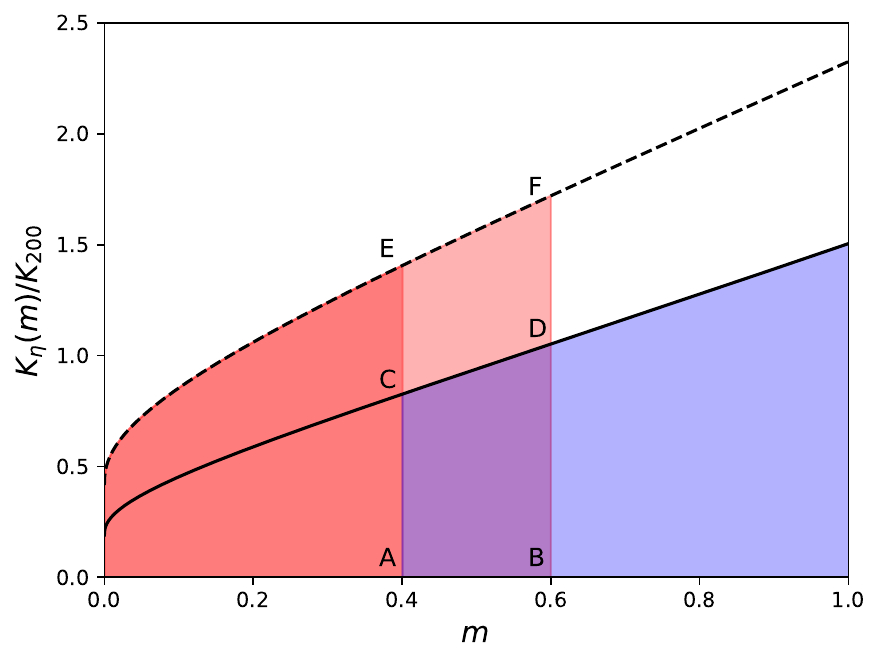} 
\end{center}
\caption{The black solid curve shows $K_0(m)/K_{200}$.
The black dashed curve shows $K_\eta(m)/K_{200}$ for $\eta=0.415$.
The red and blue shaded areas correspond to $I_\eta$ and $I_0$, respectively (Eqs.~\ref{Feta_App1} and \ref{Feta_App2}). For $f_{\rm cooled}=0.6$, the integration intervals for $I_\eta$ and $I_0$ are $(0,0.4)$ (deep red) and $(0.6,1)$ (light blue), respectively. For $f_{\rm cooled}=0.4$, the red shaded area expands up to $m=0.6$ and the blue shaded area expands down to $m=0.4$. Hence $I_\eta$ grows by the area ABFE and $I_0$ by the area ABDC. 
As $I_\eta$ grows more than $I_0$ (ABFE>ABDC), $F=I_\eta-I_0$ grows when $f_{\rm cooled}$ decreases. Increasing $\eta$ at constant $f_{\rm cooled}$ pushes the black dashed curve up while keeping the same integration intervals. The effect is to increase $I_\eta$ at constant $I_0$.
}
\label{Kex}
\end{figure}
\begin{figure}
\begin{center}
\includegraphics[width=0.99\hsize]{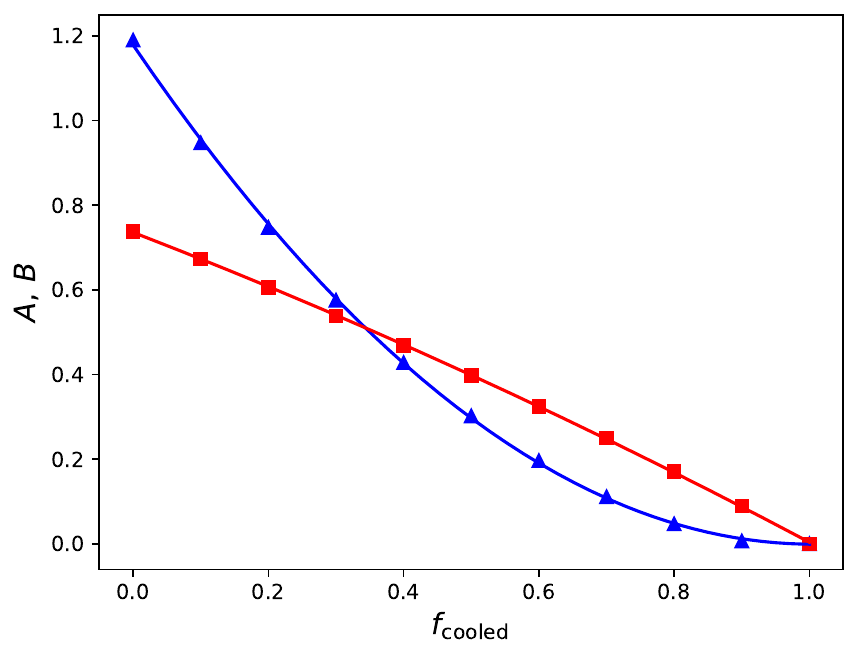} 
\end{center}
\caption{The best-fit $A$ and $B$ for six values of $f_{\rm cooled}$ (blue triangles and red squares, respectively).
The blue curve (Eq.~\ref{A_fit}) and the red curve (Eq.~\ref{B_fit}) fit the blue triangles and the red squares, respectively. 
}
\label{A_and_B_fits}
\end{figure}

We fit the dependence of $A$ and $B$ on $f_{\rm cooled}$ with second degree polynomials.
The best fits are:
\begin{equation}
A=1.1784 -2.3465f_{\rm cooled} +1.1673f_{\rm cooled}^2
\label{A_fit}
\end{equation}
and
\begin{equation}
B=0.7358 -0.6173f_{\rm cooled} -0.1147f_{\rm cooled}^2.
\label{B_fit}
\end{equation}

In this article, we start from observational data and use Eq.~(\ref{Feta_App}) to infer quantities that are not directly observable.
From $L_{\rm bol}$, we infer $\eta$, which we then use to compute $\Delta S$.
In a more theoretical approach, one could use Eq.~(\ref{Feta_App}) the other way around.
One could introduce a feedback model to compute $\Delta S$, invert Eq.~(\ref{F_fit}) to compute $\eta(F,f_{\rm cooled})$, and then use $\eta$ to predict $L_{\rm bol}$.

\section{Redshift evolution of $L_{\rm bol}(M_{vir})$}
\label{sect:calculation_lambda}

In this appendix, we investigate the evolution of the bolometric luminosity $L_{\rm bol}$ of the hot gas with redshift at a given $M_{\rm vir}$ in the GalICS semi-analytic model \citep{koutsouridou_cattaneo22,cattaneo_etal25}.
This investigation comes in four steps. 

First, we need a recipe to associate bolometric luminosities to haloes in the semi-analytic model. Bolometric luminosities are computed from Eqs.~(\ref{Lbol}), (\ref{n}) and (\ref{T}).
$L_{\rm bol}$ depends on virial quantities, which come from the N-body simulations used to construct the merger trees, on the metallicity of the hot gas, and on $\eta$, the only non-standard parameter.
We follow $\Delta S$ by assuming that ${\rm d}(\Delta S)/{\rm d}t=\epsilon_{\rm heat}\epsilon_{\rm accr}\dot{M}_\bullet c^2/T_{\rm vir}$, 
We then compute $\eta$ from $\Delta S$ and from the cooled baryon fraction $f_{\rm cooled}$ by inverting Eq.~(\ref{F_fit}):
\begin{equation}
\ln\eta={1\over B(f_{\rm cooled})} \left[{\Delta S\over f_{\rm b}M_{\rm vir}}-A(f_{\rm cooled})\right].
\end{equation}

After incorporating the calculation of $L_{\rm bol}$ into GalICS, we verified that the semi-analytic model reproduced both the slope and the normalisation of  the LMR at $z=0$.
The normalisation was a consistency check because GalICS assumed a heating efficiency $\epsilon_{\rm heat}=0.01$ that came from the local LMR,
but the slope was a genuine test, which GalICS fully passed.

Our third step was to check whether the exponent of the $L_{\rm bol}$--$M_{vir}$ relation showed any evolution with redshift.
The relation contains significant scatter and fitting it becomes difficult at redshifts $z>2$, where the number of massive haloes is small.
With this caveats in mind, the model predictions are consistent with a fixed slope $L_{\rm bol}\propto M_{\rm vir}^{1.67}$ (the expected slope for $\nu=0.17$). Only the normalisation changes with redshift.

Our final step was then to model the redshift evolution by assuming the functional form:
\begin{equation}
L_{\rm bol}(M_{\rm vir},z)=L_{\rm bol}(10^{14}{\rm\,M}_\odot,z)\left({M_{\rm vir}\over 10^{14}{\rm\,M}_\odot}\right)^{1.67}E^\lambda(z).
\end{equation}
Figure~\ref{lambda_exp} shows how the normalisation of $L_{\rm bol}(M_{\rm vir},z)$ at $M_{\rm vir}=10^{14}{\rm\,M}_\odot$ scales with $E$ (from Eq.~\ref{Ez}).
The best fit (shown by the solid line) yields an exponent $\lambda=1.999$.

For comparison, previous studies had mainly found values in the range $0.6\lesssim\lambda\lesssim 1.9$ (\citealp{sereno_etal15}, \citealp{chiu_etal22}, and references therein).
A very recent meta-analysis of a large variety of published X-ray scaling relations finds $L_{\rm bol}(M_{500},z)\propto M_{500}^{1.839}E^{2.482}$ \citep{ettori26}.

\begin{figure}
\begin{center}
\includegraphics[width=0.99\hsize]{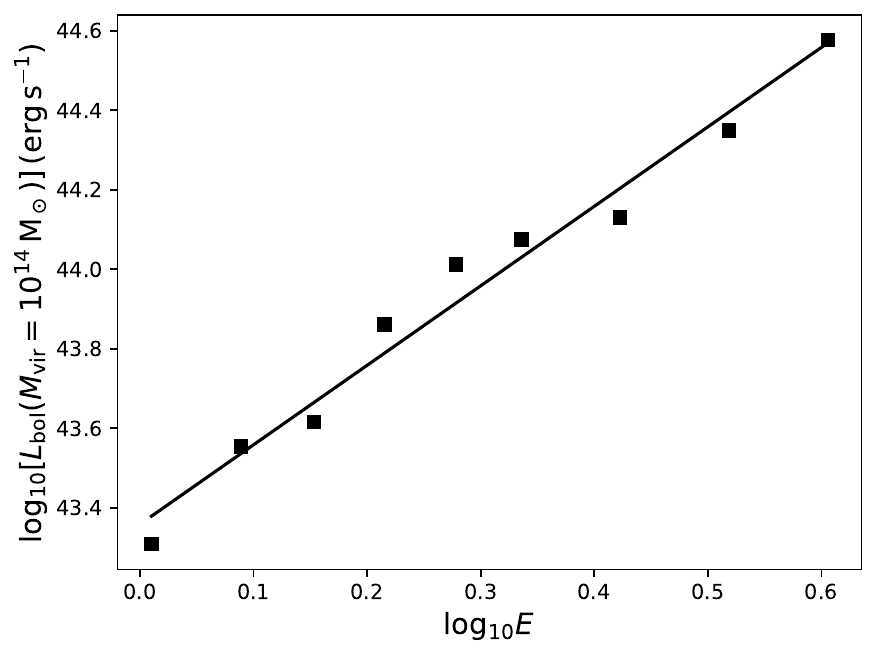} 
\end{center}
\caption{The normalisation of the $L_{\rm bol}$--$M_{\rm vir}$ relation at $M_{\rm vir}=10^{14}{\rm\,M}_\odot$ as a function of the redshift, parametrised by $E(z)$.
The symbols show the predictions of the GalICS semi-analytic model. The line is a log-log linear fit to the symbols. It corresponds to an exponent $\lambda=1.999$. }
\label{lambda_exp}
\end{figure}

\section{The calculation of $\beta_{200}$ and $\zeta$}
\label{sect:calculation_details}

In this appendix, we explain the calculation details that we had omitted in Section~\ref{sect:Qheat}
 not to interrupt the flow of the article.
Eqs.~(\ref{Qcool_int}) and (\ref{beta1}) give:
\begin{equation}
Q_{\rm cool}=\int_0^{z_{\rm q}}L_{\rm bol}(z){{\rm d}t\over{\rm d}z}{\rm d}z
\label{B1}
\end{equation}
and
\begin{equation}
    \beta_{200}={Q_{\rm cool}\over T_{200}\int_0^{z_{\rm q}}{L_{\rm bol}(z)\over T_{200}(z)}{{\rm d}t\over{\rm d}z}{\rm d}z}.
\label{Bbeta200}
\end{equation}
From the two equations above, we find:
\begin{equation}
{Q_{\rm cool}\over \beta_{200}}=\int_0^{z_{\rm q}}{T_{200}\over T_{200}(z)}L_{\rm bol}(z){{\rm d}t\over{\rm d}z}{\rm d}z.
\label{B2}
\end{equation}

Let us substitute Eq.~(\ref{Lbolz_short}) into
Eqs.~(\ref{B1}) and (\ref{B2}), evaluate
$T_{200}/T_{200}(z)$ with Eq.~(\ref{alphavir_TRINITY}), and use:
\begin{equation}
{{\rm d}t\over {\rm d}z}={t_{\rm H}\over (1+z)E(z)}
\end{equation}
where $t_{\rm H}=1/H_0$. Then Eqs.~(\ref{B1}) and (\ref{B2}) become:
\begin{equation}
Q_{\rm cool}=-L_{\rm bol}t_{\rm H}\int_0^{z_{\rm q}}\left[{M_{200}(z)\over M_{200}}\right]^{{4\over 3}+2\eta} {E^{\lambda-1}\over 1+z}{\rm\,d}z.
\label{B1bis}
\end{equation}
and
\begin{equation}
{Q_{\rm cool}\over\beta_{200}}=
-L_{\rm bol}t_{\rm H}\int_0^{z_{\rm q}}\left[{M_{200}(z)\over M_{200}}\right]^{{2\over 3}+2\eta} {E^{\lambda-{5\over 3}}\over 1+z}{\rm\,d}z.
\label{B2bis}
\end{equation}

By introducing the parameter
$\zeta\equiv-Q_{\rm cool}/(L_{\rm bol}t_{\rm H})$,
Eqs.~(\ref{B1bis}) and (\ref{B2bis})
reduce to the final forms:
\begin{equation}
\zeta=\int_0^{z_{\rm q}}\left[{M_{200}(z)\over M_{200}}\right]^{{4\over 3}+2\eta} {E^{\lambda-1}\over 1+z}{\rm\,d}z
\label{B1tris}
\end{equation}
and
\begin{equation}
{\zeta\over\beta_{200}}=
\int_0^{z_{\rm q}}\left[{M_{200}(z)\over M_{200}}\right]^{{2\over 3}+2\eta} {E^{\lambda-{5\over 3}}\over 1+z}{\rm\,d}z.
\label{B2tris}
\end{equation}
We compute $\beta_{200}$ by taking the ratio of $\zeta$
to $\zeta/\beta_{200}$.

$M_{200}(z)/M_{200}$ was read from Fig.~3 of \citet{zhang_erratum}. The number of redshifts used to sample the curves ranges from 6 for $M_{200}=8.5\times 10^{11}{\rm\,M}_\odot$ to
25 for $M_{200}=8.5\times 10^{14}{\rm\,M}_\odot$.
We fitted $M_{200}(z)/M_{200}$ with a second-degree polynomial, which we inserted into Eqs.~(\ref{B1tris}) and (\ref{B2tris}) to determine $\zeta$ and $\zeta/\beta_{200}$.
The inherent errors of reading from graphs are small compared with the  uncertainty on $M_{200}(z)/M_{200}$, given that
\citet{zhang_erratum}'s curves are systematically lower than those from
\citet{mcbride_etal09}, \citet{vandenbosch_etal14} and \citet{correa_etal15}.

\section{Comparison with previous studies}
\label{sect:comparison}

\citet{zhu_etal21} used their clusters to perform an analysis in many way similar to ours.
They also adopted the same functional form for $K(r)$ (Eq.~\ref{K}) but 
found a heating efficiency $\sim 10$ times larger than ours, although, in fairness, they had presented their estimate $\epsilon_{\rm heat}\sim 0.2$ as an upper limit.
Here we want to understand the origin of these discrepancy.
The three main differences are: i) the data used to measure $Q_{\rm heat}$,
ii) the method used to infer $Q_{\rm heat}$ from the data, and iii) the assumed BH masses.

The data used to measure  $Q_{\rm heat}$ differ but not in a way that can explain our different results.
 \citet{zhu_etal21}'s study was limited to clusters, but clusters are supposed to have the lowest $\epsilon_{\rm heat}$ (Fig.~\ref{QM}).
 Moreover,  \citet{zhu_etal21}'s clusters are systematically above our LMR (Fig.~\ref{Fig1}). Hence they should have lower entropies and require lower $\epsilon_{\rm heat}$ than we find in our analysis.
 
 The difference in $M_\bullet$ (the BH mass estimates from \citealp{phipps_etal19} are lower than those 
from \citealp{marasco_etal21} by about a factor of two)
can explain part of the discrepancy, but
the remaining factor of five to pass from $\epsilon_{\rm heat}\sim 0.2$ to
$\epsilon_{\rm heat}\sim 0.02$ must come from differences in the analysis method.
We spotted five differences but two are the significant ones.

The most significant difference with respect to the analysis is that \citet{zhu_etal21} estimated $Q_{\rm heat}=T\Delta S$ from the observed temperature $T$ of the X-ray emitting gas at $z=0$.
We considered that the gas that was heated at high redshift when $T$ was much lower.
For $\log_{10}(M_{200}/{\rm M}_\odot)=10.74$
(the mean logarithmic mass of Zhu et al.'s sample), Eq.~(\ref{alphavir_GalICS})
gives $\alpha_{200}\simeq 0.19$.
That alone can account for the factor of five that we are trying to explain.

The other main difference with respect to the analysis is that, although \citet{zhu_etal21} did not explicitly account for cooling,
 they assumed a baseline configuration with $\eta_0=0$
(instead of $\eta_0=0.19$).
That choice corresponds to assuming that
the core region was able to cool effectively before it was eventually heated.
If one lets the gas cool, it then takes more heat to bring it to the observed entropy levels.

Three other minor difference with respect to the analysis are the model for $n(r)$, \citet{zhu_etal21}'s assumption of isochoric cooling (Section~\ref{sect:method}), and the
fact that \citet{zhu_etal21}
let the parameters $a$ and $b$ in Eq.~(\ref{K}) vary instead of fixing them.
We are limited to a one-parameter model because, for a given $M_{200}$, we have only one quantity to fit: the mean bolometric luminosity of the hot gas.
\citet{zhu_etal21} dealt with individual objects, for which they had high-quality observations with Chandra, XMM-Newton and/or Suzaku.
Constraining a model with many free parameters was not a problem for them.

\citet{eckert_etal25} conducted a study similar to \citet{zhu_etal21}'s on an individual object, the fossil group SDSSTG 4436 ($M_{200}\sim 10^{14}{\rm\,M}_\odot$), presented as an extreme
case of AGN feedback.
SDSSTG 4436 displays high entropy throughout with the exception of the central few kiloparsecs, where the density matches the mean of the X-COP clusters.
For SDSSTG 4436, \citet{eckert_etal25} estimated $Q_{\rm heat}\simeq 1.5\times 10^{61}\,$erg and $M_\bullet\simeq 10^{9.5}{\rm\,M}_\odot$, which give
 $\epsilon_{\rm heat}\simeq 0.03$.

\citet{eckert_etal25}'s result agrees surprisingly well with our 
finding $0.01\lesssim\epsilon_{\rm heat}\lesssim 0.03$, especially since their method is much closer to that of \citet{zhu_etal21} than to ours.
A possible explanation is that the functional form in Eq.~(\ref{K}) is a poor fit to the ``entropy'' profile of SDSSTG 4436 (Fig.~5 of \citealp{eckert_etal25}).
Enforcing a fit of the form in Eq.~(\ref{K}) would probably lead to $\eta\sim 1$ (equivalent to ignoring the central region) and thus to a higher $Q_{\rm heat}$ than the one found by \citet{eckert_etal25}.

Our comparison with \citet{eckert_etal25} could also tell us that  SDSSTG 4436 may be extreme with respect to the norm for X-ray selected groups and clusters but  not  with respect to a broader population
of $M_{200}$-selected systems.  A truly extreme group would be so devoid of gas that it may not be detectable in X-rays.

\label{lastpage}
\end{document}